\documentclass[aps,prd,a4paper,onecolumn,amsmath,showpacs,superscriptaddress,nofootinbib,preprintnumbers,notitlepage]{revtex4-1}
 
\usepackage{verbatim}
\usepackage[T1]{fontenc}
\usepackage[utf8]{inputenc}
\usepackage[american]{babel}
\usepackage{epsfig}
\usepackage{array}
\usepackage{graphicx,subcaption,caption}
\usepackage{pgfplots}
\usepackage{booktabs}
\usepackage{multirow}
\usepackage{dcolumn}
\usepackage{amsmath}
\usepackage{mathtools}
\usepackage{amsfonts}
\usepackage{amssymb}
\usepackage{epstopdf}
\usepackage{bm}
\usepackage{siunitx}
\usepackage{braket}
\usepackage{enumitem}
\usepackage{soul}
\usepackage{color}
\usepackage{transparent}
\usepackage{pifont}
\usepackage[font={small}]{caption}

\definecolor{navyblue}{rgb}{0.0, 0.0, 0.5}
\definecolor{royalblue}{rgb}{0.25, 0.41, 0.88}
\definecolor{cadmiumgreen}{rgb}{0.0, 0.42, 0.24}
\definecolor{blue-violet}{rgb}{0.54, 0.17, 0.89}
\definecolor{darkviolet}{rgb}{0.58, 0.0, 0.83}
\definecolor{orange(colorwheel)}{rgb}{1.0, 0.5, 0.0}

\usepackage{hyperref}
\hypersetup{
    colorlinks=true, 
    linkcolor=royalblue, 
    citecolor=magenta}

\newcommand\be{\begin{equation}}
\newcommand\ee{\end{equation}}
\newcommand\bea{\begin{eqnarray}}
\newcommand\eea{\end{eqnarray}}

\newcommand\gev{\mathrm{GeV}}

\newcommand\ie{{\it i.e.}~}
\newcommand\eg{{\it e.g.}~}

\usepackage{booktabs}
\usepackage{multirow}
\usepackage{dcolumn}
\usepackage{colortbl}

\definecolor{magenta(process)}{rgb}{1.0, 0.0, 0.56}

\definecolor{darkspringgreen}{rgb}{0.09, 0.45, 0.27}

\definecolor{royalblue(web)}{rgb}{0.25, 0.41, 0.88}

\begin{document}


\title{Comparing Inflationary Models in Extended Metric-Affine Theories of Gravity}

\author{Salvatore Capozziello}
\email{capozziello@na.infn.it}
\affiliation{Dipartimento di Fisica "E. Pancini",Universit\`a degli Studi di Napoli Federico II", Via Cinthia, I-80126, Napoli, Italy}
\affiliation{Istituto Nazionale di Fisica Nucleare (INFN), sez. di Napoli, Via Cinthia 9, I-80126 Napoli, Italy}
\affiliation{Scuola Superiore Meridionale, Largo S. Marcellino 10, I-80138, Napoli, Italy}

\author{Mehdi Shokri}
\email{mehdishokriphysics@gmail.com}
\affiliation{School of Physics, Damghan University, P. O. Box 3671641167, Damghan, Iran}
\affiliation{Canadian Quantum Research Center, 204-3002 32 Ave, Vernon, BC V1T 2L7, Canada}

\preprint{}
\begin{abstract}
    We study slow-roll inflation  as a common feature in  Metric-affine Theories of Gravity. In particular, we take into account  extended metric, teleparallel and symmetric-teleparallel theories of gravity, based on different geometric invariants,  discussing analogies and differences. The analysis for each model is performed in two approaches. First, we focus on the {\it potential-slow-roll approach} by studying the reconstructed potentials for  different forms of the extended models related to the considered gravitational theory in the Einstein frame. Secondly, we investigate the {\it Hubble-slow-roll approach} for some conventional inflationary potentials related to  the specific extended model in the Jordan frame. We compare all results with cosmic microwave background anisotropy observations coming from Planck 2018 and BICEP2/Keck array satellites in order to find the observational constraints on the parameters space of the models as well as their prediction from the spectral parameters. Eventually, we attempt to present a qualitative comparison between three classes of considered modified gravities using the obtained inflationary results. The aim is to select, in principle,  cosmological signatures capable of discriminating among concurrent models in view to point out the  representation of gravity, according with geometric invariants,  which better addresses the early universe dynamics.
\end{abstract}
\date{\today}
\maketitle
\tableofcontents
\section{Introduction}
General Relativity (GR) is a successful theory describing gravitational interaction with extreme accuracy, as recently probed by black hole and gravitational wave physics. However, it presents several shortcomings at infra-red (IR) and ultra-violet (UV)  scales pointing out that the Einstein formulation is not the final picture capable of describing  gravity both at quantum and cosmic level.
For example, the late-time accelerating phase of the universe, the so-called Dark Energy (DE) epoch \cite{SupernovaSearchTeam:1998fmf,SupernovaCosmologyProject:1998vns}, remains an important open issue escaping   the standard cosmological description, built on GR. An attempt to avoid this shortcoming is adding the cosmological constant $\Lambda$ to the  matter components  in the context of the $\Lambda$-Cold Dark Matter ($\Lambda$CDM) model. This scenario claims the universe is formed by $\sim72\%$ DE, $\sim24\%$ Dark Matter (DM), and $\sim4\%$ visible baryonic matter. Despite the successes of the $\Lambda$CDM model as a realistic snapshot of the today observed universe, it suffers   ambiguities corresponding to the big difference between the predicted value of $\Lambda$, coming from theoretical physics as the vacuum value of the gravitational field, and its present observational value \cite{Sahni:1999gb,Carroll:2000fy}. Furthermore, there is no final experimental evidence of new particles capable of explaining DM and DE at a fundamental level.

Therefore,  extending the  gravitational counterpart could be the solution to address the dark side puzzle affecting astrophysics and cosmology at IR scales. Such a viewpoint navigates us towards a broad range of modified theories of gravity with interesting consequences in cosmology \cite{Capozziello:2002rd,Clifton:2011jh,Capozziello:2011et,Faraoni:2010pgm,Nojiri:2017ncd, Saridakis:2021vue, Cai:2015emx, Capozziello:2012ie, Heisenberg:2023lru}. The general idea is that  dark material components, which cannot be detected at fundamental level, could be supplemented/integrated by enlarging or modifying the gravitational sector which, in many cases, seems more suitable to correctly describe the phenomenology.

On the other hand, GR presents singularities and inconsistencies at quantum level so it is not capable of describing the gravitational interaction at UV scales. The main issue is  that there is no final Quantum Gravity theory available up to now and so effective approaches, mainly based on modified and extended gravity, seem a viable, approximate solution to the problem.   

In this context, it is worth noticing that modified/extended  gravity models describe  inflationary phases of the early universe by considering some geometrical modifications of  GR (e.g. the Starobinsky {\it the scalaron})  or some scalar field, (the  \textit{inflaton}),  dominating  dynamics  during the inflationary period. Hence, we expect that modified gravity can represent a very natural mechanism to unify DE and inflation, and then IR and UV scales,  as discussed in Ref. \cite{Nojiri:2010wj}. Besides the mentioned properties, modified gravity should be able to describe DM in the framework of the corresponding cosmological model \cite{Nojiri:2009pf,Capozziello:2017rvz}. Contrary to the above possible achievements, the price we pay for modified gravity is sometimes high through critical issues such as matter instability due to the further degrees of freedom, incompatibility with  classical  local tests, and the renormalization issue at  quantum level.

In any case, it is possible to go beyond GR  when assumptions of  the \textit{Lovelock theorem} are considered.  Such a theorem states  that  the Einstein equations are the only second-order local equations of motion for a single metric derivable from the covariant action in four-dimensional spacetime \cite{Lovelock:1971yv,Lovelock:1972vz}.  However, such a statement can be improved as soon as  further degrees of freedom, coming from extended/modified gravity are "reduced" to some effective scalar field. For example, $f(R)$ gravity, in metric formalism, has fourth-order field equations, but they can be reduced to second-order equations dealing with the further curvature components as an effective scalar field governed by a Klein-Gordon equation \cite{Faraoni:2010pgm, Capozziello:2011et}. 

Summarizing, extended/modified theories of gravity can be grouped, according to Ref. \cite{Saridakis:2021vue}, in some main families:
   \begin{itemize}
   \item \textbf{Adding Geometric Invariants.} A straightforward generalization of GR is achieved  as soon as we deal with  higher-than-two derivatives terms obtained by substituting the Ricci scalar $R$ in the Hilbert-Einstein action by  functions of curvature invariants such as $R^2,R_{\alpha\beta}R^{\alpha\beta}, R^{\alpha\beta\mu\nu}R_{\alpha\beta\mu\nu}, R \Box R, R \Box^{k}R$. Considering such terms, new degrees of freedom have to be taken into account  and they can be, in principle, capable of  explaining cosmological and astrophysical phenomenology. They emerges as effective components in the gravitational action when quantum fields in curved spaces are taken into account \cite{Birrell:1982ix,Capozziello:2011et}. In contrast, we might face some instabilities coming from the \textit{Ostrogradsky theorem}  which states that instabilities arise for theories described by  Lagrangians depending on higher order derivatives \cite{Ostrogradsky:1850fid}.   
   As the most well-known candidate, we refer to $f(R)$ gravity considering an arbitrary function of the Ricci scalar $R$ in the gravitational action and including  fourth-order derivatives of the metric where  the Ostrogradsky ghosts are avoided. Modified Gauss-Bonnet gravity or  $f(\mathcal{G})$ gravity is another possible modification to GR where  arbitrary function of the Gauss-Bonnet invariant, defined as $\mathcal{G}=R^2-4R_{\mu\nu}R^{\mu\nu}+R_{\alpha\beta\mu\nu}R^{\alpha\beta\mu\nu}$, are involved \cite{Chiba:2006jp, Bajardi:2020osh}. For a comprehensive discussion on higher-order theories of gravity,  see Ref.\cite{Faraoni:2010pgm}. For possible applications,  see \cite{Capozziello:2017xla, Capozziello:2021wwv}.

   \item \textbf{Changing Geometry.} 
   Modified theories of gravity, which are not extension of GR, emerge when we represent the dynamics of gravitational field by other geometric quantities, different from curvature, like the torsion scalar $T$ and the non-metricity scalar  $Q$. They can be dynamically equivalent to GR but with some fundamental differences in their foundation
   \cite{Capozziello:2022zzh,Capozziello:2023vne}. For example, the Teleparallel Equivalent of General Relativity (TEGR) considers torsion as the field describing gravity with zero curvature and zero non-metricity. Here the gravitational field is described by  tetrad ({\it vierbein}) fields and the  Weitzenb\"ock connection represents affinities. This means that, instead of a geodesic structure, the affine structure is relevant for dynamics \cite{Aldrovandi:2013wha}. The Symmetric Teleparallel Equivalent of General Relativity (STEGR) is another  formulation of GR dealing with non-metricity in flat spacetime with zero torsion and zero curvature. Although these three versions of GR are fully equivalent from a dynamical point of view, the basic principles on which they are formulated are very different. Furthermore, their modifications are not equivalent at all being the theories formulated with different geometries. Hence, besides $f(R)$ gravity, we can take into account $f(T)$ and $f(Q)$ gravities as modified versions of GR, TEGR, and STEGR respectively, giving the possibility that they can  be  compared by observations and experiments \cite{Cai:2015emx}. It is worth noticing, as we will see below,  that field equations derived   from  extended  theories in different geometric invariants do not  coincide as for GR, TEGR and STEGR\footnote{We have to  stress again that field equations  derived from  gravitational  actions linear in  $R$, $T$, and $Q$ give equivalent dynamics \cite{BeltranJimenez:2019esp}.} so an accurate analysis of dynamics and degrees of freedom is  necessary \cite{Capozziello:2023vne}.
   
   \item
   \textbf{Changing Dimension.} Besides the Lovelock theorem, one can generalize Einstein's gravity by introducing extra spatial dimensions. The first attempt was  the Kaluza-Klein (KK) theory where one deals with a compactified extra spatial dimension in order to unify the two fundamental forces, gravity and electromagnetism, in a five-dimensional spacetime. Despite the KK theory assuming a small size extra dimension, braneworld gravity is another class of extra-dimensional models dealing with large or infinite extra dimensions. In such a theory, our 4-dimensional universe (brane) is embedded in a bulk of extra dimensions. One of the most well-known braneworld models is the Randall-Sundrum (RS)  model \cite{Randall:1999ee} where our brane,  the matter brane, has a negative tension and one can measure the Planck mass as $10^{-19}$ GeV on the matter brane. The large distance in size  between the hidden brane, as the electroweak scale (TeV), and the matter brane,  as the Planck scale, can be explained as a curvature effect of the anti-de Sitter (AdS) bulk. Another widely-used braneworlds model is the Dvali-Gabadadze-Porrati (DGP) model \cite{Dvali:2000hr}  with  one brane, like in the case of  RS model. The main difference  with respect to the RS model is that  the bulk is Minkowski (flat), not warped, and also without the cosmological constant. Hence, there is no tension on the brane. Conventional 4-dimensional gravity is recovered on scales smaller than a crossover scale.
   
   \item \textbf{Adding New Fields.} Another possible modification of GR is adding new degrees of freedom by considering scalar, vector, or tensor fields (or even a combination of them) to the action of GR. Scalar-tensor theories are one of the most well-known modified gravity models where a scalar field is non-minimally coupled to gravity \cite{Faraoni:2004pi}. Another well-studied theory is the massive gravity \cite{deRham:2014zqa} where a fiducial metric is introduced, while in bi-gravity models \cite{Hassan:2011zd}, an additional dynamical metric is introduced. In all these picture, inflationary paradigm can be realized. Realistic models are those that better match the observational data.
   \end{itemize}

Due to the several issues that, in principle, can be addressed  by   modified theories of gravity, a wide literature  is dedicated  to  astrophysical and cosmological phenomenology stemming out from them. This shows that there is a problem of degeneration   because several models have the same aim to improve  GR at various scales. An approach to remove this shortcoming is to start from some characteristics well supported by data and then restrict the classes of viable models. This approach is successfully pursued in the recent gravitational  astronomy where  amplitude,  frequency, and polarization of gravitational waves are used to fix viable classes of theories \cite{BenAchour:2024zzk}. 

From this point of view, the inflationary   paradigm can be used: we know that early universe is well-described by an accelerated phase which solves several problems of the Standard Cosmological Model. However, the final inflationary model is not fully available also if fine data from Planck 2018 and BICEP2/Keck are restricting the classes of possible models. For example,  it seems that the Starobinsky model is one of the most reliable.

In this paper, we attempt to answer this question for three widely-used modified theories of gravity, that is  $f(R)$, $f(T)$ and $f(Q)$, the respective extensions of GR, TEGR, and STEGR. We adopt   cosmic inflation to discriminate among them in view to select the most  reliable geometric picture. Let us briefly sketch their main features.

$f(R)$ gravity is known as one of the simplest extension of GR, where we substitute the Ricci scalar $R$ with an arbitrary function $f(R)$ to explain the cosmological phenomena by introducing new degrees of freedom \cite{Capozziello:2002rd,DeFelice:2010aj}. This class of modified gravity provides, in metric formalism,  fourth-order field equations, free from the Ostrogradski ghosts. The theory can be expressed in two formalisms. The first  is the metric formalism where the field equations are obtained by varying the  $f(R)$ action   with respect to the metric $g_{\mu\nu}$. Therefore, the affine connection $\Gamma^{\alpha}{}_{\mu\nu}$ is the Levi-Civita connection depending on the metric. The second  is the Palatini formalism where $g_{\mu\nu}$ and $\Gamma^{\alpha}{}_{\mu\nu}$ are  independent variables in varying the action \cite{Palatini:1919ffw}. Although the two formalisms give the same results in GR, they produce different field equations in the context of generic $f(R)$ gravity, that is for $f(R)\neq R$. Moreover, $f(R)$ gravity, in the metric formalism, corresponds to a generalized Brans–Dicke (BD) theory
\cite{Brans:1961sx} with the BD parameter $w_{\text{BD}}=0$ \cite{OHanlon:1972xqa,Teyssandier:1983zz}. On the other hand, in the Palatini formalism, it is $w_{\text{BD}}=-3/2$
\cite{Olmo:2011uz}. 

The most known model is $f(R)=R+\alpha R^2$, where the sign of $\alpha$ depends on the metric signature to be physically compatible   (see Ref.\cite{Capozziello:2011gw} for details). It was proposed by Starobinsky in 1980 in order to describe the inflation \cite{Starobinsky:1980te}. Therefore, the correction term $\alpha R^2$ is  responsible for driving the inflationary period. Using the conformal transformation, one can reconsider the model in the Einstein frame, where the correction term $\alpha R^2$ gives rise to  a scalar field. It is the so-called Starobinsky \textit{scalaron}. Due to the excellent consistency of the Starobinsky  model with the current Cosmic Microwave Background (CMB) observations, it is now considered as a \textit{target} model for future CMB experiments as, for example, the Simons Observatory \cite{Ade:2018sbj}, CMB-S4~\cite{CMB-S4}, and the LiteBIRD satellite experiment \cite{LBIRD}. Also, $R^{2}$ model has been proposed as one of the possible alternatives to the cosmological constant of the $\Lambda$CDM model \cite{Tsujikawa:2007xu,Gannouji:2008wt,Motohashi:2009qn,Tsujikawa:2009ku,Motohashi:2010sj,Motohashi:2010tb,Motohashi:2011wy,Motohashi:2012wc}. Despite the mentioned successes, this model predicts a tiny tensor-to-scalar ratio $r\simeq0.003$ for 60 \textit{e}-folds compared with the observational value. To remove this shortcoming, a generalized form of the $R^2$ Starobinsky model, the $R^{2p}$ model, has been  proposed in Refs.\cite{Maeda:1988ab,Schmidt:1990dh}. Actually, it  provides a  generalization of  $R^{2}$ inflation as demonstrated in  \cite{Muller:1989rp,Motohashi:2014tra,Chakravarty:2014yda,Cheong:2020rao,Odintsov:2022bpg}. Moreover, the amount of gravitational waves predicted by $R^2$ Starobinski and $f(R)= R+\alpha R^{2p}$ models has been studied by considering the current uncertainties  and the possibility of an extension to the $\Lambda$CDM model  \cite{Renzi:2019ewp}. Clearly, the dimension of $\alpha$  depends on $p$. 

Another generalization  of the  Starobinsky model consists in adding a logarithmic correction, i.e.   $f(R)= R+\alpha R^{2}+\beta R^{2}\ln R$. It  can be considered a prototype model involving quantum corrections capable of describing primordial and current accelerated expansions under the same standard. Inflationary models derived from logarithmic $f(R)$ gravity have been studied in  Refs.\cite{Cognola:2007zu,Elizalde:2010ts,Ivanov:2011np,Motohashi:2012tt,Sebastiani:2013eqa,Artymowski:2014gea,Ellis:2014cma,Amin:2015lnh,Rinaldi:2014gha,Broy:2014xwa,Ben-Dayan:2014isa, Sadeghi:2015nda, Elizalde:2018now,Waeming:2020rir,Inagaki:2023mxv}. For other related cosmological topics  see Refs.\cite{Meng:2003en,Appleby:2008tv,Girones:2009nc,Guo:2013swa,Astashenok:2013vza,Alavirad:2013paa,Sadeghi:2015yyr}. Although $f(R)$ inflationary model has been  studied under the slow-roll approximation, it could be investigated also in the constant-roll regime by going beyond the slow-roll conditions \cite{Motohashi:2017vdc,Nojiri:2017qvx,Shokri:2021iqp}. Apart from the above ones, other $f(R)$ models have been  used to explain different cosmological situations. For example, the function $f(R)=R-\alpha/R^n$, with $n$ a real number, has been suggested to describe also DE in the metric formalism \cite{Capozziello:2002rd,Capozziello:2003tk,Capozziello:2003gx}. However, it is unable to fulfill some local gravity constraints as shown in \cite{Olmo:2005hc,Olmo:2005zr,Faraoni:2006hx,Erickcek:2006vf,Chiba:2006jp,Navarro:2006mw,Capozziello:2007eu}. Also in this case, the sign of $\alpha$ depends on the metric signature and its dimension on $n$. Moreover, it suffers  by matter instability \cite{Dolgov:2003px,Faraoni:2006sy} as well as  the lack of a standard matter-dominated era through a large coupling between DE and DM \cite{Amendola:2006kh,Amendola:2006eh}. The unification of DE and inflation epochs, in the context of $f(R)$ gravity, was first proposed in Ref.\cite{Nojiri:2003ft} where the positive (negative) power of curvature refers to the inflationary (DE) epoch. Moreover, the issue has been studied in
the context of modified $f(R)$ Horava-Lifshitz gravity in Ref.\cite{Elizalde:2010ep}. See also \cite{Odintsov:2019evb,Oikonomou:2020oex,Oikonomou:2020qah} for the latest developments.

$f(T)$ gravity is another widely-used modified gravity in which  the torsion $T$ scalar is considered  instead of curvature $R$  \cite{Cai:2015emx,Bahamonde:2021gfp}. Despite some similarities, $f(T)$ and $f(R)$ theories  have some important differences. First of all,  dynamical equations of $f(T)$ gravity remain second order rather than  fourth order as in  $f(R)$ gravity. Secondly, in $f(T)$ gravity, we encounter  more degrees of freedom, in comparison with $f(R)$ theory, due to the existence of a massive vector field \cite{Li:2011rn,Li:2011wu}. Thirdly, under the conformal transformation,  $f(T)$ gravity cannot be simply reformulated as the TEGR action plus a scalar field. Here the appearance of an additional scalar-torsion coupling reflects the violation of local Lorentz invariance \cite{Yang:2010ji,Bamba:2013jqa}. Moreover, the Hamiltonian analysis of $f(T)$ theory has been  discussed in Refs.\cite{Ferraro:2018tpu,Blixt:2020ekl,Blagojevic:2020dyq,Bajardi:2024qbi}. For the cosmological perturbations of $f(T)$ gravity, see Ref.\cite{Chen:2010va,Bahamonde:2020lsm, Bahamonde:2022ohm}. In \cite{Wu:2010xk}, the dynamical behaviour of $f(T)$ gravity was studied. The Born-Infeld modified teleparallel gravity was the first attempt to explain the inflationary period in the context of modified teleparallel gravity \cite{Ferraro:2006jd,Ferraro:2008ey}. It was able to solve the horizon problem in a spatially Friedman-Lema\^itre-Robertson-Walker (FLRW) metric without considering any scalar field. To describe DE, Linder proposed two interesting $f(T)$ models, \ie $f(T)=T+\alpha(-T)^p$ and 
$f(T)=T+\alpha T(1-e^{\beta/T})$ referring to a de Sitter universe \cite{Linder:2010py}. In Ref.\cite{Wu:2010mn}, the authors studied the observational constraints on the models using the Type Ia Supernovae (SneIa) set, the Baryonic Acoustic Oscillation (BAO), and the CMB photons. Also, they found that a crossing of phantom divide  is impossible for both models. In addition to the mentioned functions, some other  models of $f(T)$ gravity have been proposed in Ref.\cite{Yang:2010hw} to describe late-time cosmic acceleration. See Ref.\cite{Li:2011wu} for the cosmological results in $f(T)$ gravity. Big-Bang Nucleosynthesis (BBN) can give useful constraints on the shape of $f(T)$ function as discussed in Refs. \cite{Benetti:2020hxp, Capozziello:2017bxm}.

Besides the two discussed theories of gravity, another possible extension of GR, the so-called $f(Q)$ gravity, has been remarkably considered in  recent   literature \cite{BeltranJimenez:2019tme,Dialektopoulos:2019mtr,Dimakis:2021gby,Bajardi:2023vcc,Heisenberg:2023lru,Heisenberg:2023wgk,Nojiri:2024zab}. In this theory, the gravitational field is associated to  the non-metricity scalar $Q$ with zero torsion and curvature. Such a theory, not requiring, {\it a priori}, the Equivalence Principle, is suitable to be dealt  under the standard of gauge theories and presents other advantages. For example, it seemingly shows no strong coupling problems because of additional scalar modes \cite{Golovnev:2018wbh}, while the $f(T)$ gravity presents these problems when perturbations around the FLRW metric are considered. Also, the linear perturbations of scalar, vector, and tensor modes in $f(Q)$ gravity have been studied \cite{BeltranJimenez:2019tme, Capozziello:2024jir, Capozziello:2024vix}. The model  $f(Q)=Q+\alpha Q^p$ has been  proposed to explain early and late-time accelerating phases of the universe depending on the value of $p$ \cite{BeltranJimenez:2019tme} with $\alpha>1$ for high energy regimes (like inflation) whereas   $\alpha<1$ works for low energy regimes (like DE). In Ref.\cite{Capozziello:2022tvv},  it has been studied the slow-roll inflation in the context of $f(Q)$ gravity in both Potential-Slow-Roll (PSR) and Hubble-Slow-Roll (HSR) approaches. The present cosmic acceleration in $f(Q)$ gravity has  been investigated  in Refs.\cite{Anagnostopoulos:2021ydo,Frusciante:2021sio,Bajardi:2020fxh,Capozziello:2022wgl,Lymperis:2022oyo,Paul:2022jnq,Sokoliuk:2023ccw,Narawade:2023tnn}. 

The main purpose of the present work is   performing a detailed analysis of    $f(R)$, $f(T)$ and $f(Q)$ inflationary behaviors  and comparing  results   with the CMB anisotropies observations coming from Planck and BICEP2/Keck array datasets. The analysis of each gravity model is accomplished from the two viewpoints PSR and HSR, separately. In the PSR approach, we are going to reconstruct potentials using reliable  $f$ forms of the above models   in the Einstein frame. In the HSR approach, we deal with the Hubble parameter associated to   inflationary potentials for  specific forms of $f$ in the Jordan frame.

The layout of the paper is the following.  In Sec.\ref{emag}, we briefly present the main properties of the extended metric-affine  $f(R)$, $f(T)$ and $f(Q)$ theories considering the related cosmological models. Moreover, we discuss the conformal transformations for these gravity theories pointing out their similarities and differences. Sec.\ref{fr} is devoted to  the slow-roll inflation in $f(R)$ gravity from the  PSR and HSR viewpoints. The strategy of the previous section is developed also for $f(T)$ and $f(Q)$ in Secs.\ref{ft} and \ref{fq}, respectively. In Sec.\ref{comparison}, considering the previous results, we provide a comparison of inflationary behaviors in  the three  extended theories of gravity. Conclusions are reported in Sec.\ref{conclusion}. 

\section{Extended Metric-Affine Gravities}\label{emag}
Metric-Affine  Gravity implies  a wide class of theories where affine connection plays a prominent role to define dynamics and kinematics. For a comprehensive discussion, see Ref.\cite{Capozziello:2022zzh}. To start, we have to define 
the most general affine connection
\begin{equation}
\Gamma^{\alpha}{}_{\mu\nu}=\{^{\alpha}{}_{\mu\nu}\}+K^{\alpha}{}_{\mu\nu}+L^{\alpha}{}_{\mu\nu},
\label{1}    
\end{equation}
where the contorsion tensor $K^{\alpha}{}_{\mu\nu}$ and the disformation tensor $L^{\alpha}{}_{\mu\nu}$  are defined as follows
\begin{equation}
K^{\alpha}{}_{\mu\nu}=\frac{1}{2}g^{\alpha\lambda}(T_{\mu\lambda\nu}+T_{\nu\lambda\mu}+T_{\lambda\mu\nu}),\hspace{1cm}L^{\alpha}{}_{\mu\nu}=\frac{1}{2}g^{\alpha\lambda}(Q_{\lambda\mu\nu}-Q_{\mu\lambda\nu}-Q_{\nu\lambda\mu}).
\label{2}    
\end{equation}
The Christoffel symbols takes the   form 
\begin{equation}
\{^{\alpha}{}_{\mu\nu}\}=\frac{1}{2}g^{\alpha\lambda}(g_{\mu\lambda,\nu}+g_{\lambda\nu,\mu}-g_{\mu\nu,\lambda})\,. 
\end{equation}
Clearly, the three components define  metric, torsion and non-metric contributions to the affine connection.

Using the above definitions, we can introduce three primary tensors \ie  the Riemann, the torsion and the non-metricity tensor as
\begin{equation}
R^{\alpha}{}_{\lambda\mu\nu}\equiv2\partial_{[\mu}\Gamma^{\alpha}{}_{\nu]\lambda}+2\Gamma^{\alpha}{}_{[\mu|\beta|}\Gamma^{\beta}{}_{\nu]\lambda},\hspace{1cm}T^{\alpha}{}_{\mu\nu}\equiv\Gamma^{\alpha}{}_{\mu\nu}-\Gamma^{\alpha}{}_{\nu\mu},\hspace{1cm}
Q_{\alpha\mu\nu}\equiv\nabla_{\alpha}g_{\mu\nu},
\label{3}    
\end{equation}
where the covariant derivative of the metric takes the form 
\begin{equation}
\nabla_{\alpha}g_{\mu\nu}=\partial_{\alpha}g_{\mu\nu}-\Gamma^{\lambda}{}_{\alpha\mu}g_{\lambda\nu}-\Gamma^{\lambda}{}_{\alpha\nu}g_{\mu\lambda}.
\label{4}
\end{equation}
Also, we define parenthesis and brackets as $B_{(\alpha\mu\nu...)}=\frac{1}{n!}\sum_{p}B_{p(\alpha\mu\nu...)}$ and $B_{[\alpha\mu\nu...]}=\frac{1}{n!}\sum_{p}(-1)^{n_{p}}
B_{p(\alpha\mu\nu...)}$, respectively for symmetric and antisymmetric objects. See \cite{Capozziello:2022zzh} for details. With this formalism in mind,  let us review now the three equivalent formulations of Einstein's gravity.

In GR, the  Ricci curvature scalar $R$ is the fundamental geometric object to describe dynamics of gravity with zero torsion and non-metricity. Here the Christoffel symbols, derived from the metric,  play the role of affine connection (the Levi-Civita connection). The relation between metric and connection points out that metric (causal) structure and geodesic structure coincide. This is the main consequence of the Equivalence Principle. Contracting once and then twice, we find the corresponding definitions of the Ricci tensor
$R_{\mu\nu}=R^{\alpha}{}_{\mu\alpha\nu}$ and then of the Ricci scalar 
\begin{equation}
R=g^{\mu\nu}R_{\mu\nu}.
\label{5}
\end{equation}
Thus, the  GR action, the so-called  Hilbert-Einstein action, is given by
\begin{equation}
S_{R}=\int{d^{4}x\sqrt{-g}\Big(\frac{R}{2}+\mathcal{L}_{m}(g_{\mu\nu},\Psi_{m})\Big)},
\label{6}    
\end{equation}
where $g$ is the determinant of the metric $g_{\mu\nu}$ and $\mathcal{L}_{m}$ is the Lagrangian for   matter fields $\Psi_{m}$ assumed as  perfect fluids. Here we choose $8\pi G=c=1$.

In TEGR representation, we have  zero curvature and non-metricity. The torsion scalar is given by  
\begin{equation}
T=S_{\rho}{}^{\mu\nu}T^{\rho}{}_{\mu\nu},
\label{7}
\end{equation}
where the tensor $S_{\rho}{}^{\mu\nu}$ is the {\it torsion superpotential}. It  is given by a combination of contorsion and torsion tensors as
\begin{equation}
S_{\rho}{}^{\mu\nu}=\frac{1}{2}(K^{\mu\nu}{}_{\rho}+\delta^{\mu}_{\rho}T^{\lambda\nu}{}_{\lambda}-\delta^{\nu}_{\rho}T^{\lambda\mu}{}_{\lambda}).
\label{8}
\end{equation}
The connection is  the Weitzenb\"{o}ck one: $\Gamma^{\alpha}{}_{\mu\nu}=e^{\alpha}_{i}\partial_{\mu}e^{i}_{\nu}$ given as the affine connection where $e^{\mu}_{i} (\mu=0,1,2,3)$ are the components of the {\it vierbein} field $e_{i}(x^{\mu}) (i=0,1,2,3)$. Here the Greek letters indicate the spacetime indices while the Latin ones are the tetrad indices. In this picture, the affine structure is more fundamental than the geodesic structure.
The total action in TEGR  is 
\begin{equation}
S_{T}=\int{d^{4}x e\Big(\frac{T}{2}+\mathcal{L}_{m}(e^i_{\mu},\Psi_{m})\Big)},
\label{9}    
\end{equation}
where $e=det(e^i_{\mu})=\sqrt{-g}$. 

Finally,  STEGR is an equivalent viewpoint where gravitational interaction is given by  non-metricvity with zero curvature and torsion. Here, we deal with the non-metricity scalar 
\begin{equation}
Q=-Q_{\alpha\mu\nu}P^{\alpha\mu\nu}
,
\label{10}
\end{equation}
where the non-metricity conjecture can be related to the {\it  non-metricity conjugate tensor} 
\begin{equation}
P^{\alpha}{}_{\mu\nu}=-\frac{1}{2}L^{\alpha}{}_{\mu\nu}+\frac{1}{4}(Q^{\alpha}-\hat{Q}^{\alpha})g_{\mu\nu}-\frac{1}{4}\delta^{\alpha}_{(\mu}Q_{\nu)}.
\label{11}
\end{equation}
We can define two independent traces of non-metricity tensor  as $Q_{\alpha}=g^{\mu\nu}Q_{\alpha\mu\nu}$ and $\hat{Q}_{\alpha}=g^{\mu\nu}Q_{\mu\alpha\nu}$.  STEGR assumes the general connection  
\begin{equation}
\Gamma^{\alpha}{}_{\mu\nu}:=\frac{\partial x^{\alpha}}{\partial \xi^{\lambda}}\frac{\partial^{2}\xi^{\lambda}}{\partial x^{\mu}\partial x^{\nu}},
\label{12}
\end{equation}
where $\xi^{\lambda}=\xi^{\lambda}(x)$ is an arbitrary function of spacetime.  In principle, connection can vanish  under a generic transformation $x^{\mu}\rightarrow\xi^{\mu}(x^{\nu})$. Adapting to the so-called {\it coincident gauge}, it is always possible to determine a coordinate transformation where the affine connection vanishes ($\Gamma^{\alpha}{}_{\mu\nu}=0$) and then the non-metricity tensor reduces to $Q_{\alpha\mu\nu}=\partial_{\alpha}g_{\mu\nu}$ which is due to the reduction of the covariant derivative to the ordinary partial derivative $(\nabla_{\mu} \rightarrow\partial_{\mu})$. Moreover, the disformation tensor turns out to be
the Christoffel symbols with a negative sign as $L^{\alpha}{}_{\mu\nu}=-\Gamma^{\alpha}{}_{\mu\nu}$. The total action in STEGR takes the  form 
\begin{equation}
S_{Q}=\int{d^{4}x\sqrt{-g}\Big(\frac{Q}{2}+\mathcal{L}_{m}(g_{\mu\nu},\Psi_{m})\Big)}.
\label{13}    
\end{equation}
Despite the equivalence of the three mentioned viewpoints (the so-called Geometrical Trinity of Gravity \cite{BeltranJimenez:2019esp}), their extensions do not show the same dynamics and degrees of freedom. For example, $f(R)$ gravity, in metric formalism, presents fourth-order field equations while $f(T)$ and $f(Q)$ remain of second order. To be compared, a prominent role is played by the boundary terms \cite{Capozziello:2023vne}. In the rest of this section, we present  properties of $f(R)$, $f(T)$, and $f(Q)$ gravities as  extended versions of GR, TEGR, and STEGR, respectively and the related cosmological models.

\subsection{$f(R)$ metric gravity}
A straightforward generalization of GR is realized  when we deal with  higher-order dynamics obtained by   substituting the Ricci scalar $R$ in the Hilbert-Einstein action by a function of curvature invariants. The most well-known example is $f(R)$ gravity which considers an arbitrary function of the Ricci scalar $R$ in the gravitational
action, including the fourth-order derivatives of the metric in the field equation without the Ostrogradski ghosts. Therefore, in $f(R)$ gravity, the 4-dimensional Hilbert-Einstein action (\ref{6}) can be extended as follows \cite{Capozziello:2002rd, Capozziello:2011et, DeFelice:2010aj}
\begin{equation}
S=\int{d^{4}x\sqrt{-g} \Big( \frac{f(R)}{2}+\mathcal{L}_{m}(g_{\mu\nu},\Psi_{m})\Big)}.
\label{a1}    
\end{equation}
By varying this action with respect to the metric $g_{\mu\nu}$, the field equations are
\begin{equation}
\mathcal{G}_{\mu\nu}\equiv FR_{\mu\nu}-\frac{1}{2}g_{\mu\nu}f-\nabla_{\mu}\nabla_{\nu}F+g_{\mu\nu}\Box F=T_{\mu\nu}^{(m)},
\label{a2}
\end{equation}
where $F$, d'Alembert $\Box$ and the energy-momentum tensor of matter fields $T_{\mu\nu}^{(m)}$ are respectively given by \begin{equation}
F\equiv\frac{df(R)}{dR},\hspace{1cm}\Box\equiv\frac{1}{\sqrt{-g}}\partial_{\mu}[\sqrt{-g}g^{\mu\nu}\partial_{\nu}],\hspace{1cm} T_{\mu\nu}^{(m)}=-\frac{2}{\sqrt{-g}}\frac{\delta\mathcal{L}_{m}}{\delta g^{\mu\nu}}.
\label{a3}
\end{equation} 
By setting $\nabla^{\mu}\mathcal{G}_{\mu\nu}=0$, the conservation law of the energy-momentum tensor $\nabla^{\mu} T_{\mu\nu}^{(m)}=0$ is valid. Also,  field Eqs. (\ref{a2}) can be rewritten in the standard  form \cite{Capozziello:2002rd,Starobinsky:2007hu}
\begin{equation}
G_{\mu\nu}\equiv R_{\mu\nu}-\frac{1}{2}g_{\mu\nu}R=T_{\mu\nu}^{m}+T_{\mu\nu}^{c},
\label{a4}
\end{equation}
where 
\begin{equation}
T_{\mu\nu}^{c}\equiv (1-F)R_{\mu\nu}+\frac{1}{2}(f-R)g_{\mu\nu}+\nabla_{\mu}\nabla_{\nu}F-g_{\mu\nu}\Box F,
\label{a5}
\end{equation}
is the curvature contribution to the energy-momentum tensor.
In such a case, the conservation law of the energy-momentum tensor $\nabla^{\mu} T_{\mu\nu}^{m}=0$ is valid as a consequence of the Bianchi identities  $\nabla^{\mu} G_{\mu\nu}=0$ if $\nabla^{\mu} T_{\mu\nu}^{c}=0$. Note that in the case $f(R)=R$, we reduce to the standard GR. 

To obtain the cosmological  equations of $f(R)$ gravity, we assume a spatially flat universe ($k=0$) described by the FLRW metric 
\begin{equation}
ds^{2}=-dt^{2}+a(t)^{2}(dx^{2}+dy^{2}+dz^{2}),
\label{a6}
\end{equation}
where $a$ and $t$ are the cosmic  scale factor and cosmic time, respectively. Now, plugging the FLRW metric and  the definition of the perfect-fluid energy-momentum tensor  into  field Eqs. (\ref{a2}), we find the dynamical equations 
\begin{equation}
3FH^{2} = \frac{(F R-f)}{2}-3H\dot{F}+\rho,\hspace{1cm} 2F\dot{H}=-\ddot{F}+H\dot{F}-(\rho+p), 
\label{a7}
\end{equation}
where $ H\equiv\left(\frac{\dot{a}}{a}\right)$ is the Hubble parameter and dot is the derivative with respect to the cosmic time $t$. Moreover, $\rho$ and $p$ represents energy density and pressure of the perfect fluid, respectively. In the presence of a barotropic fluid, the above expressions can be rewritten as
\begin{equation}
H^{2}=\frac{1}{3}\rho_{eff},\hspace{1cm}2\dot{H}+3H^{2}=-p_{eff},
\label{a8}   
\end{equation}
where the effective energy density $\rho_{eff}$ and effective pressure $p_{eff}$ of the fluid are defined as
\begin{equation}
\rho_{eff}=\frac{2(\rho_{m}+\rho_{\varphi})-6H\dot{F}+(FR-f)}{2F},\hspace{1cm}p_{eff}=\frac{2(p_{m}+p_{\varphi})+4H\dot{F}+2\ddot{F}-(FR-f)}{2F}.
\label{a9}
\end{equation}
Here we are considering the contribution of standard matter $m$, any scalar field $\varphi$ and curvature. 
Then, we can define   the effective equation of state (EoS) as
\begin{equation}
w_{eff}=\frac{p_{eff}}{\rho_{eff}}=\frac{2(p_{m}+p_{\varphi})+4H\dot{F}+2\ddot{F}-(FR-f)}{2(\rho_{m}+\rho_{\varphi})-6H\dot{F}+(FR-f)}.
\label{a10}    
\end{equation}
The higher-order  $f(R)$ gravity, in the {\it Jordan frame}, can be reduced to the   {\it Einstein frame}, using the conformal transformation
\begin{equation}
\tilde{g}_{\mu\nu}=\Omega^{2}g_{\mu\nu}.
\label{a11}
\end{equation}
 It maps the field equations into a mathematically equivalent sets of equations where further degrees of freedom are decoupled \cite{Faraoni:2010pgm}. The conformal factor $\Omega=\Omega(\varphi(x))$ is a differentiable and non-zero function. The tilde indicates  parameters written in the Einstein frame. 

Under  conformal transformation (\ref{a11}), we define the following identities \cite{Maeda:1988ab,Faraoni:1998qx}
\begin{equation}
\tilde{g}^{\mu\nu}=\Omega^{-2}g^{\mu\nu},\hspace{0.75cm}\sqrt{-\tilde{g}}=\Omega^{4}\sqrt{-g}.
\label{a12}    
\end{equation}
For scalar functions, we have $\nabla_{\mu}\Omega=\partial_{\mu}\Omega$, and so $\tilde{\nabla}_{\mu}\Omega=\nabla_{\mu}\Omega$. Recall that $\tilde{\partial_{\mu}}=\partial_{\mu}$ since $x^{\mu}$ is unaffected by  conformal transformations. Using the above relations, the conformal version of the Christoffel symbols (here, $\Gamma^{\alpha}{}_{\mu\nu}=\{^{\alpha}_{\mu\nu}\}$), Riemann and Ricci tensors are given by
\begin{equation}
\tilde{\Gamma}^{\alpha}{}_{\mu\nu}=\Gamma^{\alpha}{}_{\mu\nu}+\Big(\delta^{\alpha}_{\mu}\partial_{\nu}\ln\Omega+\delta^{\alpha}_{\nu}\partial_{\mu}\ln\Omega-g_{\mu\nu}\partial^{\alpha}\ln\Omega\Big),
\label{a13}    
\end{equation}
\begin{eqnarray}
\tilde{R}_{\lambda\mu\nu}{}^{\alpha}=R_{\lambda\mu\nu}{}^{\alpha}+2\delta^{\alpha}_{[\lambda}\partial_{\mu]}\partial_{\nu}\ln\Omega-2g^{\alpha\sigma}g_{\nu[\lambda}\partial_{\mu]}\partial_{\sigma}\ln\Omega+\hspace{6cm}\nonumber\\
+2\partial_{[\lambda}\ln\Omega\delta^{\alpha}_{\mu]}\partial_{\nu}\ln\Omega-2\partial_{[\lambda}\ln\Omega g_{\mu]\nu}g^{\alpha\sigma}\partial_{\sigma}\ln\Omega-2g_{\nu[\lambda}\delta^{\alpha}_{\mu]}g^{\sigma\rho}\partial_{\sigma}\ln\Omega\partial_{\rho}\ln\Omega,\hspace{0.05cm}
\label{a14}    
\end{eqnarray}
and
\begin{equation}
\tilde{R}_{\mu\nu}=R_{\mu\nu}-2\partial_{\mu}\partial_{\nu}\ln\Omega-g_{\mu\nu}g^{\rho\sigma}\partial_{\rho}\partial_{\sigma}\ln\Omega+2\partial_{\mu}\ln\Omega\partial_{\nu}\ln\Omega-2g_{\mu\nu}g^{\rho\sigma}\partial_{\rho}\ln\Omega\partial_{\sigma}\ln\Omega.
\label{a15}    
\end{equation}
Then, the Ricci scalar in the Einstein frame  takes the  form
\begin{equation}
\tilde{R}=\Omega^{-2}\Big(R-\frac{12\Box\sqrt{\Omega}}{\sqrt{\Omega}}-3g^{\mu\nu}\partial_{\mu}\ln\Omega\partial_{\nu}\ln\Omega\Big).
\label{a16}
\end{equation}
Also, the inverse transformation of  relation (\ref{a16}) can be derived. It is
\begin{equation}
R=\Omega^{2}\Big(\tilde{R}-6\tilde{g}^{\mu\nu}\partial_{\mu}w\partial_{\nu}w+6\tilde{\Box}w\Big),
\label{a17}
\end{equation}
where $w\equiv\ln\Omega$ and ${\displaystyle \tilde{\Box}\equiv\frac{1}{\sqrt{-\tilde{g}}}\partial_{\mu}[\sqrt{-\tilde{g}}\tilde{g}^{\mu\nu}\partial_{\nu}]}$. 

Now let us rewrite the action in the Jordan frame (\ref{a1}) as follows
\begin{equation}
S_{E}=\int{d^{4}x\sqrt{-g}\Big(\frac{FR}{2}-U\Big)}+\int{d^{4}x\mathcal{L}_{m}(g_{\mu\nu},\Psi_{M})},\hspace{1cm}\text{with}\hspace{1cm}U=\frac{FR-f}{2}.
\label{a18}    
\end{equation}
Using  relation (\ref{a17}) and definitions (\ref{a12}), the above action is transformed as
\begin{equation}
S_{E}=\int{d^{4}x\sqrt{-\tilde{g}}\bigg(\frac{1}{2}F\Omega^{-2}\Big[\tilde{R}-6\tilde{g}^{\mu\nu}\partial_{\mu}w\partial_{\nu}w+6\tilde{\Box}w\Big]-\Omega^{-4}U\bigg)}+\int{d^{4}x\mathcal{L}_{m}(\Omega^{-2}\tilde{g}_{\mu\nu},\Psi_{M})}.
\label{a19}    
\end{equation}
Then, by choosing $F=\Omega^{2}$, we find the action as
\begin{equation}
S_{E}=\frac{1}{2}\int{d^{4}x\sqrt{-\tilde{g}}\bigg(\tilde{R}-6\tilde{g}^{\mu\nu}\partial_{\mu}w\partial_{\nu}w+6\tilde{\Box}w-2F^{-2}U\bigg)}+\int{d^{4}x\mathcal{L}_{m}(F^{-1}\tilde{g}_{\mu\nu},\Psi_{M})}.
\label{a20}    
\end{equation}
Compared to the standard Hilbert-Einstein action, there exists an additional term $\tilde{\Box}w$ that  vanishes thanks to the Gauss theorem. Therefore, by using the definition $\varphi\equiv\sqrt{3/2}\ln F$, the action in the Einstein frame (\ref{a20}) is given by
\begin{equation}
S_{E}=\int{d^{4}x\sqrt{-\tilde{g}}\bigg(\frac{1}{2}\tilde{R}-\frac{1}{2}\tilde{g}^{\mu\nu}\partial_{\mu}\varphi\partial_{\nu}\varphi-V(\varphi)\bigg)}+\int{d^{4}x\mathcal{L}_{m}(F^{-1}\tilde{g}_{\mu\nu},\Psi_{m})},\hspace{0.15cm}
\label{a21}    
\end{equation}
where the  potential is 
\begin{equation}
V(\varphi)=\frac{FR-f}{2F^{2}},\hspace{1cm}\mbox{with}\hspace{1cm}F(R)\equiv f'(R)\equiv e^{\sqrt{\frac{2}{3}}\varphi}.
\label{a22}    
\end{equation}
In this case, the scalar field $\varphi$ can be identified starting from $f'(R)$. For $f(R)=R+\alpha R^2$, we have the Starobinsky scalaron.

\subsection{$f(T)$ teleparallel gravity}
Another class of extended theories of gravity can be derived from TEGR considering the torsion scalar $T$. Similar to $f(R)$ gravity, we can take into account  $f(T)$ gravity and  the action of TEGR (\ref{9}) extends as follows \cite{Cai:2015emx} 
\begin{equation}
S_{T}=\int{d^{4}x\Big(e\frac{f(T)}{2}+\mathcal{L}_{m}(e^i_{\mu},\Psi_{m})\Big)}.
\label{b1}    
\end{equation}
Variation of the above action with respect to the vierbein $e^i_{\mu}$ leads to the following field equations  
\begin{equation}
\big[e^{-1}\partial_{\mu}(eS_{i}{}^{\mu\nu})-e^\alpha_{i}T^{\rho}{}_{\mu\alpha}S_{\rho}{}^{\nu\mu}\big]f'+S_{i}{}^{\mu\nu}\partial_{\mu}Tf''+\frac{1}{4}e^{\nu}_{i}f=\frac{1}{2}e^{\rho}_{i}T_{\rho}^{(m)\nu},
\label{b2}
\end{equation}
where $'$ and $''$ represent the first and second derivative with respect to $T$, respectively. Also, $T_{\rho}^{(m)\nu}$ is the energy-momentum tensor related to the matter field. By assuming a spatially flat universe described by the FLRW metric related to the orthonormal tetrad components as
\begin{equation}
g_{\mu\nu}=\eta_{ij}e^{i}_{\mu}e^{j}_{\nu},
\label{b2+}    
\end{equation}
the cosmological equations for $f(T)$ gravity can be obtained as
\begin{equation}
12f'H^2+f=2\rho,\hspace{1cm}(12H^2f''-f')\dot{H}=\frac{1}{2}(\rho+p),
\label{b3}
\end{equation}
where $\rho$ and $p$ are the energy density and pressure of the perfect fluid, respectively. Here, dot denotes the derivative with respect to the cosmic time $t$. In the presence of a barotropic
fluid, the effective EoS is given by
\begin{equation}
w_{eff}=\frac{p_{eff}}{\rho_{eff}}=\frac{2(p_{m}+p_{\varphi})-4\dot{H}f'+f-48H^2\dot{H}f''}{2(\rho_{m}+\rho_{\varphi})-f}.
\label{b4}    
\end{equation}
Now, under the conformal transformation (\ref{a11}) and using the following identities
\begin{equation}
\tilde{g}^{\mu\nu}=\Omega^{-2}g^{\mu\nu},\hspace{0.5cm},\tilde{e}^{i}_{\mu}=\Omega e^{i}_{\mu},\hspace{0.5cm}\tilde{e}^{\mu}_{i}=\Omega^{-1}e^{\mu}_{i},\hspace{0.5cm}\tilde{e}=\Omega^4e,
\label{b5}
\end{equation}
the conformal version of the torsion and $S_{\rho}{}^{\mu\nu}$ tensors are expressed as
\begin{equation}
\tilde{T}^\rho{}_{\mu\nu}=T^\rho{}_{\mu\nu}+\delta^{\rho}_{\nu}\partial_{\mu}\ln\Omega-\delta^{\rho}_{\mu}\partial_{\nu}\ln\Omega,
\label{b6}
\end{equation}
and
\begin{equation}
\tilde{S}_\rho{}^{\mu\nu}=\Omega^{-2}\Big(S_\rho{}^{\mu\nu}+\delta^{\mu}_{\rho}\partial^{\nu}\ln\Omega-\delta^{\nu}_{\rho}\partial^{\mu}\ln\Omega\Big).
\label{b7}
\end{equation}
Then, the torsion scalar in the Einstein frame is given by
\begin{equation}
\tilde{T}=\Omega^{-2}\Big(T-6g^{\mu\nu}\partial_{\mu}\ln\Omega\partial_{\nu}\ln\Omega+4g^{\mu\nu}\partial_{\nu}\ln\Omega T^\rho{}_{\rho\mu}\Big).
\label{b8}
\end{equation}
The inverse transformation of the scalar (\ref{b8}) is
\begin{equation}
T=\Omega^{2}\Big(\tilde{T}-6\tilde{g}^{\mu\nu}\partial_{\mu}w\partial_{\nu}w-4\tilde{g}^{\mu\nu}\partial_{\nu}w\tilde{T}^\rho{}_{\rho\mu}\Big),
\label{b9}
\end{equation}
where $w\equiv\ln\Omega$.
To obtain the action in the Einstein frame, we require to rewrite the action (\ref{b1}) as
\begin{equation}
S_{E}=\int{d^{4}x e\Big(\frac{FT}{2}-U\Big)}+\int{d^{4}x\mathcal{L}_{m}(e^i_{\mu},\Psi_{m})\Big)},\hspace{1cm}\text{with}\hspace{1cm}U=\frac{FT-f}{2}.
\label{b10}    
\end{equation}
Plugging Eq.(\ref{b9}) into  action (\ref{b10}) and then using the identities (\ref{b5}), the transformed action can be found as
\begin{equation}
S_{E}=\int{d^{4}x\tilde{e}\bigg(\frac{1}{2}F\Omega^{-2}\Big[\tilde{T}-6\tilde{g}^{\mu\nu}\partial_{\mu}w\partial_{\nu}w-4\tilde{g}^{\mu\nu}\partial_{\nu}w\tilde{T}^\rho{}_{\rho\mu}\Big]-\Omega^{-4}U\bigg)}+\int{d^{4}x\mathcal{L}_{m}(\Omega^{-1}\tilde{e}^i_{\mu},\Psi_{m})}.
\label{b11}    
\end{equation}
By introducing $F=\Omega^2$, we find
\begin{equation}
S_{E}=\frac{1}{2}\int{d^{4}x\tilde{e}\bigg(\tilde{T}-6\tilde{g}^{\mu\nu}\partial_{\mu}w\partial_{\nu}w-4\tilde{g}^{\mu\nu}\partial_{\nu}w\tilde{T}^\rho{}_{\rho\mu}-2F^{-2}U\bigg)}+\int{d^{4}x\mathcal{L}_{m}(F^{-\frac{1}{2}}\tilde{e}^i_{\mu},\Psi_{m})}.
\label{b12}    
\end{equation}
As we see, $f(T)$ gravity cannot be represented as a teleparallel action
plus a scalar field under  conformal transformation, due to the appearance of an additional scalar-torsion coupling
term $\partial^{\mu}w \tilde{T}_{\mu}$. In fact, this term clearly reflects the violation of local Lorentz invariance in the $f(T)$ gravity as a crucial difference with respect to  $f(R)$ gravity \cite{Li:2010cg}. It is worth noticing that, in some special frames, we can neglect the scalar-torsion coupling  in order to find easily the related solutions, however, in general,   the problem
of Lorentz invariance violation  remains.   

With these considerations in mind,  we can introduce a new scalar field $\varphi\equiv\sqrt{3/2}\ln F$, and then reduce the action in the Einstein frame  as
\begin{equation}
S_{E}=\int{d^{4}x\tilde{e}\bigg(\frac{1}{2}\tilde{T}-\frac{1}{2}\tilde{g}^{\mu\nu}\partial_{\mu}\varphi\partial_{\nu}\varphi-V(\varphi)\bigg)}+\int{d^{4}x\mathcal{L}_{m}(F^{-\frac{1}{2}}\tilde{e}^i_{\mu},\Psi_{m})},
\label{b13}    
\end{equation}
where the  potential is 
\begin{equation}
V(\varphi)=\frac{FT-f}{2F^{2}},\hspace{1cm}\mbox{with}\hspace{1cm}F(T)\equiv f'(T)\equiv e^{\sqrt{\frac{2}{3}}\varphi}.
\label{b14}    
\end{equation}
In this framework, it is possible compare results with those obtained from action \eqref{a21}.

\subsection{$f(Q)$ non-metric gravity}
Inspired by the above examples of extended gravity, \ie  $f(R)$ and $f(T)$, one can define an extended version of STEGR by the action \cite{Heisenberg:2023lru}
\begin{equation}
S=\int{d^{4}x\sqrt{-g}\Big(-\frac{f(Q)}{2}+\mathcal{L}_{m}(g_{\mu\nu},\Psi_{m})\Big)}.
\label{c1}   
\end{equation}
By varying the above action with respect to the metric $g_{\mu\nu}$, the field equations can be found as
\begin{equation}
\frac{2}{\sqrt{-g}}\nabla_{\alpha}(\sqrt{-g}f'P^{\alpha}{}{}_{\mu\nu})+\frac{1}{2}g_{\mu\nu}f+f'(P_{\mu\alpha\lambda} Q_{\nu}{}^{\alpha\lambda}-2Q_{\alpha\lambda\mu}P^{\alpha\lambda}{}_{\nu})=T^{(m)}_{\mu\nu},
\label{c2}
\end{equation}
where $'$ denotes the derivative with respect to $Q$. It is 
\begin{equation}
f'=\frac{df(Q)}{dQ},\hspace{1cm}\mbox{and}\hspace{1cm}T^{(m)}_{\mu\nu}=-\frac{2}{\sqrt{-g}}\frac{\delta(\sqrt{-g}\mathcal{L}_{m})}{\delta g^{\mu\nu}}.
\end{equation}
The field equation of  connection is given by 
\begin{equation}
\nabla_{\mu}\nabla_{\nu}(\sqrt{-g}f'P^{\mu\nu}{}_{\alpha})=0.
\end{equation}
To formulate the dynamical equations, we assume a spatially flat universe ($k=0$) described by the  above FLRW metric. Then, the cosmological background equations can be written as 
\begin{equation}
12f'H^{2}-f=2\rho,\hspace{1cm}(12f''H^{2}+f')\dot{H}=-\frac{1}{2}(\rho+p),
\label{c3}
\end{equation}
where $''$ denotes the second derivatives with respect to $Q$ and the dot is the derivative with respect to the cosmic time $t$. Also, $\rho$ and $p$ are known as the energy density and pressure of the perfect fluid. In the presence of a barotropic fluid, the effective EoS takes the form 
\begin{equation}
w_{eff}=\frac{p_{eff}}{\rho_{eff}}=\frac{2(p_{m}+p_{\varphi})-4\dot{H}f'-f+48H^2\dot{H}f''}{2(\rho_{m}+\rho_{\varphi})+f}.
\label{c4}    
\end{equation}
Moreover, under the conformal transformation (\ref{a11}) and using the associated identities (\ref{a12}), the conformal version of non-metricity tensor and non-metricity conjecture in the coincident gauge are obtained as
\begin{equation}
\tilde{Q}_{\alpha\mu\nu}=\Omega^{2}\Big(Q_{\alpha\mu\nu}+2g_{\mu\nu}\partial_{\alpha}\ln{\Omega}\Big),
\label{c5}    
\end{equation}
and
\begin{equation}
\tilde{P}^{\alpha\mu\nu}=-\Omega^{-4}\bigg[P^{\alpha\mu\nu}+\Omega^{2}\partial_{\beta}\ln{\Omega}\Big(-\frac{3}{2}g^{\alpha\beta}g^{\mu\nu}-g^{(\mu\beta}g^{\nu)\alpha}+3g^{\mu\beta}g^{\alpha\nu}+g^{\nu\beta}g^{\alpha\mu}\Big)\bigg].
\label{c6}    
\end{equation}
Then, the non-metricity scalar in the Einstein frame is
\begin{equation}
\tilde{Q}=-\Omega^{-2}Q+20g^{\mu\nu}\partial_{\mu}\ln\Omega\partial_{\nu}\ln\Omega+\frac{7}{2}(\Omega^{-2}-1)Q^{\mu}\partial_{\mu}\ln\Omega.
\label{c7}
\end{equation}\\
The inverse transformation of  relation (\ref{c7}) is given by
\begin{equation}
Q=-\Omega^{2}\tilde{Q}+20\Omega^{2}\tilde{g}^{\mu\nu}\partial_{\mu}w\partial_{\nu}w+\frac{7}{2}(1-\Omega^{2})\tilde{Q}^{\mu}\partial_{\mu}w,
\label{c8}    
\end{equation}
where  $w\equiv\ln\Omega$. In order to find the action in the Einstein frame, we rewrite the action in the Jordan frame (\ref{c1}) as
\begin{equation}
S_{E}=\int{d^{4}x\sqrt{-g}\Big(-\frac{FQ}{2}+U\Big)}+\int{d^{4}x\mathcal{L}_{m}(g_{\mu\nu},\Psi_{M})},\hspace{1cm}\text{with}\hspace{1cm}U=\frac{FQ-f}{2}.
\label{c9}    
\end{equation}
Using the relation (\ref{c8}) and $\sqrt{-g}=\Omega^{-4}\sqrt{-\tilde{g}}$, the action (\ref{c9}) is rewritten as
\begin{equation}
S_{E}=\int{d^{4}x\sqrt{-\tilde{g}}\bigg(\frac{1}{2}F\Omega^{-2}\Big[\tilde{Q}-20\tilde{g}^{\mu\nu}\partial_{\nu}w\partial_{\mu}w-\frac{7}{2}\frac{(1-\Omega^{2})}{\Omega^{2}}\tilde{Q}^{\mu}\partial_{\mu}w\Big]+\Omega^{-4}U\bigg)}+\int{d^{4}x\mathcal{L}_{m}(\Omega^{-2}\tilde{g}_{\mu\nu},\Psi_{M})}.
\label{c10}    
\end{equation}
And, by choosing $F=\Omega^{2}$, we have
\begin{equation}
S_{E}=\frac{1}{2}\int{d^{4}x\sqrt{-\tilde{g}}\bigg(\tilde{Q}-20\tilde{g}^{\mu\nu}\partial_{\nu}w\partial_{\mu}w-\frac{7}{2}\frac{(1-F)}{F}\tilde{Q}^{\mu}\partial_{\mu}w+2F^{-2}U\bigg)}+\int{d^{4}x\mathcal{L}_{m}(F^{-1}\tilde{g}_{\mu\nu},\Psi_{M})}.
\label{c11}    
\end{equation}
Similar to the $f(T)$ gravity, we cannot rewrite the above action as a symmetric teleparallel action plus a scalar field under the conformal transformation since there is an additional scalar-non-metricity coupling  term $\tilde{Q}^{\mu}\partial_{\mu}w$. However, as discussed in \cite{Yang:2010ji} for $f(T)$, the further coupling can be neglected in particular regimes. It is easy to see that, being $\partial_{\mu}w=\partial_{\mu} \ln F= \partial_{\mu} F/F$, then $(1-F)/F^{3/2}\rightarrow 0$ for $F\rightarrow \infty$. In  this regime also $\ln F$ is slowly varying and then also $\partial_\mu w$ is negligible. In other words,  being $F$  large in strong field regime, the third term in (\ref{c11}) can be neglected. This is an important feature because the transition between the strong to the weak field regime can be related to the behavior of the function $F$, that is the conformal factor $\Omega^2$. We are in the strong field for $F\rightarrow\infty$, while we are in the weak field for  $F\rightarrow 0$. It means that conformal transformations work in strong field, while in the weak field there is a breaking of conformal structure  which is led by  the third term in (\ref{c11}). In other words, the behavior of function $F$, in the framework of non-metric gravity,  constitutes a natural mechanism to exit from inflation \cite{Capozziello:2022tvv}.

Motivated by these considerations, we define a new scalar field $\varphi\equiv\sqrt{5}\ln F$ so that the action in the Einstein frame takes the following form
\begin{equation}
S_{E}=\int{d^{4}x\sqrt{-\tilde{g}}\bigg(\frac{1}{2}\tilde{Q}-\frac{1}{2}\tilde{g}^{\mu\nu}\partial_{\mu}\varphi\partial_{\nu}\varphi-V(\varphi)\bigg)}+\int{d^{4}x\mathcal{L}_{m}(F^{-1}\tilde{g}_{\mu\nu},\Psi_{m})},\hspace{0.15cm}
\label{c12}    
\end{equation}
where the  potential is 
\begin{equation}
V(\varphi)=\frac{f-FQ}{2F^{2}},\hspace{1cm}\mbox{with}\hspace{1cm}F(Q)\equiv f'(Q)\equiv e^{\sqrt{\frac{1}{5}}\varphi}.
\label{c13}    
\end{equation}
Clearly, the three extended theories can be compared in the Einstein frame when we define suitable   regimes for the couplings in $f(T)$ and in $f(Q)$ gravity.
In the next section, we shall study how the inflationary period can be described in the considered three  theories of gravity. Results will be   compared with the recent observations of CMB anisotropies.

\section{Inflation in $f(R)$ gravity}\label{fr}
In this section, we discuss cosmological  inflation  in $f(R)$ gravity in the two perspectives of PSR and HSR. This will be the paradigm with respect to compare $f(T)$ and $f(Q)$ inflation.

\subsection{The Potential-Slow-Roll Inflation}

Let us investigate the PSR approach to inflation by considering power-law and logarithmic functions of $R$. The corresponding potentials can be derived after a conformal transformation. We study dynamics in the Einstein frame when  ordinary matter can  neglected, that is  $\mathcal{L}_m=0$.

\subsubsection{The $f(R)=R+\alpha R^{2p}$  model}

We start with the well-known $f(R)$ function \cite{Maeda:1988ab,Schmidt:1990dh}
\begin{equation}
f(R)=R+\alpha R^{2p},
\label{d1}
\end{equation}
\begin{figure*}[!hbtp]
	\centering
    \includegraphics[width=.55\textwidth,keepaspectratio]{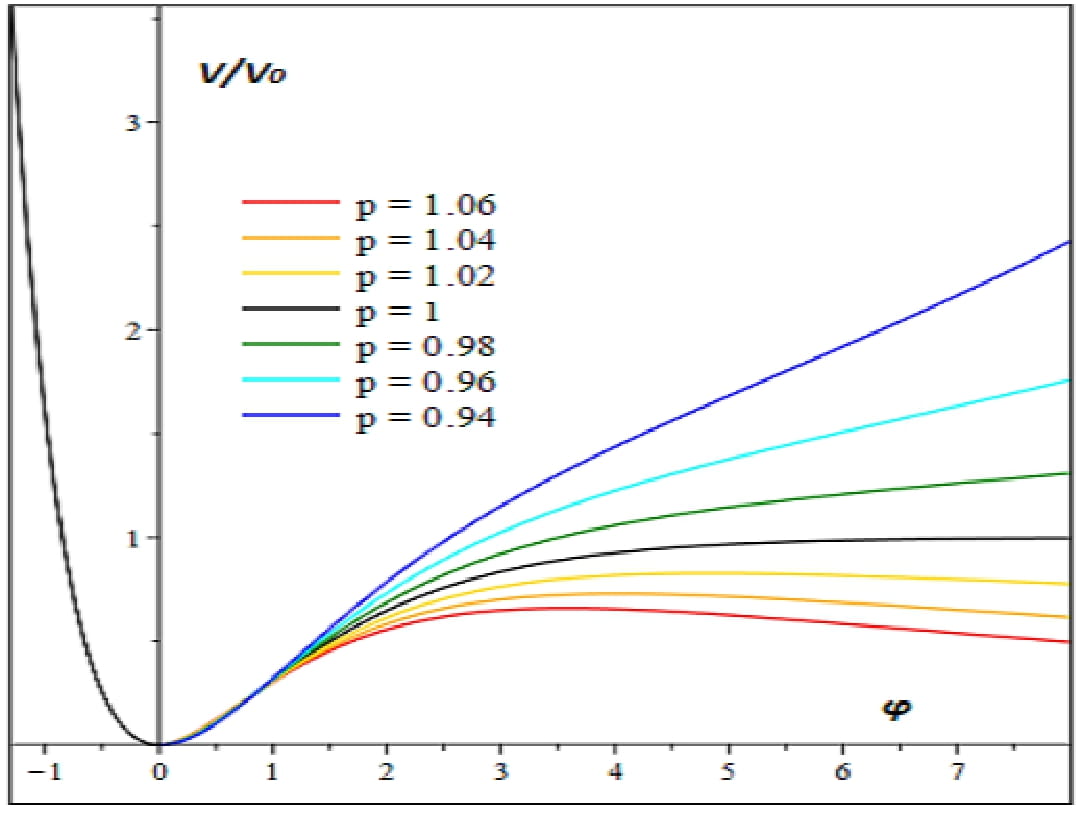}
	\caption{The potential of the $R^{2p}$ model (\ref{d1}) for different powers of $p$.}
	\label{fig1}
\end{figure*}
where $\alpha=\frac{1}{M^{4p-2}}$ with $p$  a real number which has to be close to the unity for inflation. Here, ${\displaystyle M\simeq 10^{13}\,\gev}$ is a normalized energy scale from the amplitude of observed power spectrum for the primordial perturbations \cite{Starobinsky:1980te}. From Eq.(\ref{a22}),  the scalaron potential  is given by
\begin{equation}
V(\varphi)=V_{0}e^{-2\sqrt{\frac{2}{3}}\varphi}(e^{\sqrt{\frac{2}{3}}\varphi}-1)^{\frac{2p}{2p-1}},
\label{d2}
\end{equation}
where $ V_{0}=(\frac{2p-1}{4p})M^{2}(\frac{1}{2p})^{\frac{1}{2p-1}}$. In Fig.\ref{fig1},  we present the behaviour of  potential (\ref{d2}) for different values of the power $p$. Two main classes of behaviours have to be considered: i) For $p>1$ the potential shows a maximum around $ \varphi_m = \sqrt{3/2}\ln{\frac{2p -1}{p-1}}$ but then approaches asymptotically to zero for large value of scalaron. Inflation therefore can happen both for $0 \leq \varphi \leq \varphi_m$ and $\varphi > \varphi_m$. We predict that the $R^{2p}$  model shows a tiny deviation from the $R^{2}$ Starobinsky model. Therefore, we neglect the latter region and are interested only the region $0\leq\varphi\leq\varphi_{m}$. ii) For $p < 1$ the potential increases but its decreasing towards zero is  steeper than in $R^2$ Starobinsky model. This leads to larger tensor-to-scalar ratios. iii) For $p = 1 $ we reproduce the potential of the $R^{2}$ Starobinsky inflation which asymptotically reaches to a constant value.

The slow-roll parameters in the PSR approach are
\begin{equation}
\epsilon_{V}=\frac{1}{2}\bigg(\frac{V'(\varphi)}{V(\varphi)}\bigg)^{2},\qquad\quad \eta_{V}=\frac{V''(\varphi)}{V(\varphi)},
\label{d3}
\end{equation}
where $'$ represents the derivative with respect to the scalaron $\varphi$. The number of \textit{e}-folds $N$ is given by
\begin{equation}
N\equiv\frac{1}{\sqrt{2}}\int^{\varphi_{i}}_{\varphi_{f}}{\frac{V}{V'}d\varphi} \equiv\int^{\varphi_{i}}_{\varphi_{f}}{\frac{1}{\sqrt{2\epsilon_{V}}}d\varphi},
\label{d4}
\end{equation}
where subscribes $i$ and $f$ indicate  the initial  and final period of inflation, respectively. During the slow-roll inflation, it is $\epsilon_V\ll1$ and $\eta_V\ll1$, while inflation terminates when $\epsilon_V=1$. The spectral parameters, \ie the spectral index and the tensor-to-scalar ratio in the PSR approach are
\begin{equation}
n_{s}=1-6\epsilon_{V}+2\eta_{V},\quad\quad\quad r=16\epsilon_{V}.
\label{d5}  
\end{equation}

Let us now calculate the above parameters for the $R^{2p}$ model for the general case $p\neq1$. The slow-roll parameters (\ref{d3})  are 
\begin{equation}
\epsilon_V=\frac{4\big((p-1)e^{\sqrt{\frac{2}{3}}\varphi}-(2p-1)\big)^{2}}{3(2p-1)^{2}\big(e^{\sqrt{\frac{2}{3}}\varphi}-1\big)^{2}},\hspace{1cm}\eta_V=\frac{(-40p^{2}+52p-16)e^{\sqrt{\frac{2}{3}}\varphi}+8(p-1)^{2}e^{2\sqrt{\frac{2}{3}}\varphi}+8(2p-1)^{2}}{3(2p-1)^{2}\big(e^{\sqrt{\frac{2}{3}}\varphi}-1\big)^{2}}.
\label{d6}
\end{equation}
\begin{figure*}[!hbtp]
	\centering
	\includegraphics[width=.70\textwidth,keepaspectratio]{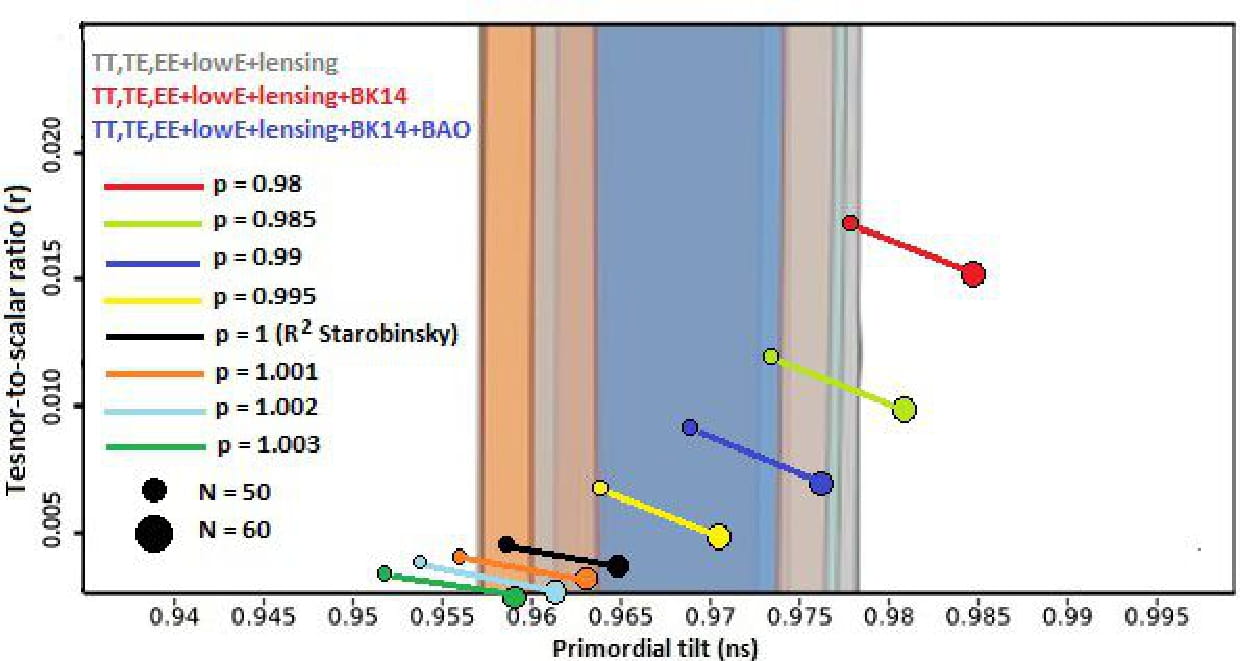}
\caption{The marginalized joint 68\% and 95\% CL regions for $n_{s}$ and $r$ at $k = 0.002$ Mpc$^{-1}$ from Planck alone and in combination with BK14 or BK14+BAO \cite{Planck:2018jri} and the $n_{s}-r$ constraints on the $R^{2p}$ model (\ref{d1}) for two cases $p=1$ ($R^2$ Starobinsky inflation) and $p\neq1$ (near-Starobinsky inflation).}
	\label{fig2}
\end{figure*}
From the first slow-roll parameter and setting $\epsilon_V=1$, the value of the scalaron at the end of inflation $\varphi_{f}$ takes the following from
\begin{equation}
\varphi_{f}=\sqrt{\frac{3}{2}}\ln\bigg[\frac{(2p-1)(\sqrt{3}-2)}{\sqrt{3}(2p-1)-2(p-1)}\bigg].
\label{d7}
\end{equation}
Moreover, using Eq.(\ref{d4}), the number of \textit{e}-folds of the model is 
\begin{equation}
N\simeq-\frac{3p}{4(p-1)}\ln\bigg(\frac{(p-1)e^{\sqrt{\frac{2}{3}}\varphi_{i}}}{(1-2p)}+1\bigg).
\label{d8}
\end{equation}
Thus, from Eqs. (\ref{d5}), the spectral parameters of the models are 
\begin{equation}
n_{s}\simeq\frac{(4\mathcal{A}^2-4\mathcal{A}-5)p^2+(4\mathcal{A}^2- 2\mathcal{A}+8)p-5\mathcal{A}^2}{3(2\mathcal{A}p-\mathcal{A}-p)^2},\hspace{1cm}r\simeq\frac{64(p- 1)^2\mathcal{A}^2}{3(2\mathcal{A}p-\mathcal{A}- p)^2},
\label{d9}
\end{equation}
where $\mathcal{A}= e^{-\frac{4(p-1)N}{3p}}$.
For the well-known $R^{2}$ Starobinsky model \cite{Starobinsky:1980te} and by  setting $p=1$, the slow-roll parameters (\ref{d3}) are
\begin{equation}
\epsilon_V=\frac{4}{3\big(e^{\sqrt{\frac{2}{3}}\varphi}-1\big)^{2}},\hspace{1cm}\eta_V=\frac{4\big(2-e^{\sqrt{\frac{2}{3}}\varphi}\big)}{3\big(e^{\sqrt{\frac{2}{3}}\varphi}-1\big)^{2}}.
\label{d10}    
\end{equation}
By setting $\epsilon_V=1$, the value of the scalaron at the end of inflation is 
\begin{equation}
\varphi_{f}=\sqrt{\frac{3}{2}}\ln\Big(1+\sqrt{\frac{4}{3}}\Big).
\label{d11}
\end{equation}
Also, the number of \textit{e}-folds (\ref{d4}) of the Starobinsky model is given by
\begin{equation}
N\simeq\frac{3}{4}e^{\sqrt{\frac{2}{3}}\varphi_{i}}.
\label{d12}    
\end{equation}
\begin{figure*}[!hbtp]
	\centering
	\includegraphics[width=.55\textwidth,keepaspectratio]{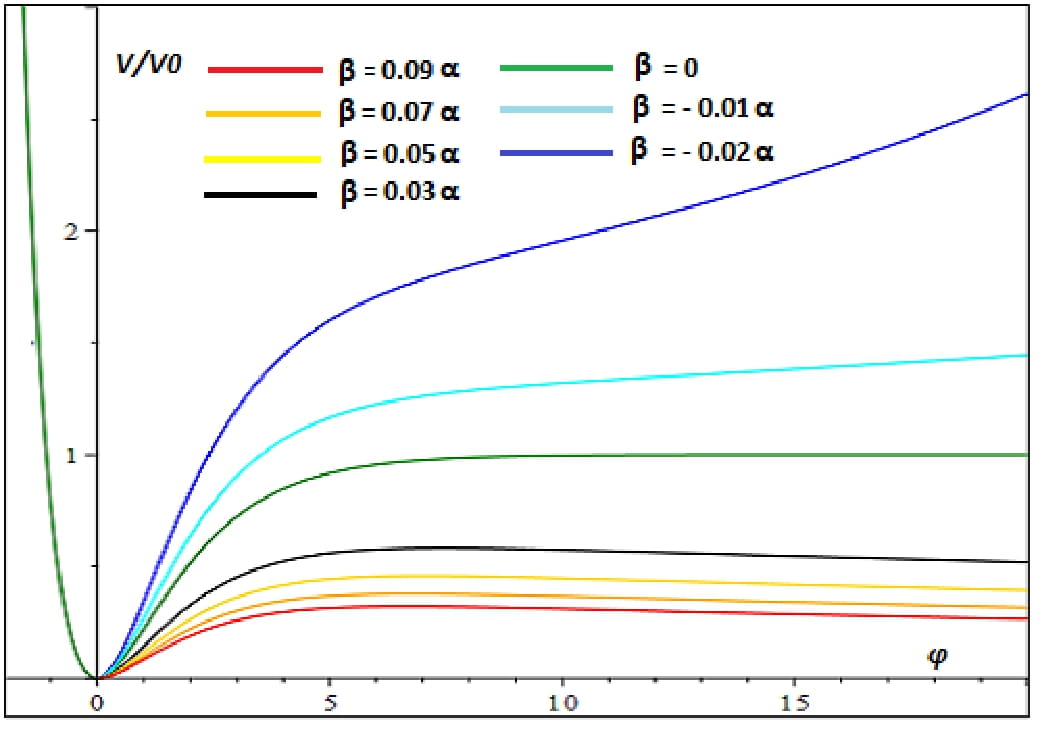}
	\caption{The potential of the logarithmic corrected $f(R)$ model (\ref{d14}) for different values of $\beta$ and $\alpha\sim10^{-9}$.}
	\label{fig3}
\end{figure*}
The spectral parameters (\ref{d5})  are, in this case,
\begin{equation}
n_{s}\simeq\frac{16N^{2}-56N-15}{(4N-3)^{2}},\hspace{1cm}r\simeq\frac{192}{(4N-3)^{2}}.
\label{d13}    
\end{equation}
In Fig.\ref{fig2} we show the consistency relation $r=r(n_{s}) $  coming from the marginalized joint 68\% and 95\% CL regions of Planck 2018 alone and in combination with the BK14 or BK14+BAO datasets \cite{Planck:2018jri} on the $R^{2p}$ model (\ref{d1}) for two cases $p=1$ ($R^2$ Starobinsky inflation) and $p\neq1$ (near-Starobinsky inflation)

For the case $p=1$ (solid black line), we reproduce the results of the $R^2$ Starobinsky model  $(n_{s}=0.95815$, $r=0.0049473$) and  $(n_{s}=0.96539$, $r=0.0034183)$ for $N=50$ and $N=60$, respectively. By considering a generalized form of the Starobinsky model \ie the $R^{2p}$ model, we find different results of the spectral index $n_{s}$ and the tensor-to-scalar ratio $r$ depending on the value of $p$. For the Planck  dataset alone, we see that the power $0.985\leq p\leq1.002$ predicts the values of $n_{s}$ and $r$ which are in good agreement with the observations, in particular, at 68\% CL, while it reduces to $0.99\leq p\leq1.001$ at 95\% CL. For a combination of Planck and BK14, the results are almost similar to the results coming from Planck  case alone. For the full consideration of the CMB observations Planck+BK14+BAO, one can find that the obtained values of $n_{s}$ and $r$ for the powers $0.985\leq p\leq1.001$ and $0.99\leq p\leq0.995$ are compatible with the observations at 68\% and 95\% CL, respectively.

Besides the mentioned results, the figure tells us that the $R^{2p}$ model with $p<1$ ($p>1$) offers larger (smaller) values of $n_{s}$ and $r$ compared to the $R^2$ Starobinsky model, respectively. Notice that a large deviation of $p$ from the Starobinsky model ($p=1$) is ruled out by the observations.

\subsubsection{The $f(R)=R+\alpha R^2+\beta R^{2}\ln R$  model}
A generalized version of the $R^{2}$ Starobinsky model is  the logarithmic model \cite{Ben-Dayan:2014isa}
\begin{equation}
f(R)=R+\alpha R^2+\beta R^{2}\ln R,
\label{d14}
\end{equation}
with the corrections coming from quantum gravity effects. We can fix the phenomenological parameters $\alpha$, $\beta$ for the moment. Recall that in the case $\alpha=\frac{1}{6M^2}$ and $\beta\simeq0$, we recover exactly the $R^{2}$ Starobinsky model. From Eq.(\ref{a22}), the corresponding potential of the scalaron field $\varphi$ is give by
\begin{equation}
V(\varphi)=\frac{(\alpha+\beta)R^{2}(1+\frac{\beta}{\alpha+\beta}\ln R)}{2\Big(1+(2\alpha+\beta)R(1+\frac{2\beta}{2\alpha+\beta}\ln R)\Big)^{2}}.
\label{d15}    
\end{equation}
Inverting $F$ in Eq.(\ref{a22}), we find 
\begin{equation}
R=\frac{e^{\sqrt{\frac{2}{3}}\varphi}-1}{2\beta W_{k}(X)},\hspace{1cm}\mbox{with}\hspace{1cm}X\equiv\left(\frac{e^{\sqrt{\frac{2}{3}}\varphi}-1}{2\beta}\right)e^{\frac{2\alpha+\beta}{2\beta}},
\label{d16}  
\end{equation}
where $W_{k}$ is the Lambert function of branch $k = 0$ for $\beta>0$, and  $k = -1$ for $\beta < 0$. Using the iterative method and considering only the leading order in $R$, the potential for $|\beta|\ll\alpha$ becomes
\begin{equation}
V(\varphi)\simeq V_{0}\frac{(1-e^{-\sqrt{\frac{2}{3}}\varphi})^{2}}{\bigg[1+\frac{\beta}{2\alpha}+\frac{\beta}{\alpha}\ln\Big(\frac{e^{\sqrt{\frac{2}{3}}\varphi}-1}{2\alpha}\Big)\bigg]},
\label{d17}  
\end{equation}
where $V_{0}=\frac{1}{8\alpha}$. In Fig.\ref{fig3}, we present the behaviour of the potential for different values of $\beta$ and $\alpha\sim10^{-9}$. For $\beta=0$, we have the   $R^{2}$ Starobinsky model. For $\beta<0$, the potential goes upward showing a lager value of $r$ because of a larger slope. For $\beta>0$, the potential presents a maximum value and then it shows a decreasing behaviour towards zero for larger values of $\varphi$.

From Eq.(\ref{d3}), the slow-roll parameters are 
\begin{equation}
\epsilon_V=\frac{4\Big(2\beta\ln(\frac{e^{\sqrt{\frac{2}{3}}\varphi}-1}{2\alpha})+(1-e^{\sqrt{\frac{2}{3}}\varphi})\beta+2\alpha\Big)^{2}}{3\Big(2\beta\ln(\frac{e^{\sqrt{\frac{2}{3}}\varphi}-1}{2\alpha})+\beta+2\alpha\Big)^{2}\big(e^{\sqrt{\frac{2}{3}}\varphi}-1\big)^{2}},
\label{d18}    
\end{equation}
\begin{figure*}[!hbtp]
	\centering
	\includegraphics[width=.70\textwidth,keepaspectratio]{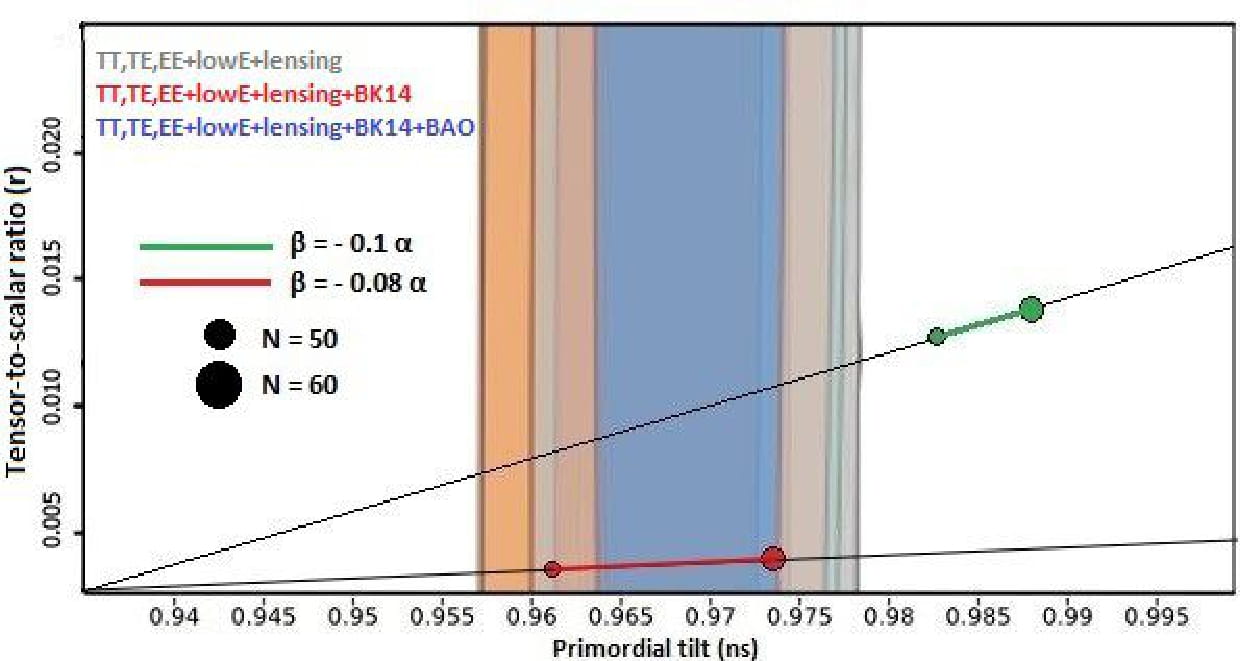}
\caption{The marginalized joint 68\% and 95\% CL regions for $n_{s}$ and $r$ at $k = 0.002$ Mpc$^{-1}$ from Planck alone and in combination with BK14 or BK14+BAO \cite{Planck:2018jri} and the $n_{s}-r$ constraints on the logarithmic corrected $f(R)$ model (\ref{d14}) for allowed values $\beta=-0.08\alpha, -0.1\alpha$ when $\alpha\sim10^{-9}$ is considered.}
	\label{fig4}
\end{figure*}
\begin{eqnarray}
&\!&\!\eta_V=\frac{1}{3\Big(2\beta\ln(\frac{e^{\sqrt{\frac{2}{3}}\varphi}-1}{2\alpha})+\beta+2\alpha\Big)^{2}\big(e^{\sqrt{\frac{2}{3}}\varphi}-1\big)^{2}}\Bigg\{-16\beta^{2}(e^{\sqrt{\frac{2}{3}}\varphi}-2)\ln\Big(\frac{e^{\sqrt{\frac{2}{3}}\varphi}-1}{\alpha}\Big)^{2}+8\beta\Big((4\beta\ln({2})-4\alpha-5\beta)\times
\nonumber\\&\!&\!
\times e^{\sqrt{\frac{2}{3}}\varphi}-8\beta\ln({2})+8\alpha+4\beta\Big)\ln\big(\frac{e^{\sqrt{\frac{2}{3}}\varphi}-1}{\alpha}\big)-8\big(\beta\ln(2)-\alpha-2\beta\big)\big(2\beta\ln(2)-2\alpha-\beta\big)e^{\sqrt{\frac{2}{3}}\varphi}+16e^{2\sqrt{\frac{2}{3}}\varphi}\beta^{2}+\nonumber\\&\!&\!
+8\big(2\beta\ln(2)-2\alpha-\beta\big)^{2}\Bigg\}.
\label{d19}    
\end{eqnarray}
For an alternative notation, we can rewrite the above slow-roll parameters as deviation from the slow-roll parameters of the $R^{2}$ Starobinsky model (\ref{d10}). In such a picture, the slow-roll parameters of the logarithmic corrected model are given by
\begin{equation}
\epsilon_V=\epsilon_{s}-\delta\sqrt{\frac{4\epsilon_{s}}{3}}+\frac{\delta^{2}}{3},\hspace{1cm}\eta_V=\eta_{s}-\delta\sqrt{\frac{16\epsilon_{s}}{3}}+\frac{4\delta^2}{3},
\label{d20}
\end{equation}
where $\delta\equiv\beta\alpha^{-1}\big[1+\frac{\beta}{2\alpha}+\frac{\beta}{\alpha}\ln\big(\frac{e^{\sqrt{\frac{2}{3}}\varphi_{i}}-1}{2\alpha}\big)\big]^{-1}$. Also, $\epsilon_{s}$ and $\eta_{s}$ are the first and second slow-roll parameters of the $R^2$ Starobinsky model (\ref{d10}). 
Considering $\epsilon=1$, choosing the positive sign and then using the Taylor expansion for $\big|\frac{e^{\sqrt{\frac{2}{3}}\varphi_{f}}-1}{2\alpha}-1\big|\leq1$, the value of the scalar field $\varphi$ at the end of inflation for $|\beta|\ll\alpha$ is 
\begin{equation}
\varphi_{f}\simeq\sqrt{\frac{3}{2}}\ln\Bigg\{\frac{1}{6}\Bigg(3-\frac{6\alpha^2}{\beta}+\sqrt{3}\Bigg[2(1-\alpha)\pm\sqrt{7+4\sqrt{3}+\frac{4\alpha}{\beta^2}\Big(3\alpha^3+\alpha\beta\big(3+2\sqrt{3}(1+\alpha)\big)+\beta^2(\alpha-2-\sqrt{3})\Big)}\Bigg]\Bigg)\Bigg\}.
\label{d21}
\end{equation}
The number of \textit{e}-folds (\ref{d4}) of this model is
\begin{equation}
N\simeq-\frac{3\ln\big(\frac{1-\frac{\delta^{2}}{3\epsilon_{s}}}{1-\frac{\delta^{2}}{3}}\big)-6\tanh^{-1}(\frac{\delta}{\sqrt{3\epsilon_{s}}})}{\delta(2+\delta)}.
\label{d22}    
\end{equation}

Eventually, the spectral parameters (\ref{d5}) are
\begin{eqnarray}
&\!&\!n_{s}\simeq1-\frac{8\Big(2\beta\ln(\frac{\mathcal{A}-1}{2\alpha})+(1-\mathcal{A})\beta+2\alpha\Big)^{2}}{\Big(2\beta\ln(\frac{\mathcal{A}-1}{2\alpha})+\beta+2\alpha\Big)^{2}\big(\mathcal{A}-1\big)^{2}}+\frac{2}{3\Big(2\beta\ln(\frac{\mathcal{A}-1}{2\alpha})+\beta+2\alpha\Big)^{2}\big(\mathcal{A}-1\big)^{2}}\times\nonumber\\&\!&\!
\times\Bigg\{-16\beta^{2}(\mathcal{A}-2)\ln\Big(\frac{\mathcal{A}-1}{\alpha}\Big)^{2}+8\beta\Big((4\beta\ln({2})-4\alpha-5\beta)\mathcal{A}-8\beta\ln({2})+8\alpha+4\beta\Big)\ln\big(\frac{\mathcal{A}-1}{\alpha}\big)-\nonumber\\&\!&\!
-8\big(\beta\ln(2)-\alpha-2\beta\big)\big(2\beta\ln(2)-2\alpha-\beta\big)\mathcal{A}+16\mathcal{A}^2\beta^2+16\big(2\beta\ln(2)-2\alpha-\beta\big)^{2}\Bigg\},
\label{d23}    
\end{eqnarray}
\begin{figure*}[!hbtp]
	\centering
	\includegraphics[width=.55\textwidth,keepaspectratio]{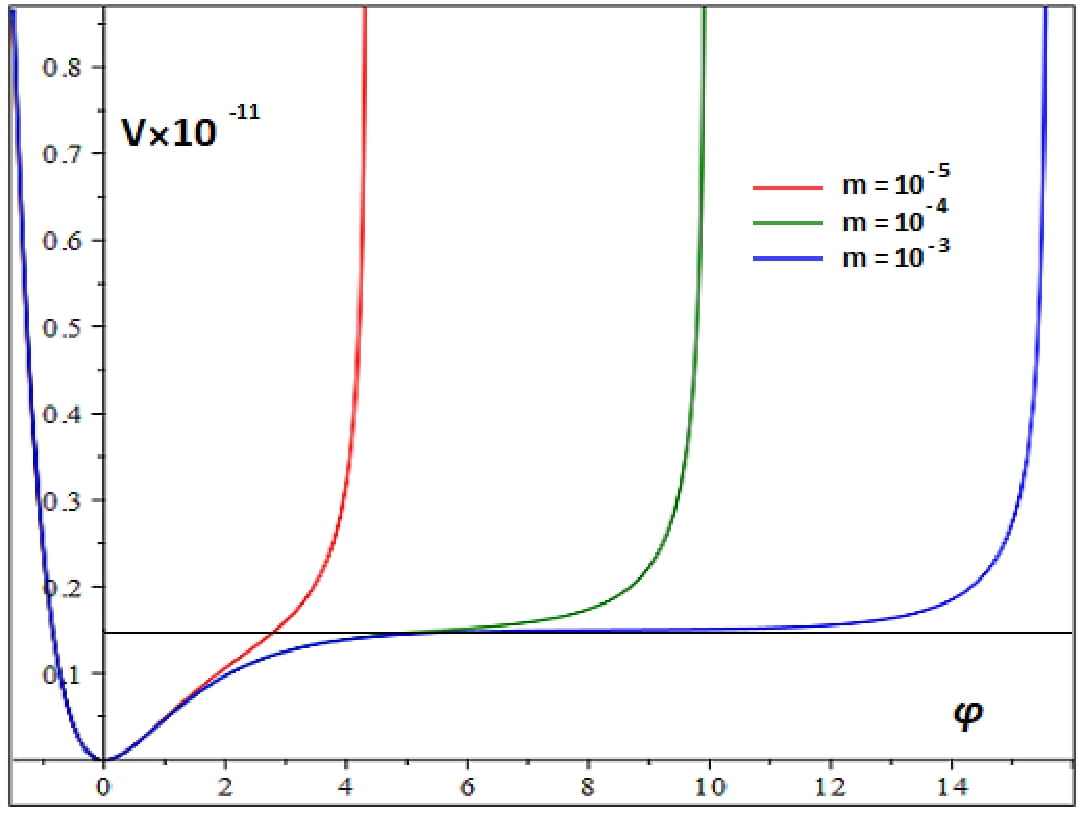}
	\caption{The potential of the viable logarithmic $f(R)$ model (\ref{e1}) for different values of $m=10^{-5}, 10^{-4}, 10^{-3}$ and $M\sim10^{-6}$.}
	\label{fig5}
\end{figure*}
\begin{equation}
r\simeq\frac{64\Big(2\beta\ln(\frac{\mathcal{A}-1}{2\alpha})+(1-\mathcal{A})\beta+2\alpha\Big)^{2}}{3\Big(2\beta\ln(\frac{\mathcal{A}-1}{2\alpha})+\beta+2\alpha\Big)^{2}\big(\mathcal{A}-1\big)^{2}},
\label{d24}    
\end{equation}
where the quantities are defined as
\begin{equation}
\mathcal{A}=\frac{2\alpha}{\beta}\Big(-e^{-\frac{N\beta}{3\alpha}}\mathcal{B}\mp1\Big)+1,\hspace{0.7cm}\mathcal{B}=1+\frac{\beta}{2\alpha}\Big(e^{\sqrt{\frac{2}{3}}\varphi_{f}}-1\Big).
\label{d25}
\end{equation}
Notice that the above spectral parameters can be rewritten as
\begin{equation}
n_{s}=(n_{s})_{s}+\delta\sqrt{\frac{16\epsilon_{s}}{3}},\hspace{1cm}r=r_{s}+16(-\delta\sqrt{\frac{4\epsilon_{s}}{3}}+\frac{\delta^{2}}{3}),
\label{d26}
\end{equation}  
where $(n_{s})_{s}$ and $r_{s}$ are the spectral index and the tensor-to-scalar ratio of the $R^2$ Starobinsky model (\ref{d13}), respectively. In Fig.\ref{fig4} we show the consistency relation $r=r(n_{s}) $ coming from the marginalized joint 68\% and 95\% CL regions of Planck 2018 alone and in combination with the BK14 or BK14+BAO datasets \cite{Planck:2018jri} on the logarithmic corrected model (\ref{d14}) for allowed values $\beta=-0.08\alpha, -0.1\alpha$ when $\alpha\sim10^{-9}$ is considered. 

In the case $\beta=-0.08\alpha$ (solid red line), the logarithmic model predicts $n_{s}=0.96324$, $r=0.0027253$ for $N=50$. In comparison with the $R^2$ Starobinsky model (for $N=50$), we realize that the logarithmic model presents a more favoured value of $n_{s}$ with a smaller value of $r$. Hence, the smallness of $r$ in the $R^2$ Starobinsky model is still valid even in the presence of the logarithmic corrections. For $N=60$, we obtain $n_{s}=0.97482$, $r=0.0049489$ which shows a less favoured value of $n_{s}$ with a larger value of $r$  compared to the predictions of the $R^{2}$ Starobinsky model (for $N=60$). As an important result, we find that improving one of two parameters $n_{s}$ and $r$ leads to losing the validity of the other. This point can be emphasized by taking a look at the predictions of the model for $\beta=-0.1\alpha$ (solid green line) in which the logarithmic model shows a more acceptable value of $r\sim10^{-2}$  with the excluded value of $n_{s}\sim0.98$ for both $N=50$ and $N=60$. 

\begin{figure*}[!hbtp]
	\centering
	\includegraphics[width=.70\textwidth,keepaspectratio]{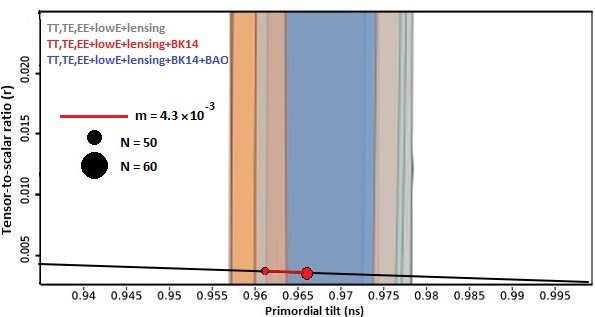}
\caption{The marginalized joint 68\% and 95\% CL regions for $n_{s}$ and $r$ at $k = 0.002$ Mpc$^{-1}$ from Planck alone and in combination with BK14 or BK14+BAO \cite{Planck:2018jri} and the $n_{s}-r$ constraints on the viable logarithmic $f(R)$ model (\ref{e1}) for allowed values $m=4.3\times10^{-3}$ and $M=4.5\times10^{-6}$.}
	\label{fig6}
\end{figure*}

\subsubsection{The $f(R)=R+\alpha\big(\beta R-\ln(1+\beta R)\big)$ model}

Another viable logarithmic $f(R)$ model, proposed in \cite{Amin:2015lnh},  is
\begin{equation}
f(R)=R+\alpha\big(\beta R-\ln(1+\beta R)\big),
\label{e1}
\end{equation}
where $\alpha=\frac{m^4}{3M^2}$ and $\beta=\frac{1}{m^2}$. This model leads to an accredited inflationary model with no singularity. Also, in the weak field limit, the model reduces to GR. Here, $m$ defines the scale below which inflation
starts and will be determined from the observations, while $M$ is the mass of scalaron in the Einstein frame, implying the end of inflation. From Eq.(\ref{a22}), the associated potential is given by
\begin{equation}
V(\varphi)=m^2e^{-2\sqrt{\frac{2}{3}}\varphi}\Bigg(1-e^{\sqrt{\frac{2}{3}}\varphi}-\frac{m^2}{3M^2}\ln\Big[1+\frac{3M^2}{m^2}\Big(1-e^{\sqrt{\frac{2}{3}}\varphi}\Big)\Big]\Bigg).
\label{e2}
\end{equation}
Fig.\ref{fig5} presents the behaviour of the viable logarithmic $f(R)$ model (\ref{e1}) for different values of $m=10^{-5}, 10^{-4}, 10^{-3}$ and $M\sim10^{-6}$ in which the potential starts to be flat at around $0.15\times10^{-11}$. Also, the potential tends to the infinity at $\varphi\rightarrow\infty$ as a theoretical requirement of the model \cite{Amin:2015lnh}.

The slow-roll parameters  are
\begin{equation}
\epsilon_V=\frac{\Big[2m^2\big(M^2(1-e^{\sqrt{\frac{2}{3}}\varphi})+\frac{m^2}{3}\big)\ln\big(1+\frac{3M^2}{m^2}(1-e^{\sqrt{\frac{2}{3}}\varphi})\big)+M^2\big((9M^2 +2m^2)e^{\sqrt{\frac{2}{3}}x}-3M^2(2+e^{2\sqrt{\frac{2}{3}}\varphi})-2m^2\big)\Big]^2}{3\big(M^2(1-e^{\sqrt{\frac{2}{3}}\varphi})+\frac{m^2}{3}\big)^2\Big[3M^2(1-e^{\sqrt{\frac{2}{3}}\varphi})- m^2\ln\big(1+\frac{3M^2}{m^2}(1-e^{\sqrt{\frac{2}{3}}\varphi})\big)\Big]^2},
\label{e3}
\end{equation}
\begin{eqnarray}
&\!&\!\eta_V=-\frac{2}{\big(M^2(1-e^{\sqrt{\frac{2}{3}}\varphi})+\frac{m^2}{3}\big)^2\Big[3M^2(1-e^{\sqrt{\frac{2}{3}}\varphi})- m^2\ln\big(1+\frac{3M^2}{m^2}(1-e^{\sqrt{\frac{2}{3}}\varphi})\big)\Big]}\Bigg\{-\frac{4}{3}m^2\times\nonumber\\&\!&\!
\times\big(2M^2(M^2+\frac{m^2}{3})e^{\sqrt{\frac{2}{3}}\varphi}- M^4e^{2\sqrt{\frac{2}{3}}\varphi}-(M^2+\frac{m^2}{3})^2\big)\ln\big(\frac{3M^2(1-e^{\sqrt{\frac{2}{3}}\varphi})}{m^2}+ 1\big)+M^2\Big(\big(9M^4+\nonumber\\&\!&\!
+\frac{13}{3}M^2m^2+\frac{4}{9}m^4\big)e^{\sqrt{\frac{2}{3}}\varphi}-(6M^4+ 2M^2m^2)e^{2\sqrt{\frac{2}{3}}\varphi}+ M^4e^{\sqrt{6}\varphi}-4(M^2+\frac{m^2}{3})^2\Big)\Bigg\}.
\label{e4}
\end{eqnarray}
By setting $\epsilon_V=1$ and choosing the positive sign and using the Taylor expansion for $\big|(3M^2/m^2)e^{\sqrt{\frac{2}{3}}\varphi_{f}}\big|\leq1$, the value of the scalar field $\varphi$ at the end of inflation for $M\ll m$ is given by
\begin{equation}
\varphi_{f}\simeq\sqrt{\frac{3}{2}}\ln\Bigg\{\frac{2m^2(1-m^2)-3\sqrt{3}M^2
\pm\sqrt{4m^8+8m^6(-1+3M^2)+4m^4(1+6(2+\sqrt{3})M^2)+12\sqrt{3}m^2M^2+27M^4}}{6(2+m^2)M^2}\Bigg\}.
\label{e5}
\end{equation}
Now, from the first slow-roll parameter (\ref{e3}), the number of \textit{e}-folds (\ref{d4}) can be obtained as
\begin{equation}
N\simeq\frac{\ln(\sigma)-\zeta\tanh^{-1}(\delta)-2\sqrt{\frac{2}{3}}\varphi}{\frac{8m^2}{3}},
\label{e6}
\end{equation}
where 
\begin{equation}
\delta\equiv\frac{m^2(3M^2e^{\sqrt{\frac{2}{3}}\varphi}-1)+6M^2e^{\sqrt{\frac{2}{3}}\varphi}+m^4}{m^2\sqrt{m^4+m^2(6M^2-2)+12M^2+1}},\hspace{0.5cm}\zeta\equiv\frac{2(m^2-6M^2-1)}{\sqrt{m^4+m^2(6M^2-2)+12M^2+1}},
\label{e7}
\end{equation}
\begin{equation}
\sigma\equiv3(m^2+2)M^2e^{2\sqrt{\frac{2}{3}}\varphi}+2m^2e^{\sqrt{\frac{2}{3}}\varphi}(m^2-1)-2m^4.
\label{e8}
\end{equation}
Moreover, the spectral parameters (\ref{d5}) in the limit of $M\ll m$, are 
\begin{eqnarray}
&\!&\!
n_{s}\simeq\frac{1}{3\big(3M^2(\mathcal{A}-1)- m^2\big)^2\Big(m^2\ln\Big[3(1-\mathcal{A})\frac{M^2}{m^2}+1\Big]+3M^2(\mathcal{A}-1)\Big)^2}\Bigg\{-5m^4\big(3M^2(\mathcal{A} -1)-m^2\big)^2\times
\nonumber\\&\!&\!\times\ln\Big[3(1-\mathcal{A})\frac{M^2}{m^2}+1\Big]-6M^2m^2\Big(-18M^4e^{\sqrt{6}\mathcal{B}}+9(\mathcal{A}^3-3\mathcal{A}^2+9\mathcal{A}-5)M^4-6m^2(\mathcal{A}^2-7\mathcal{A} +5)M^2+\nonumber\\&\!&\!
+5m^4(\mathcal{A}-1)\Big)\ln\Big[3(1-\mathcal{A})\frac{M^2}{m^2}+1\Big]-9M^4(\mathcal{A}-1)\Big(-36M^4e^{\sqrt{6}\mathcal{B}}+9(3\mathcal{A}^3+3\mathcal{A}^2+3\mathcal{A}-5)M^4+6m^2\times\nonumber\\&\!&\!
\times(3\mathcal{A}^2 +4\mathcal{A}-5)M^2+5m^4(\mathcal{A}-1)\Big)\Bigg\},
\label{e9}
\end{eqnarray}
\begin{equation}
r\simeq\frac{16\Big(2m^2\big(3M^2(\mathcal{A}-1)- m^2\big)\Big[3(1-\mathcal{A})\frac{M^2}{m^2}+1\Big]+3M^2\big(3(\mathcal{A}-2)M^2-2m^2\big)(\mathcal{A}-1)\Big)^2}{3\big(3M^2(\mathcal{A}-1)- m^2\big)^2\Big(m^2\ln\Big[3(1-\mathcal{A})\frac{M^2}{m^2}+1\Big]+3M^2(\mathcal{A}-1)\Big)^2},
\label{e10}
\end{equation}
where $\mathcal{A}=\frac{1+\sqrt{1+8m}}{4m}+e^{\frac{8Nm^2}{3}}$ and $\mathcal{B}=\sqrt{\frac{3}{2}}\ln(\mathcal{A})$.
Fig.\ref{fig6} presents the consistency relation $r=r(n_{s})$ coming from the marginalized joint 68\% and 95\% CL regions of Planck 2018 alone and in combination with the BK14 or BK14+BAO datasets \cite{Planck:2018jri} on the viable logarithmic model (\ref{e1}) for allowed values $m=4.3\times10^{-3}$ and $M=4.5\times10^{-6}$. As we see, the model predicts the observationally acceptable values of $n_{s}=0.96126$, $r=0.002808$ and  $n_{s}=0.96612$, $r=0.0023928$ for $N=50$ and $N=60$, respectively. As a result, in comparison with the $R^2$ Starobinsky model, we find that the viable logarithmic model (\ref{e1}) predicts more favoured values of $n_{s}$ with smaller values of $r$ for both $N=50$ and $N=60$. Hence, the smallness of $r$ in $f(R)$ gravity models remains an ambiguity even when such logarithmic corrections are added in the Hilbert-Einstein  action (\ref{6}).

\subsection{The Hubble-Slow-Roll Inflation}
Here, we present the inflationary analysis in the Jordan frame using the action (\ref{a1}) in which a single scalar field is responsible for driving the inflationary epoch. To this aim, we consider the HSR formalism using the well-known function of $R$
\begin{equation}
f(R)=\alpha R+\beta R^2, 
\label{g1}
\end{equation}
where $\alpha$ and $\beta$ are the parameters of the model. Using the definition $\dot{F}=\dot{R}F'$, the Friedman Eq. (\ref{a7}) can be rewritten as
\begin{equation}
H^{2}=\frac{\frac{\beta R^2}{2}-6\beta H\dot{R}+\rho}{3(\alpha+2\beta R)},
\label{g2}
\end{equation}
where again dot is the derivative with respect to cosmic time $t$. Notice that by inserting the slow-roll conditions $|\dot{H}/H^2|\ll1$ and $|\ddot{H}/H\dot{H}|\ll1$ in the relation $R=6(2H^2+\dot{H})$, we have $R\simeq12H^2$  and $\dot{R}\simeq24H\dot{H}$. Assuming a single scalar field as the fluid filling the universe during inflation, the energy density and pressure of the prefect fluid are given by 
\begin{equation}
\rho_{\varphi}=\frac{\dot{\varphi}^{2}}{2}+V(\varphi),\hspace{1cm}p_{\varphi}=\frac{\dot{\varphi}^{2}}{2}-V(\varphi),
\label{g3}
\end{equation}
where $V$ is the potential of the scalar field. Note that under the slow-roll condition $\dot{\varphi}^{2}/2\ll V(\varphi)$, inflaton behaves like the cosmological constant $\Lambda$ since EoS of inflaton is $w\simeq-1$. Moreover, the slow-roll parameters in the HSR formalism are introduced as
\begin{equation}
\epsilon_{H}=-\frac{\dot{H}}{H^{2}},\hspace{1cm}\eta_{H}=-\frac{\ddot{H}}{2H\dot{H}}.
\label{g4}
\end{equation}
During the inflationary era, $\epsilon_H\ll1$ and $\eta_H\ll1$, while inflation ends when the condition $\epsilon_H=1$ is fulfilled. The number of \textit{e}-folds of the model can be found as
\begin{equation}
N=\int^{t_{f}}_{t_{i}}Hdt,
\label{g5}
\end{equation}
where the subscribes $f$ and $i$ denote  the end and the beginning of inflation, respectively. Also, the spectral index and the tensor-to-scalar ratio in the HSR formalism are defined by
\begin{equation}
n_{s}=1-4\epsilon_{H}+2\eta_{H},\hspace{1cm}r=16\epsilon_{H}.
\label{g6}
\end{equation}
Let us now specify the study for some well-known inflationary models.
\subsubsection{Monomial Potential}
As the first case, we consider models characterized by a monomial potential like 
\begin{equation}
V(\varphi)=\lambda\varphi^{n},
\label{g7}
\end{equation}
where $\lambda$ is the parameter of the model giving a mass scale. This class of potentials comes from particle physics and describe the interaction with other fields. The number $n$ is usually a positive integer \eg $n=2$ belongs to the well-behaved potential $V=\frac{1}{2}M^{2}\varphi^{2}$ where $M$ is the mass of the inflaton. Also, in the non-minimal coupling (NMC) models involved with the extra term $\xi\varphi^{2}R$, we deal with an effective mass $m_{eff}=\sqrt{M^{2}+\xi R}$. The case $n=4$ corresponds to the chaotic inflation described with the potential $V=\lambda\varphi^{4}$ where   $\lambda$ is a self-interacting constant \cite{Linde:1983gd,Linde:2007fr}. Furthermore, some fractional powers e.g $n=2/3$ and $n=4/3$ could arise in axion monodromy inflation \cite{Flauger:2009ab,McAllister:2008hb,Kaloper:2008fb,Kaloper:2011jz,Kaloper:2014zba}. Let us now perform the inflationary analysis for the interesting cases $n=2,4/3,1,2/3$, separately. 
\begin{itemize}
\item
\textbf{Case \boldmath$n=2$.}
From Eq.(\ref{g2}), the Friedman equation  for  $n=2$ gives
\begin{equation}
H^{2}=\frac{\lambda\varphi^2+3C\varphi^{\frac{\alpha}{16\beta\lambda}}\big(\alpha-32\beta\lambda\big)}{3\big(\alpha-32\beta\lambda\big)},
\label{g8}
\end{equation}
where $C$ is an integration constant. Now by choosing a unitary $C$  and  using the definition of the slow-roll parameters (\ref{g4}), we have
\begin{equation}
\epsilon_{H}=\frac{(\alpha-32\beta\lambda)\Big(32\lambda^2\varphi^2\beta+3\alpha \varphi^{\frac{\alpha}{16\beta\lambda}}(\alpha-32\beta\lambda)\Big)}{16\beta\Big(\lambda\varphi^{2}+3\varphi^{\frac{\alpha}{16\beta\lambda}}(\alpha-32\beta\lambda)\Big)^2},
\label{g9}
\end{equation}
\begin{equation}
\eta_{H}=\frac{3\lambda  \varphi^{2+\frac{\alpha}{16\beta\lambda}}\big(\alpha-32\beta\lambda\big)^4}{16\beta\Big(3\alpha (\alpha-32\beta\lambda)\varphi^{\frac{\alpha}{16\beta\lambda}}+ 32\lambda^2\varphi^2\beta\Big)\Big(3(\alpha-32\beta\lambda)\varphi^{\frac{\alpha}{16\beta\lambda}}+\lambda\varphi^2\Big)^2}.
\label{g10}
\end{equation}
By setting the condition $\epsilon_{H}=1$, the value of inflaton at the end of inflation $\varphi_{f}$ is obtained by the equation
\begin{equation}
1=\frac{(\alpha-32\beta\lambda)\Big(32\lambda^2\varphi_{f}^2\beta+3\alpha \varphi_{f}^{\frac{\alpha}{16\beta\lambda}}(\alpha-32\beta\lambda)\Big)}{16\beta\Big(\lambda\varphi_{f}^{2}+3\varphi_{f}^{\frac{\alpha}{16\beta\lambda}}(\alpha-32\beta\lambda)\Big)^2}.
\label{g11}
\end{equation}
From Eq.(\ref{g5}) and using the Hubble parameter (\ref{g8}) and the reduced Klein-Gordon equation $3H\dot{\varphi}\simeq-V'$, the number of \textit{e}-folds
is
\begin{equation}
N=\frac{96\beta(\alpha-32\beta\lambda)\varphi^{\frac{\alpha}{16\beta\lambda}}+\alpha\varphi^{2}}{4\alpha(\alpha-32\beta\lambda)}\Big|^{\varphi_{i}}_{\varphi_{f}}.
\label{g12}
\end{equation}
To find an expression for $\varphi_i$, we use the following binomial series representation 
\begin{equation}
1+ax^n=1+a+\sum_{\nu=1}^{\infty}\frac{n^{\nu} a\ln^{\nu}(x)}{\nu!}.
\label{g13}    
\end{equation}
\begin{figure*}[!hbtp]
     \centering	\includegraphics[width=.45\textwidth,keepaspectratio]{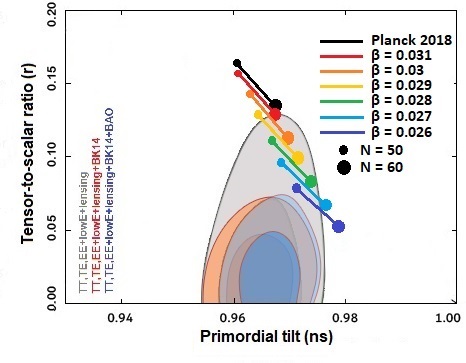}
     \includegraphics[width=.45\textwidth,keepaspectratio]{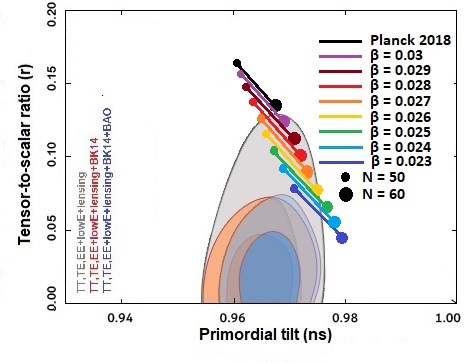}
\caption{The marginalized joint 68\% and 95\% CL regions for $n_{s}$ and $r$ at $k = 0.002$ Mpc$^{-1}$ from Planck alone and in combination with BK14 or BK14+BAO \cite{Planck:2018jri} and the $n_{s}-r$ constraints on the parameters space of the monomial potential (\ref{g7}) in $f(R)$ gravity with $n=2$ associated to the first (left panel) and second (right panel) orders of the binomial series of the number of \textit{e}-folds (\ref{g12}). The results are obtained for the case that the parameters $C$, $\alpha$ and $\lambda$ are assumed  unit.}
\label{fig7}
\end{figure*}
Then, by keeping the terms up to the first and second orders of the binomial series, the number of \textit{e}-folds takes the two following forms
\begin{equation}
N\simeq\frac{\varphi_i^2}{4(\alpha-32\beta\lambda)},\hspace{1cm}N\simeq\frac{\varphi_i^2}{4(\alpha-32\beta\lambda)}\Big(1+\frac{96(\alpha-32\beta\lambda)}{\alpha}\Big),
\label{g14}
\end{equation}
respectively. Now, using the slow-roll parameters (\ref{g9}) and (\ref{g10}) and also the number of \textit{e}-folds (\ref{g14}), the spectral parameters (\ref{g6}) are calculated by 
\begin{eqnarray}
n_{s}\simeq1-\frac{(\alpha-32\beta\lambda)\Big(32\lambda^2\mathcal{A}\beta+3\alpha\mathcal{A}^{\frac{\alpha}{32\beta\lambda}}(\alpha-32\beta\lambda)\Big)}{4\beta\Big(\lambda\mathcal{A}+3\mathcal{A}^{\frac{\alpha}{32\beta\lambda}}(\alpha-32\beta\lambda)\Big)^2}\hspace{5cm}\nonumber\\
+\frac{3\lambda\mathcal{A}^{1+\frac{\alpha}{32\beta\lambda}}\big(\alpha-32\beta\lambda\big)^4}{8\beta\Big(3\alpha (\alpha-32\beta\lambda)\mathcal{A}^{\frac{\alpha}{32\beta\lambda}}+ 32\lambda^2\mathcal{A}\beta\Big)\Big(3(\alpha-32\beta\lambda)\mathcal{A}^{\frac{\alpha}{32\beta\lambda}}+\lambda\mathcal{A}\Big)^2},\hspace{0.2cm}
\label{g15}
\end{eqnarray}
\begin{equation}
r\simeq\frac{(\alpha-32\beta\lambda)\Big(32\lambda^2\mathcal{A}\beta+3\alpha\mathcal{A}^{\frac{\alpha}{32\beta\lambda}}(\alpha-32\beta\lambda)\Big)}{\beta\Big(\lambda\mathcal{A}+3\mathcal{A}^{\frac{\alpha}{32\beta\lambda}}(\alpha-32\beta\lambda)\Big)^2},
\label{g16}
\end{equation}
where $\mathcal{A}=4N(\alpha-32\beta\lambda)$ and $\mathcal{A}=\frac{4N\alpha(\alpha-32\beta\lambda)}{\alpha+96(\alpha-32\beta\lambda)}$ associated to the first and second orders of the binomial series of the number of \textit{e}-folds (\ref{g12}).

\item
\textbf{Case \boldmath$n=\frac{4}{3}$.} From Eq.(\ref{g2}), the Friedman equation  for the power $n=\frac{4}{3}$ gives
\begin{equation}
H^{2}=\frac{e^{\frac{9\alpha\varphi^{2/3}}{64\lambda\beta}}\Big(243\alpha^3C+4096\lambda^3\beta^2\Gamma[3,\frac{9\alpha\varphi^{2/3}}{64\lambda\beta}]\Big)}{243\alpha^{3}},
\label{h1}
\end{equation}
where $\Gamma$ is the incomplete Gamma function and $C$ is an integration constant. Using Eq.(\ref{g4}), the slow-roll parameters of the model for $C=1$ take the following form
\begin{equation}
\epsilon_{H}=\frac{81\alpha^4e^{-\frac{9\alpha\varphi^{2/3}}{32\lambda\beta}}\bigg(81\alpha^2\Big(3\alpha e^{\frac{9\alpha\varphi^{2/3}}{64\lambda\beta}}-\lambda\varphi^{4/3}\Big)+4096\lambda^3\beta^2e^{\frac{9\alpha\varphi^{2/3}}{64\lambda\beta}}\Gamma[3,\frac{9\alpha\varphi^{2/3}}{64\lambda\beta}]\bigg)}{16\beta\bigg(243\alpha^3+4096\lambda^3\beta^2\Gamma[3,\frac{9\alpha\varphi^{2/3}}{64\lambda\beta}]\bigg)^2},
\label{h2}
\end{equation}
\begin{figure*}[!hbtp]
     \centering	\includegraphics[width=.45\textwidth,keepaspectratio]{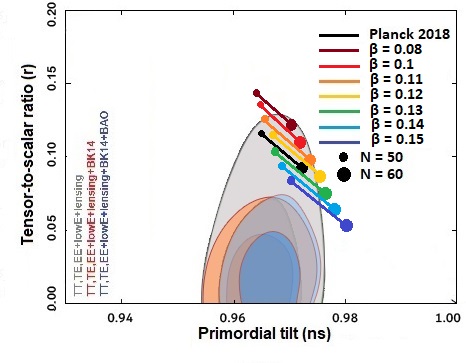}
     \includegraphics[width=.45\textwidth,keepaspectratio]{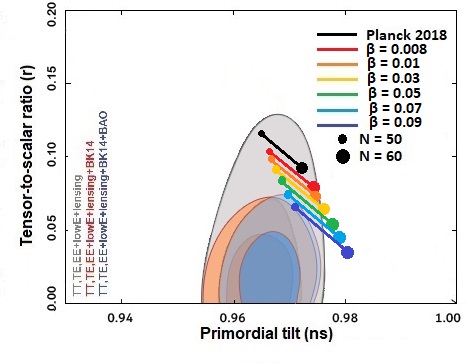}
\caption{The marginalized joint 68\% and 95\% CL regions for $n_{s}$ and $r$ at $k = 0.002$ Mpc$^{-1}$ from Planck alone and in combination with BK14 or BK14+BAO \cite{Planck:2018jri} and the $n_{s}-r$ constraints on the parameters space of the monomial potential (\ref{g7}) in $f(R)$ gravity with $n=4/3$ associated to a non-zero (left panel) and zero (right panel) integration constant $C$. The results are obtained for the case that the parameters $\alpha$ and $\lambda$ are considered as the unit.}
\label{fig8}
\end{figure*}
\begin{eqnarray}
&\!&\!\eta_{H}=-\frac{729\lambda\alpha^5\varphi^{2/3}e^{-\frac{9\alpha\varphi^{2/3}}{32\lambda\beta}}\Big(243\alpha^3+4096\lambda^3\beta^2\Gamma[3,\frac{9\alpha\varphi^{2/3}}{64\lambda\beta}]\Big)^{-2}}{16\Big(81\alpha^2\beta\big(3\alpha e^{\frac{9\alpha\varphi^{2/3}}{64\lambda\beta}}-\lambda\varphi^{4/3}\big)+4096\lambda^3\beta^3e^{\frac{9\alpha\varphi^{2/3}}{64\lambda\beta}}\Gamma[3,\frac{9\alpha\varphi^{2/3}}{64\lambda\beta}]\Big)}\Bigg\{243\alpha^3\Big(128\lambda\beta e^{\frac{9\alpha\varphi^{2/3}}{64\lambda\beta}}-\nonumber\\&\!&\!
-9\alpha\varphi^{2/3}e^{\frac{9\alpha \varphi^{2/3}}{64\lambda\beta}}+3\lambda\varphi^2\Big)+4096\lambda^3\beta^2e^{\frac{9\alpha\varphi^{2/3}}{64\lambda\beta}}(128\lambda\beta-9\alpha\varphi^{2/3})\Gamma[3,\frac{9\alpha\varphi^{2/3}}{64\lambda\beta}]
\Bigg\}.
\label{h3}
\end{eqnarray}
For the end of inflation ($\epsilon_H=1$), the value of the scalar field $\varphi$ can be found by
\begin{equation}
1=\frac{81\alpha^4e^{-\frac{9\alpha\varphi_{f}^{2/3}}{32\lambda\beta}}\bigg(81\alpha^2\Big(3\alpha e^{\frac{9\alpha\varphi_{f}^{2/3}}{64\lambda\beta}}-\lambda\varphi_{f}^{4/3}\Big)+4096\lambda^3\beta^2e^{\frac{9\alpha\varphi_{f}^{2/3}}{64\lambda\beta}}\Gamma[3,\frac{9\alpha\varphi_{f}^{2/3}}{64\lambda\beta}]\bigg)}{16\beta\bigg(243\alpha^3+4096\lambda^3\beta^2\Gamma[3,\frac{9\alpha\varphi_{f}^{2/3}}{64\lambda\beta}]\bigg)^2}.
\label{h4}
\end{equation}
Using Eq.(\ref{g5}), the number of \textit{e}-folds is given by 
\begin{equation}
N=\frac{1}{648\alpha^4}\Bigg[243\alpha^3\Big(\varphi^2+64\beta e^{\frac{9\alpha\varphi^{2/3}}{64\lambda\beta}}\Big)+262144\lambda^3\beta^3e^{{\frac{9\alpha\varphi^{2/3}}{64\lambda\beta}}}\Gamma(3,\frac{9\alpha\varphi^{2/3}}{64\lambda\beta})\Bigg]^{\varphi_{i}}_{\varphi_{f}}.
\label{h5}
\end{equation}
By hiring the slow-roll parameters (\ref{h2}) and (\ref{h3}) and the above number of \textit{e}-folds,
the spectral parameters (\ref{g6}) 
are obtained as
\begin{eqnarray}
n_{s}\simeq1-\frac{81\alpha^4e^{-\frac{9\alpha\mathcal{A}}{32\lambda\beta}}\bigg(81\alpha^2\Big(3\alpha e^{\frac{9\alpha\mathcal{A}}{64\lambda\beta}}-\lambda\mathcal{A}^2\Big)+4096\lambda^3\beta^2e^{\frac{9\alpha\mathcal{A}}{64\lambda\beta}}\Gamma[3,\frac{9\alpha\mathcal{A}}{64\lambda\beta}]\bigg)}{4\beta\bigg(243\alpha^3+4096\lambda^3\beta^2\Gamma[3,\frac{9\alpha\mathcal{A}}{64\lambda\beta}]\bigg)^2}-\hspace{4cm}\nonumber\\-\frac{729\lambda\alpha^5\mathcal{A}e^{-\frac{9\alpha\mathcal{A}}{32\lambda\beta}}\Big(243\alpha^3+4096\lambda^3\beta^2\Gamma[3,\frac{9\alpha\mathcal{A}}{64\lambda\beta}]\Big)^{-2}}{8\Big(81\alpha^2\beta\big(3\alpha e^{\frac{9\alpha\mathcal{A}}{64\lambda\beta}}-\lambda\mathcal{A}^2\big)+4096\lambda^3\beta^3e^{\frac{9\alpha\mathcal{A}}{64\lambda\beta}}\Gamma[3,\frac{9\alpha\mathcal{A}}{64\lambda\beta}]\Big)}\Bigg\{243\alpha^3\Big(128\lambda\beta e^{\frac{9\alpha\mathcal{A}}{64\lambda\beta}}-\hspace{2cm}\nonumber\\
-9\alpha\mathcal{A}e^{\frac{9\alpha \mathcal{A}}{64\lambda\beta}}+3\lambda\mathcal{A}^3\Big)+4096\lambda^3\beta^2e^{\frac{9\alpha\mathcal{A}}{64\lambda\beta}}(128\lambda\beta-9\alpha\mathcal{A})\Gamma[3,\frac{9\alpha\mathcal{A}}{64\lambda\beta}]
\Bigg\},\hspace{1cm}
\label{h6}
\end{eqnarray}
\begin{figure*}[!hbtp]
     \centering	\includegraphics[width=.45\textwidth,keepaspectratio]{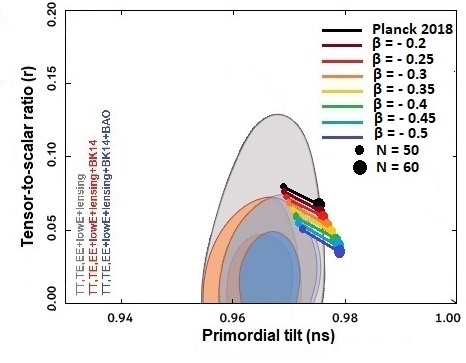}
     \includegraphics[width=.45\textwidth,keepaspectratio]{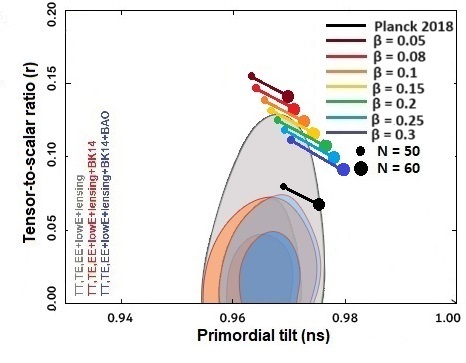}
\caption{The marginalized joint 68\% and 95\% CL regions for $n_{s}$ and $r$ at $k = 0.002$ Mpc$^{-1}$ from Planck alone and in combination with BK14 or BK14+BAO \cite{Planck:2018jri} and the $n_{s}-r$ constraints on the parameters space of the monomial potential (\ref{g7}) in $f(R)$ gravity with $n=1$ associated to the negative (left panel) and positive (right panel) values of $\beta$. The results are obtained for the case that the parameters $C$, $\alpha$ and $\lambda$ are considered as the unit.}
\label{fig9}
\end{figure*}
\begin{equation}
r\simeq\frac{81\alpha^4e^{-\frac{9\alpha\mathcal{A}}{32\lambda\beta}}\bigg(81\alpha^2\Big(3\alpha e^{\frac{9\alpha\mathcal{A}}{64\lambda\beta}}-\lambda\mathcal{A}^2\Big)+4096\lambda^3\beta^2e^{\frac{9\alpha\mathcal{A}}{64\lambda\beta}}\Gamma[3,\frac{9\alpha\mathcal{A}}{64\lambda\beta}]\bigg)}{\beta\bigg(243\alpha^3+4096\lambda^3\beta^2\Gamma[3,\frac{9\alpha\mathcal{A}}{64\lambda\beta}]\bigg)^2},
\label{h7}
\end{equation}
where $\mathcal{A}=\frac{64\lambda\beta}{9\alpha}\ln\Big(\frac{N\alpha}{24\beta}\Big)$. In the case of $C=0$, the expressions of the spectral index $n_s$ and the tensor-to-scalar ratio $r$  reduce to
\begin{eqnarray}
n_{s}\simeq1-\frac{81\alpha^4e^{-\frac{9\alpha\mathcal{A}}{32\lambda\beta}}\bigg(-81\lambda\alpha^2\mathcal{A}^2+4096\lambda^3\beta^2e^{\frac{9\alpha\mathcal{A}}{64\lambda\beta}}\Gamma[3,\frac{9\alpha\mathcal{A}}{64\lambda\beta}]\bigg)}{4\beta\bigg(4096\lambda^3\beta^2\Gamma[3,\frac{9\alpha\mathcal{A}}{64\lambda\beta}]\bigg)^2}-\hspace{5cm}\nonumber\\-\frac{729\lambda\alpha^5\mathcal{A}e^{-\frac{9\alpha\mathcal{A}}{32\lambda\beta}}\Big(4096\lambda^3\beta^2\Gamma[3,\frac{9\alpha\mathcal{A}}{64\lambda\beta}]\Big)^{-2}}{8\Big(-81\lambda\alpha^2\beta\mathcal{A}^2+4096\lambda^3\beta^3e^{\frac{9\alpha\mathcal{A}}{64\lambda\beta}}\Gamma[3,\frac{9\alpha\mathcal{A}}{64\lambda\beta}]\Big)}\Bigg\{729\lambda\alpha^3\mathcal{A}^3+\hspace{1.5cm}\nonumber\\+4096\lambda^3\beta^2e^{\frac{9\alpha\mathcal{A}}{64\lambda\beta}}(128\lambda\beta-9\alpha\mathcal{A})\Gamma[3,\frac{9\alpha\mathcal{A}}{64\lambda\beta}]
\Bigg\},\hspace{0.7cm}
\label{h8}
\end{eqnarray}
\begin{equation}
r\simeq\frac{81\alpha^4e^{-\frac{9\alpha\mathcal{A}}{32\lambda\beta}}\bigg(-81\lambda\alpha^2\mathcal{A}^2+4096\lambda^3\beta^2e^{\frac{9\alpha\mathcal{A}}{64\lambda\beta}}\Gamma[3,\frac{9\alpha\mathcal{A}}{64\lambda\beta}]\bigg)}{\beta\bigg(4096\lambda^3\beta^2\Gamma[3,\frac{9\alpha\mathcal{A}}{64\lambda\beta}]\bigg)^2},
\label{h9}
\end{equation}
where $\mathcal{A}=\sqrt[3]{\frac{8}{3}N\alpha}$.
\item
\textbf{Case \boldmath$n=1$.} From Eq.(\ref{g2}), the Friedman equation  for the power $n=1$ gives
\begin{equation}
H^{2}=\frac{\alpha\lambda\varphi+8\beta\lambda^2+3Ce^{\frac{\alpha\varphi}{8\lambda\beta}}\alpha^2}{3\alpha^2}.
\label{i1}
\end{equation}
\begin{figure*}[!hbtp]
     \centering	\includegraphics[width=.45\textwidth,keepaspectratio]{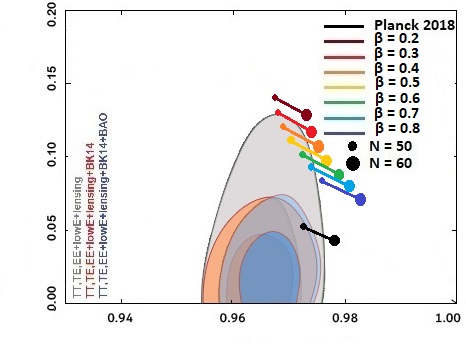}
     \includegraphics[width=.45\textwidth,keepaspectratio]{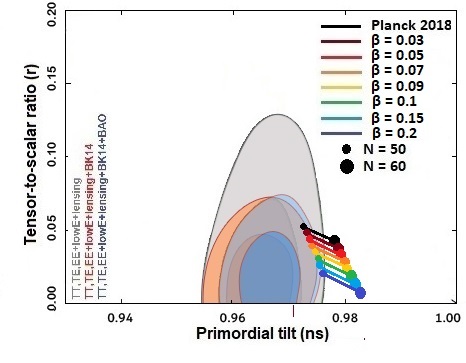}
\caption{The marginalized joint 68\% and 95\% CL regions for $n_{s}$ and $r$ at $k = 0.002$ Mpc$^{-1}$ from Planck alone and in combination with BK14 or BK14+BAO \cite{Planck:2018jri} and the $n_{s}-r$ constraints on the parameters space of the monomial potential (\ref{g7}) in $f(R)$ gravity with $n=2/3$ associated to a non-zero (left panel) and zero (right panel) integration constant $C$. The results are obtained for the case that the parameters $\alpha$ and $\lambda$ are considered as the unit.}
\label{fig10}
\end{figure*}
As above, $C$ is an  integration constant. Then, by fixing $C$ as the unit, we find the slow-roll parameters of the model as follows
\begin{equation}
\epsilon_{H}=\frac{\alpha^3\Big(8\beta\lambda^2+3e^{\frac{\alpha\varphi}{8\lambda\beta}}\alpha^2\Big)}{16\beta\Big(\alpha\lambda\varphi+8\beta\lambda^2+3e^{\frac{\alpha\varphi}{8\lambda\beta}}\alpha^2\Big)^2},
\label{i2}
\end{equation}
\begin{equation}
\eta_{H}=\frac{\lambda\alpha^3\Big(3\alpha^2(\alpha\varphi-8\beta\lambda)e^{\frac{\alpha\varphi}{8\lambda\beta}}-64\lambda^3\beta^2\Big)}{16\beta\big(8\beta\lambda^2+3e^{\frac{\alpha\varphi}{8\lambda\beta}}\alpha^2\big)\Big(\alpha\lambda\varphi+8\beta\lambda^2+3e^{\frac{\alpha\varphi}{8\lambda\beta}}\alpha^2\Big)^2}.
\label{i3}
\end{equation}
At the end of inflation ($\epsilon_{H}=1$), the value of inflaton can be derived by the equation
\begin{equation}
1=\frac{\alpha^3\Big(8\beta\lambda^2+3e^{\frac{\alpha\varphi_{f}}{8\lambda\beta}}\alpha^2\Big)}{16\beta\Big(\alpha\lambda\varphi_{f}+8\beta\lambda^2+3e^{\frac{\alpha\varphi_{f}}{8\lambda\beta}}\alpha^2\Big)^2}.
\label{i4}
\end{equation}
Also, using Eq.(\ref{g5}), the number of \textit{e}-folds of the model is 
\begin{equation}
N=
\frac{\alpha\varphi^2+16\beta\lambda\varphi+48\beta e^{\frac{\alpha\varphi}{8\lambda\beta}}\alpha}{2\alpha^2}\Big|^{\varphi_{i}}_{\varphi_{f}}.
\label{i5}
\end{equation}
Plugging the slow-roll parameters (\ref{i2}) and (\ref{i3}) into Eq.(\ref{g6}) and then using the above number of \textit{e}-folds,
the spectral parameters
can be found as
\begin{eqnarray}
n_{s}\simeq1-\frac{\alpha^3\Big(8\beta\lambda^2+3e^{\frac{\alpha\mathcal{A}}{8\lambda\beta}}\alpha^2\Big)}{4\beta\Big(\alpha\lambda\mathcal{A}+8\beta\lambda^2+3e^{\frac{\alpha\varphi}{8\lambda\beta}}\alpha^2\Big)^2}+\frac{\lambda\alpha^3\Big(3\alpha^2(\alpha\mathcal{A}-8\beta\lambda)e^{\frac{\alpha\mathcal{A}}{8\lambda\beta}}-64\lambda^3\beta^2\Big)}{8\beta\big(8\beta\lambda^2+3e^{\frac{\alpha\mathcal{A}}{8\lambda\beta}}\alpha^2\big)\Big(\alpha\lambda\mathcal{A}+8\beta\lambda^2+3e^{\frac{\alpha\mathcal{A}}{8\lambda\beta}}\alpha^2\Big)^2},
\label{i6}
\end{eqnarray}
\begin{eqnarray}
r\simeq\frac{\alpha^3\Big(8\beta\lambda^2+3e^{\frac{\alpha\mathcal{A}}{8\lambda\beta}}\alpha^2\Big)}{\beta\Big(\alpha\lambda\mathcal{A}+8\beta\lambda^2+3e^{\frac{\alpha\mathcal{A}}{8\lambda\beta}}\alpha^2\Big)^2},
\label{i7}
\end{eqnarray}
where $\mathcal{A}=-\frac{8\lambda\beta}{\alpha}+\frac{1}{\alpha}\sqrt{2N\alpha^3+64\beta^2\lambda^2}$ or $\mathcal{A}=\frac{8\lambda\beta}{\alpha}\ln(\frac{N\alpha}{24\beta})$ associated to the negative and positive values of $\beta$.
\item
\textbf{Case \boldmath$n=\frac{2}{3}$.}  From Eq.(\ref{g2}), the Friedman equation  for the power $n=\frac{2}{3}$ can be solved as
\begin{equation}
H^{2}=e^{\frac{9\alpha\varphi^{4/3}}{64\lambda\beta}}\Bigg(C+\frac{8\varphi^2\Gamma[\frac{3}{2}, \frac{9\alpha\varphi^{4/3}}{64\lambda\beta}]}{9\beta(\frac{\alpha\varphi^{4/3}}{\lambda\beta})^{3/2}}\Bigg),
\label{j1}
\end{equation}
where $C$ is the integration constant. By choosing the integration constant $C$ as the unit and using the definition of the slow-roll parameters (\ref{g4}), we have
\begin{equation}
\epsilon_{H}=\frac{3\alpha^3\varphi^{4/3}e^{-\frac{9\alpha\varphi^{4/3}}{32\lambda\beta}}\bigg(-3\lambda\varphi^{2/3}+e^{\frac{9\alpha\varphi^{4/3}}{64\lambda\beta}}\Big(9\alpha+\frac{8\lambda\varphi^{2/3}\Gamma[\frac{3}{2}, \frac{9\alpha\varphi^{4/3}}{64\lambda\beta}]}{\sqrt{\frac{\alpha\varphi^{4/3}}{\lambda\beta}}}\Big)\bigg)}{16\beta^2\lambda\Big(9\alpha\sqrt{\frac{\alpha\varphi^{4/3}}{\lambda\beta}}+8\lambda\varphi^{2/3}\Gamma[\frac{3}{2}, \frac{9\alpha\varphi^{4/3}}{64\lambda\beta}]\Big)^2},
\label{j2}
\end{equation}
\begin{eqnarray}
&\!&\!\eta_{H}=\frac{e^{-\frac{9\alpha\varphi^{4/3}}{32\lambda\beta}}\Big(9\alpha \sqrt{\frac{\alpha\varphi^{4/3}}{\lambda\beta}}+8\lambda\varphi^{2/3}\Gamma[\frac{3}{2}, \frac{9\alpha\varphi^{4/3}}{64\lambda\beta}]\Big)^{-1}\Big(9\alpha^2\varphi^{2/3}+8\beta\lambda^2\sqrt{\frac{\alpha\varphi^{4/3}}{\lambda\beta}}\Gamma[\frac{3}{2}, \frac{9\alpha\varphi^{4/3}}{64\lambda\beta}]\Big)^{-1}}{16\beta^2\Big(-3\sqrt{\frac{\alpha\varphi^{4/3}}{\lambda\beta}}\big(\lambda\varphi^{2/3}-3\alpha e^{\frac{9\alpha\varphi^{4/3}}{64\lambda\beta}}\big)+8\lambda\varphi^{2/3}e^{\frac{9\alpha\varphi^{4/3}}{64\lambda\beta}}\Gamma[\frac{3}{2}, \frac{9\alpha\varphi^{4/3}}{64\lambda\beta}]\Big)}\Bigg\{-9\alpha^4\varphi^{4/3}\times\nonumber\\&\!&\!
\times\Big(3\lambda\varphi^2+e^{\frac{9\alpha\varphi^{4/3}}{64\lambda\beta}}\big(32\lambda\beta-9\alpha\varphi^{4/3}\big)\Big)+\frac{8\lambda\alpha^3\varphi^{2}e^{\frac{9\alpha\varphi^{4/3}}{64\lambda\beta}}\big(9\alpha\varphi^{4/3}-32\lambda\beta\big)\Gamma[\frac{3}{2}, \frac{9\alpha\varphi^{4/3}}{64\lambda\beta}]}{\sqrt{\frac{\alpha\varphi^{4/3}}{\lambda\beta}}}\Bigg\}.
\label{j3}
\end{eqnarray}
For the end of inflation, the value of scalar field $\varphi$ is obtained by the following equation
\begin{equation}
1=\frac{3\alpha^3\varphi_{f}^{4/3}e^{-\frac{9\alpha\varphi_{f}^{4/3}}{32\lambda\beta}}\bigg(-3\lambda\varphi_{f}^{2/3}+e^{\frac{9\alpha\varphi_{f}^{4/3}}{64\lambda\beta}}\Big(9\alpha+\frac{8\lambda\varphi_{f}^{2/3}\Gamma[\frac{3}{2}, \frac{9\alpha\varphi_{f}^{4/3}}{64\lambda\beta}]}{\sqrt{\frac{\alpha\varphi_{f}^{4/3}}{\lambda\beta}}}\Big)\bigg)}{16\beta^2\lambda\Big(9\alpha\sqrt{\frac{\alpha\varphi_{f}^{4/3}}{\lambda\beta}}+8\lambda\varphi_{f}^{2/3}\Gamma[\frac{3}{2}, \frac{9\alpha\varphi_{f}^{4/3}}{64\lambda\beta}]\Big)^2}.
\label{j4}
\end{equation}
Also, the number of \textit{e}-folds is given by
\begin{equation}
N=\frac{9\big(\frac{\alpha\varphi^{4/3}}{\beta\lambda}\big)^{3/2}\Big(\varphi^2+32e^{\frac{9\alpha\varphi^{4/3}}{64\beta\lambda}}\beta \Big)+256e^{\frac{9\alpha\varphi^{4/3}}{64\beta\lambda}}\varphi^2\Gamma\big(\frac{3}{2},\frac{9\alpha\varphi^{4/3}}{64\beta\lambda}\big)}{12\alpha\big(\frac{\alpha\varphi^{4/3}}{\beta\lambda}\big)^{3/2}}\Big|^{\varphi_{i}}_{\varphi_{f}}.
\label{j5}
\end{equation}
Using the slow-roll parameters (\ref{j2}) and (\ref{j3}) and the obtained number of \textit{e}-folds (\ref{j5}), one can derive the spectral parameters of the model as
\begin{eqnarray}
n_{s}\simeq1-\frac{3\alpha^3\mathcal{A}e^{-\frac{9\alpha\mathcal{A}}{32\lambda\beta}}\bigg(-3\lambda\sqrt{\mathcal{A}}+e^{\frac{9\alpha\mathcal{A}}{64\lambda\beta}}\Big(9\alpha+\frac{8\lambda\sqrt{\mathcal{A}}\Gamma[\frac{3}{2}, \frac{9\alpha\mathcal{A}}{64\lambda\beta}]}{\sqrt{\frac{\alpha\mathcal{A}}{\lambda\beta}}}\Big)\bigg)}{4\beta^2\lambda\Big(9\alpha\sqrt{\frac{\alpha\mathcal{A}}{\lambda\beta}}+8\lambda\sqrt{\mathcal{A}}\Gamma[\frac{3}{2}, \frac{9\alpha\mathcal{A}}{64\lambda\beta}]\Big)^2}+\hspace{5cm}\nonumber\\
+\frac{e^{-\frac{9\alpha\mathcal{A}}{32\lambda\beta}}\Big(9\alpha \sqrt{\frac{\alpha\mathcal{A}}{\lambda\beta}}+8\lambda\sqrt{\mathcal{A}}\Gamma[\frac{3}{2}, \frac{9\alpha\mathcal{A}}{64\lambda\beta}]\Big)^{-1}\Big(9\alpha^2\sqrt{\mathcal{A}}+8\beta\lambda^2\sqrt{\frac{\alpha\mathcal{A}}{\lambda\beta}}\Gamma[\frac{3}{2}, \frac{9\alpha\mathcal{A}}{64\lambda\beta}]\Big)^{-1}}{8\beta^2\Big(-3\sqrt{\frac{\alpha\mathcal{A}}{\lambda\beta}}\big(\lambda\sqrt{\mathcal{A}}-3\alpha e^{\frac{9\alpha\mathcal{A}}{64\lambda\beta}}\big)+8\lambda\sqrt{\mathcal{A}}e^{\frac{9\alpha\mathcal{A}}{64\lambda\beta}}\Gamma[\frac{3}{2}, \frac{9\alpha\mathcal{A}}{64\lambda\beta}]\Big)}\Bigg\{-9\alpha^4\mathcal{A}\times\hspace{0.4cm}\nonumber\\
\times\Big(3\lambda\mathcal{A}^{3/2}+e^{\frac{9\alpha\mathcal{A}}{64\lambda\beta}}\big(32\lambda\beta-9\alpha\mathcal{A}\big)\Big)+\frac{8\lambda\alpha^3\mathcal{A}^{3/2}e^{\frac{9\alpha\mathcal{A}}{64\lambda\beta}}\big(9\alpha\mathcal{A}-32\lambda\beta\big)\Gamma[\frac{3}{2}, \frac{9\alpha\mathcal{A}}{64\lambda\beta}]}{\sqrt{\frac{\alpha\mathcal{A}}{\lambda\beta}}}\Bigg\},\hspace{0.5cm}
\label{j6}
\end{eqnarray}
\begin{eqnarray}
r\simeq\frac{3\alpha^3\mathcal{A}e^{-\frac{9\alpha\mathcal{A}}{32\lambda\beta}}\bigg(-3\lambda\sqrt{\mathcal{A}}+e^{\frac{9\alpha\mathcal{A}}{64\lambda\beta}}\Big(9\alpha+\frac{8\lambda\sqrt{\mathcal{A}}\Gamma[\frac{3}{2}, \frac{9\alpha\mathcal{A}}{64\lambda\beta}]}{\sqrt{\frac{\alpha\mathcal{A}}{\lambda\beta}}}\Big)\bigg)}{\beta^2\lambda\Big(9\alpha\sqrt{\frac{\alpha\mathcal{A}}{\lambda\beta}}+8\lambda\sqrt{\mathcal{A}}\Gamma[\frac{3}{2}, \frac{9\alpha\mathcal{A}}{64\lambda\beta}]\Big)^2},
\label{j7}
\end{eqnarray}
where $\mathcal{A}=\frac{64\lambda\beta}{9\alpha}\ln(\frac{12N\alpha}{288\beta})$. In the case of $C=0$, the above relations for the spectral index $n_s$ and the tensor-to-scalar ratio $r$ reduce to
\begin{eqnarray}
n_{s}\simeq1-\frac{3\alpha^3\mathcal{A}e^{-\frac{9\alpha\mathcal{A}}{32\lambda\beta}}\bigg(-3\lambda\sqrt{\mathcal{A}}+\frac{8\lambda e^{\frac{9\alpha\mathcal{A}}{64\lambda\beta}}\sqrt{\mathcal{A}}\Gamma[\frac{3}{2}, \frac{9\alpha\mathcal{A}}{64\lambda\beta}]}{\sqrt{\frac{\alpha\mathcal{A}}{\lambda\beta}}}\bigg)}{4\beta^2\lambda\Big(8\lambda \sqrt{\mathcal{A}}\Gamma[\frac{3}{2}, \frac{9\alpha\mathcal{A}}{64\lambda\beta}]\Big)^2}+\hspace{5cm}\nonumber\\
+\frac{e^{-\frac{9\alpha\mathcal{A}}{32\lambda\beta}}\Big(8\lambda\sqrt{\mathcal{A}}\Gamma[\frac{3}{2}, \frac{9\alpha\mathcal{A}}{64\lambda\beta}]\Big)^{-1}\Big(8\beta\lambda^2\sqrt{\frac{\alpha\mathcal{A}}{\lambda\beta}}\Gamma[\frac{3}{2}, \frac{9\alpha\mathcal{A}}{64\lambda\beta}]\Big)^{-1}}{8\beta^2\Big(-3\lambda\mathcal{A}\sqrt{\frac{\alpha}{\lambda\beta}}+8\lambda\sqrt{\mathcal{A}}e^{\frac{9\alpha\mathcal{A}}{64\lambda\beta}}\Gamma[\frac{3}{2}, \frac{9\alpha\mathcal{A}}{64\lambda\beta}]\Big)}\Bigg\{-27\lambda\alpha^4\mathcal{A}^{5/2}+\hspace{0.4cm}\nonumber\\
+\frac{8\lambda\alpha^3\mathcal{A}^{3/2}e^{\frac{9\alpha\mathcal{A}}{64\lambda\beta}}\big(9\alpha\mathcal{A}-32\lambda\beta\big)\Gamma[\frac{3}{2}, \frac{9\alpha\mathcal{A}}{64\lambda\beta}]}{\sqrt{\frac{\alpha\mathcal{A}}{\lambda\beta}}}\Bigg\},\hspace{0.5cm}
\label{j8}
\end{eqnarray}
\begin{eqnarray}
r\simeq\frac{3\alpha^3\mathcal{A}e^{-\frac{9\alpha\mathcal{A}}{32\lambda\beta}}\bigg(-3\lambda\sqrt{\mathcal{A}}+\frac{8\lambda e^{\frac{9\alpha\mathcal{A}}{64\lambda\beta}}\sqrt{\mathcal{A}}\Gamma[\frac{3}{2}, \frac{9\alpha\mathcal{A}}{64\lambda\beta}]}{\sqrt{\frac{\alpha\mathcal{A}}{\lambda\beta}}}\bigg)}{\beta^2\lambda\Big(8\lambda \sqrt{\mathcal{A}}\Gamma[\frac{3}{2}, \frac{9\alpha\mathcal{A}}{64\lambda\beta}]\Big)^2},
\label{j9}
\end{eqnarray}
where $\mathcal{A}=(\frac{12N\alpha}{9})^{2/3}$.
\end{itemize}
Now, we can review predictions of the monomial potential (\ref{g7}) studied in the context of $f(R)$ gravity and also the observational constraints on the parameters space of the model by comparing the obtained results with the CMB anisotropies observations. 

In Fig.\ref{fig7}, we present the $n_{s}-r$ constraints coming from the marginalized
joint 68\% and 95\% CL regions of Planck 2018 in combination with the BK14+BAO datasets \cite{Planck:2018jri} on the monomial potential (\ref{g7}) with $n=2$ studied in the context of $f(R)$ gravity when the parameters $C$, $\alpha$ and $\lambda$ are  unit. Panels reveal predictions of the model $n=2$ associated to  the first (left panel) and the second (right panel) orders of the binomial series in comparison with the values in Planck 2018 (solid black line) for $N=50$ (small black circle) and $N=60$ (big black circle). In the left panel, we present predictions of the monomial potential with $n=2$ for different values of the parameter $\beta$ by keeping only the first term of the binomial series of the number of \textit{e}-folds (\ref{g12}). The panel shows that, in the range  $0.026\leq\beta\leq0.03$, predictions of the model for $n_s$ and $r$ are compatible with the CMB observations coming from Planck 2018 alone at 68\% CL. Note that, by considering Planck 2018 combined with BK14 and BAO, the model $n=2$ is fully ruled out by the observations for any value of $\beta$. The right panel  shows predictions of the model $n=2$ for different values of $\beta$ when  terms up to the second orders of the binomial series of the number of \textit{e}-folds (\ref{g12}) are kept. As we see, the general behavior of the model is almost similar to the case shown in the left panel with a difference in the observationally allowed range $0.023\leq\beta\leq0.03$. Moreover, the right panel presents  smaller and larger values of $n_s$ and $r$ (respectively) in comparison with  values shown in the left panel. Despite the inconsistency of the model in GR with the observations, both cases shown in the panels are in good agreement with the CMB observations. Remind that all the above results are obtained for a non-zero integration constant $C=1$, while for $C=0$, we almost recover the result of Planck 2018 for power $n=2$ in GR (solid black line) $n_{s}=0.96002$, $r=0.15999$ and $n_{s}=0.96667$, $r=0.13333$ for $N=50$ and $N=60$, respectively.

In Fig.\ref{fig8}, we present the $n_{s}-r$ constraints coming from the marginalized
joint 68\% and 95\% CL regions of Planck 2018 in combination with the BK14+BAO datasets \cite{Planck:2018jri} on the monomial potential (\ref{g7}) with $n=4/3$ studied in the context of $f(R)$ gravity when the parameters $\alpha$ and $\lambda$ are considered as the unit. Panels reveal predictions of the model $n=4/3$ associated to  non-zero (left panel) and zero (right panel) integration constant $C$ in comparison with values in Planck 2018 (solid black line) for $N=50$ (small black circle) and $N=60$ (big black circle). In the left panel, we present predictions of the monomial potential with $n=4/3$ for different values of the parameter $\beta$ by assuming the integration constant $C$ as the unit. The panel reveals that the predictions of the model for $n_s$ and $r$ in the range $0.08\leq\beta\leq0.15$ are in good agreement with the CMB anisotropies observations coming from Planck 2018 alone at 68\% CL. It is clear that, by considering Planck 2018 combined with BK14 and BAO, the model is not compatible with the observations. In comparison with predictions of the model in GR (black solid line), one can see that the monomial potential with $n=4/3$ in $f(R)$ gravity shows smaller (larger) values of $n_s$ ($r$) for $\beta\leq0.11$. This situation is different in the case $\beta\geq0.13$ since we have larger values of $n_s$ associated to  the smaller values of $r$. The right panel tells us predictions of the monomial potential with $n=4/3$ in $f(R)$ gravity for different values of $\beta$ when the integration constant $C$ is assumed as zero. Compared to the case $C\neq0$ shown in the left panel, we find that the presented values of $n_s$ and $r$ in the range $0.008\leq\beta\leq0.09$ are more compatible with CMB observations since they are situated in the observational regions coming from Planck 2018 alone at 68\% CL. Also, the monomial potential with $n=4/3$ in $f(R)$ gravity predicts more favoured values of $n_s$ and $r$ in $N=50$ rather than the model in GR (black solid line). 

In Fig.\ref{fig9}, we present the $n_{s}-r$ constraints coming from the marginalized
joint 68\% and 95\% CL regions of Planck 2018 in combination with the BK14+BAO datasets \cite{Planck:2018jri} on the monomial potential (\ref{g7}) with $n=1$ studied in the context of $f(R)$ gravity when the parameters $C$, $\alpha$ and $\lambda$ are considered as the unit. Panels reveal the predictions of the model $n=1$ associated to  the negative (left panel) and positive (right panel) values of $\beta$ in comparison with the values in Planck 2018 (solid black line) for $N=50$ (small black circle) and $N=60$ (big black circle). In the left panel, we present predictions of the monomial potential with $n=1$ for the negative values of $\beta$. The panel shows that the predictions of the model for $n_s$ and $r$ in the range $-0.5\leq\beta\leq-0.2$ are situated in the regions where are consistent with the CMB observations coming from Planck 2018 alone and its combination with BK14 and BAO at 68\% CL. Also, the monomial potential with $n=1$ in $f(R)$ gravity presents more favoured values of $n_s$ and $r$ in $N=50$ compared to the the model in GR (black solid line). The right panel reveals predictions of the monomial potential with $n=1$ when the positive values of $\beta$ are assumed. As we see, the obtained values of $n_s$ and $r$ in the range $0.2\leq\beta\leq0.3$ have less consistency with the observations rather than the case $\beta<0$ shown in the left panel. Moreover, the monomial potential with $n=1$ in $f(R)$ gravity shows less favoured values of $n_s$ and $r$ in $N=50$ compared to the predictions of the model in GR (black solid line). Notice that all above information is associated to  a non-zero integration constant $C=1$, while for $C=0$, we almost reproduce the result of Planck 2018 for power $n=1$ in GR (solid black line) \ie  $n_{s}=0.97004$, $r=0.079998$ and $n_{s}=0.97501$, $r=0.066667$ for $N=50$ and $N=60$, respectively.

In Fig.\ref{fig10}, we present the $n_{s}-r$ constraints coming from the marginalized joint 68\% and 95\% CL regions of Planck 2018 in combination with the BK14+BAO datasets \cite{Planck:2018jri} on the monomial potential (\ref{g7}) with $n=2/3$ studied in the context of $f(R)$ gravity when the parameters $\alpha$ and $\lambda$ are considered as the unit. Panels reveal predictions of the model $n=2/3$ associated to a non-zero (left panel) and zero (right panel) integration constant $C$ in comparison with the values in Planck 2018 (solid black line) for $N=50$ (small black circle) and $N=60$ (big black circle). The left panel shows predictions of the monomial potential with $n=2/3$ for different values of $\beta$ by assuming the integration constant as the unit. The panel reveals predictions of the model for $n_s$ and $r$ in the range $0.3\leq\beta\leq0.7$ in $N=50$ are compatible with the CMB ansiotropies observations coming from Planck 2018 alone at 68\% CL. Clearly, the obtained values of $n_s$ and $r$ in $N=60$ for any $\beta$ are excluded by the observations. Also, note that for a combination of Planck 2018 with BK14 and BAO, the model is ruled out by the observations. The right panel presents predictions of the monomial potential with $n=2/3$ in $f(R)$ gravity for different values of $\beta$ when the integration constant $C$ is considered as zero. Compared to the case $C\neq0$ shown in the left panel, we see that the obtained values of $n_s$ and $r$ for the reduced range $0.03\leq\beta\leq0.2$ in $N=50$ are more consistent with the CMB observations since they are in the observational regions coming from Planck 2018 alone and its combination with BK14 and BAO at 68\% CL. Also, the monomial potential with $n=2/3$ in $f(R)$ gravity predicts less favoured values of $n_s$ and $r$ in $N=50$ in comparison with predictions of the model in GR (black solid line).
  
\subsubsection{Power-Law Inflation}
As the second case, we focus on the exponential potential
\begin{equation}
V=V_{0}e^{-\lambda\varphi},
\label{k1} 
\end{equation}
associated to  the power-law inflation with the  scale factor $a(t)\propto t^{q}$, $q>1$ \cite{Abbott:1984fp,Lucchin:1984yf,Sahni:1990tx}. Here, $\lambda$ has the dimension $[mass]^{-1}$ and $V_{0}$ refers to the energy scale with the dimension of $[mass]^{4}$. This class of inflationary models is usually excluded by the CMB anisotropies observations \cite{Planck:2018jri}.

From Eq.(\ref{g2}), the Friedman equation of the exponential potential (\ref{k1}) is given by
\begin{equation}
H^2=\frac{e^{-\frac{\alpha e^{\lambda\varphi}}{8\beta V_0 \lambda^2}}\Big(24C\beta \lambda^2+\mbox{ExpIntegralEi}(\frac{\alpha e^{\lambda\varphi}}{8\beta V_0 \lambda^2})\Big)}{24\beta \lambda^2},
\label{k2}
\end{equation}
\begin{figure*}[!hbtp]
     \centering	\includegraphics[width=.32\textwidth,keepaspectratio]{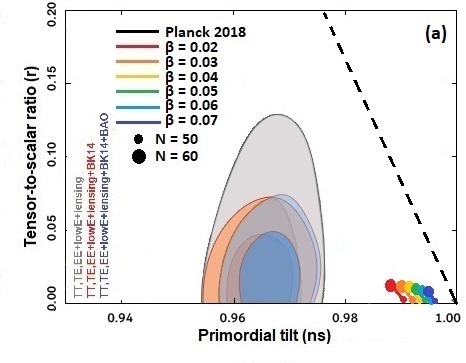}
     \includegraphics[width=.32\textwidth,keepaspectratio]{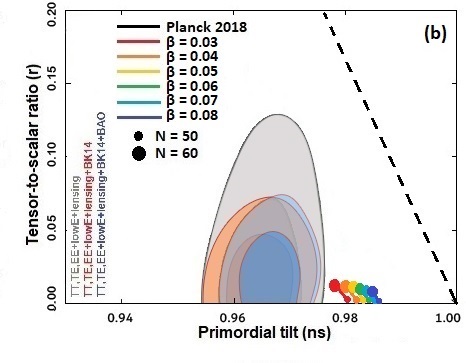}
     \includegraphics[width=.32\textwidth,keepaspectratio]{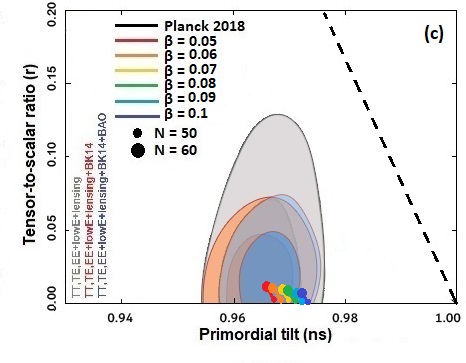}

\caption{The marginalized joint 68\% and 95\% CL regions for $n_{s}$ and $r$ at $k = 0.002$ Mpc$^{-1}$ from Planck alone and in combination with BK14 or BK14+BAO \cite{Planck:2018jri} and the $n_{s}-r$ constraints on the parameters space of the power-law inflation (\ref{k1}) in $f(R)$ gravity associated to $\lambda=0.01$ (a), $\lambda=0.02$ (b) and $\lambda=0.03$ (c). The results are obtained for the case that the parameters $C$, $\alpha$ and $V_0$ are considered as the unit.}
\label{fig11}
\end{figure*}
where $\mbox{ExpIntegralEi(x)}$ is the exponential integral function $\mbox{Ei(x)}$. Also, $C$ is the integration constant. By considering the integration constant $C$ as the unit and  using Eq.(\ref{g4}), the slow-roll parameters of the model are found as
\begin{equation}
\epsilon_H=\frac{\lambda^2 e^{\frac{\alpha e^{\lambda\varphi}}{8\beta V_0 \lambda^2}}\bigg(\alpha\Big(24\beta \lambda^2+\mbox{ExpIntegralEi}(\frac{\alpha e^{\lambda\varphi}}{8\beta V_0 \lambda^2})\Big)-8\beta \lambda^2 V_0 e^{-\lambda\varphi+\frac{\alpha e^{\lambda\varphi}}{8\beta V_0 \lambda^2}}\bigg)}
{2\Big(24\beta \lambda^2+\mbox{ExpIntegralEi}(\frac{\alpha e^{\lambda\varphi}}{8\beta V_0 \lambda^2})\Big)^2},
\label{k3}    
\end{equation}
\begin{eqnarray}
\eta_H=\frac{4\beta V_0\lambda^4 e^{-\lambda\varphi+\frac{\alpha e^{\lambda\varphi}}{4\beta V_0 \lambda^2}}\bigg(-8\beta \lambda^2V_0e^{\frac{\alpha e^{\lambda\varphi}}{8\beta V_0 \lambda^2}}+\Big(24\beta \lambda^2+\mbox{ExpIntegralEi}(\frac{\alpha e^{\lambda\varphi}}{8\beta V_0 \lambda^2})\Big)(\alpha e^{\lambda\varphi}-8\beta \lambda^2V_0)\bigg)}
{\Big(24\beta \lambda^2+\mbox{ExpIntegralEi}(\frac{\alpha e^{\lambda\varphi}}{8\beta V_0 \lambda^2})\Big)^2\Big(\alpha e^{\lambda\varphi}\big(24\beta \lambda^2+\mbox{ExpIntegralEi}(\frac{\alpha e^{\lambda\varphi}}{8\beta V_0 \lambda^2})\big)-8\beta \lambda^2V_0e^{\frac{\alpha e^{\lambda\varphi}}{8\beta V_0 \lambda^2}}\Big)}.
\label{k4}    
\end{eqnarray}
By setting $\epsilon_H=1$, the value of inflaton at the end of inflation can be derived by the following expression
\begin{equation}
1=\frac{\lambda^2 e^{\frac{\alpha e^{\lambda\varphi_f}}{8\beta V_0 \lambda^2}}\bigg(\alpha\Big(24\beta \lambda^2+\mbox{ExpIntegralEi}(\frac{\alpha e^{\lambda\varphi_f}}{8\beta V_0 \lambda^2})\Big)-8\beta \lambda^2 V_0 e^{-\lambda\varphi_f+\frac{\alpha e^{\lambda\varphi_f}}{8\beta V_0 \lambda^2}}\bigg)}
{2\Big(24\beta \lambda^2+\mbox{ExpIntegralEi}(\frac{\alpha e^{\lambda\varphi_f}}{8\beta V_0 \lambda^2})\Big)^2},
\label{k5}    
\end{equation}
Also, from Eq.(\ref{g5}), the number of \textit{e}-folds of the model is given by 
\begin{equation}
N=\frac{e^{-\frac{\alpha e^{\lambda\varphi}}{8\beta V_0 \lambda^2}}\Big(24\beta \lambda^2+\mbox{ExpIntegralEi}(\frac{\alpha e^{\lambda\varphi}}{8\beta V_0 \lambda^2})\Big)-\lambda\varphi}{\alpha \lambda^2}\Bigg|^{\varphi_i}_{\varphi_f}.
\label{k6}    
\end{equation}
By using the slow-roll parameters (\ref{k3}) and (\ref{k4}) and the obtained number of \textit{e}-folds (\ref{k6}), 
the spectral parameters are calculated as
\begin{eqnarray}
n_s\simeq1-\frac{2\lambda^2 e^{\frac{\alpha\mathcal{A}}{8\beta V_0 \lambda^2}}\bigg(\alpha\Big(24\beta \lambda^2+\mbox{ExpIntegralEi}(\frac{\alpha\mathcal{A}}{8\beta V_0 \lambda^2})\Big)-8\beta k^2 V_0 \mathcal{A}^{-1}e^{\frac{\alpha \mathcal{A}}{8\beta V_0 \lambda^2}}\bigg)}
{\Big(24\beta \lambda^2+\mbox{ExpIntegralEi}(\frac{\alpha \mathcal{A}}{8\beta V_0 \lambda^2})\Big)^2}+\hspace{4cm}\nonumber\\
+\frac{8\beta V_0\lambda^4 \mathcal{A}^{-1} e^{\frac{\alpha \mathcal{A}}{4\beta V_0 \lambda^2}}\bigg(-8\beta \lambda^2V_0e^{\frac{\alpha \mathcal{A}}{8\beta V_0 \lambda^2}}+\Big(24\beta \lambda^2+\mbox{ExpIntegralEi}(\frac{\alpha \mathcal{A}}{8\beta V_0 \lambda^2})\Big)(\alpha \mathcal{A}-8\beta \lambda^2V_0)\bigg)}
{\Big(24\beta \lambda^2+\mbox{ExpIntegralEi}(\frac{\alpha \mathcal{A}}{8\beta V_0 \lambda^2})\Big)^2\Big(\alpha \mathcal{A}\big(24\beta k^2+\mbox{ExpIntegralEi}(\frac{\alpha \mathcal{A}}{8\beta V_0 \lambda^2})\big)-8\beta \lambda^2V_0e^{\frac{\alpha \mathcal{A}}{8\beta V_0 \lambda^2}}\Big)},\hspace{0.6cm}
\label{k7}    
\end{eqnarray}
\begin{equation}
r\simeq\frac{8\lambda^2 e^{\frac{\alpha\mathcal{A}}{8\beta V_0 \lambda^2}}\bigg(\alpha\Big(24\beta \lambda^2+\mbox{ExpIntegralEi}(\frac{\alpha\mathcal{A}}{8\beta V_0 \lambda^2})\Big)-8\beta \lambda^2 V_0 \mathcal{A}^{-1}e^{\frac{\alpha \mathcal{A}}{8\beta V_0 \lambda^2}}\bigg)}
{\Big(24\beta \lambda^2+\mbox{ExpIntegralEi}(\frac{\alpha \mathcal{A}}{8\beta V_0 \lambda^2})\Big)^2},
\label{k8}    
\end{equation}
where $\mathcal{A}=-\frac{8\beta V_{0}\lambda^2}{\alpha}\ln\Big(\frac{N\alpha}{24\beta}\Big)$. 

In Fig.\ref{fig11}, we present the $n_{s}-r$ constraints coming from the marginalized
joint 68\% and 95\% CL regions of Planck 2018 in combination with the BK14+BAO datasets \cite{Planck:2018jri} on the power-law inflation (\ref{k1}) studied in the context of $f(R)$ gravity when the parameters $C$, $\alpha$ and $V_0$ are considered as the unit. Panels present the predictions of the model associated to $\lambda=0.01$ (a), $\lambda=0.02$ (b) and $\lambda=0.03$ (c) in comparison with Planck 2018. In panel (a), we show the predictions of the power-law inflation related to $\lambda=0.01$ for allowed values of $\beta\geq0.02$ in which the obtained values of $n_s$ and $r$ are fully excluded by the CMB observations analogous to predictions of Planck 2018 (dashed line). As we see in panel (b), the values of $n_s$ and $r$ for $\lambda=0.02$ are close to the observational regions but still ruled out for allowed values of the parameter $\beta\geq0.03$. In panel (c), we present predictions of the power-law inflation in the case of $\lambda=0.03$ in which the obtained values of $n_s$ and $r$ for allowed $0.05\leq\beta\leq0.1$ are in good agreement with the CMB observations coming from Planck 2018 alone and its combination with BK14 and BAO at 68\% CL and 95\% CL. In summary, although the power-law inflation in GR is excluded by the observations, it comes back to the playground in $f(R)$ theory regime for some allowed values of $\beta$ when $\lambda\geq0.03$ is considered.

\subsubsection{Natural Inflation}
Natural inflation (NI) overcomes the flatness problem of the hot Big Bang theory using a Pseudo-Nambu-Goldstone boson as inflaton with a flat potential \cite{Adams:1992bn,Freese:1990rb}
\begin{equation}
V(\varphi)=V_{0}\Big(1+\cos(\frac{\varphi}{f})\Big),
\label{l1}
\end{equation}
in which a global $U(1)$ symmetry is spontaneously
broken at scale $f$, with  explicit soft symmetry breaking
at a lower scale $\Lambda$. Also, $V_{0}$ and $f$ both represent the energy scale with the dimension $[mass]^{4}$ and $[mass]$, respectively. 
For the small angle approximation $f\ll\varphi$, the potential reduces to $V\simeq2\Lambda^{4}$, while for large angle approximation $f\gg m_{pl}$, its recovers chaotic inflation $ V(\psi)=m^{2}\psi^{2}/2$ where $\psi=\varphi-\sigma$ and $\sigma=constant$.

From Eq.(\ref{g2}), the Friedman equation of NI (\ref{l1}) is given by
\begin{eqnarray}
H^2=\frac{1}{3\alpha f^2-48\beta V_0}\Bigg(V_0f^2\big(1+\cos(\frac{\varphi}{f})\big)\sin^{-\frac{\alpha f^2}{8\beta V_0}}\big(\frac{\varphi}{2f}\big)\times\hspace{6cm}\nonumber\\
\times{}_2F_1\Big[1-\frac{\alpha f^2}{16\beta V_0}, 1-\frac{\alpha f^2}{16\beta V_0}, 2-\frac{\alpha f^2}{16\beta V_0}, \cos^2\big(\frac{\varphi}{2f}\big)\Big]+3C(\alpha f^2-16\beta V_0)\cot^{\frac{\alpha f^2}{8\beta V_0}}\big(\frac{\varphi}{2f}\big)\Bigg),\hspace{0.5cm}
\label{l2}
\end{eqnarray}
where ${}_2F_1$ is the hypergeometric function and $C$ is the integration constant. Using the following property of such special functions 
\begin{equation}
{}_2F_1\Big[\frac{1}{2}, \frac{1}{2}, \frac{3}{2}, x^2\Big]=\frac{\sin^{-1}(x)}{x},
\label{l3}
\end{equation}
the Hubble parameter (\ref{l2}) reduces to 
\begin{equation}
H^2=\frac{\cot(\frac{\varphi}{2f})\Big(12C\beta-f^2\sin^{-1}\big(\cos(\frac{\varphi}{2f})\big)\Big)}{12\beta}.
\label{l4}    
\end{equation}
In such a case, we encounter with two regimes: i) $\beta/\alpha\ll m_{pl}$ so that $f\sim m_{pl}$. ii) $\beta/\alpha\sim m_{pl}$ or $\beta/\alpha\gg m_{pl}$ , so that  $f\gg m_{pl}$. As we see, the second regime reminds us the large angle approximation.  
Using Eq.(\ref{g4}) for $C=1$, the slow-roll parameters of the model are
\begin{equation}
\epsilon_H=\frac{\beta V_0\tan(\frac{\varphi}{2f})\Bigg(24\beta-f^2\Big(2\sin^{-1}\big(\cos(\frac{\varphi}{2f})\big)+\cos(\frac{\varphi}{2f})\sqrt{2\big(1-\cos(\frac{\varphi}{f})\big)}\Big)\Bigg)}{f\Big(f^2\sin^{-1}\big(\cos(\frac{\varphi}{2f})\big)-12\beta\Big)^2},
\label{l5}
\end{equation}
\begin{eqnarray}
\eta_H=\frac{2\beta V_0\sin^3(\frac{\varphi}{2f})\Big(f^2\sin^{-1}\big(\cos(\frac{\varphi}{2f})\big)-12\beta\Big)^{-2}}{\sqrt{2\big(1-\cos(\frac{\varphi}{f})\big)}\Big(f^2\sin^{-1}\big(\cos(\frac{\varphi}{2f})\big)-12\beta\Big)+f^2\sin(\frac{\varphi}{2f})\sin(\frac{\varphi}{f})}\Bigg\{f^2\cos(\frac{\varphi}{2f})\times\hspace{3cm}\nonumber\\
\times\sqrt{2\big(1-\cos(\frac{\varphi}{f})\big)}+2\cos(\frac{\varphi}{f})\Big(f^2\sin^{-1}\big(\cos(\frac{\varphi}{2f})\big)-12\beta\Big)\Bigg\}.\hspace{0.5cm}
\label{l6}
\end{eqnarray}
\begin{figure*}[!hbtp]
     \centering	\includegraphics[width=.32\textwidth,keepaspectratio]{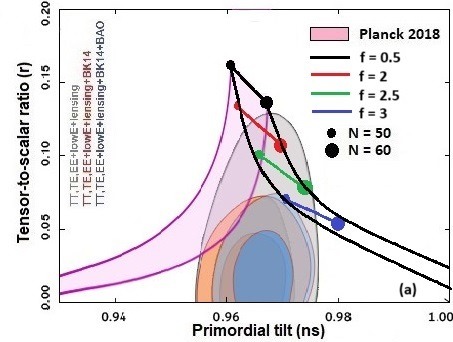}
     \includegraphics[width=.32\textwidth,keepaspectratio]{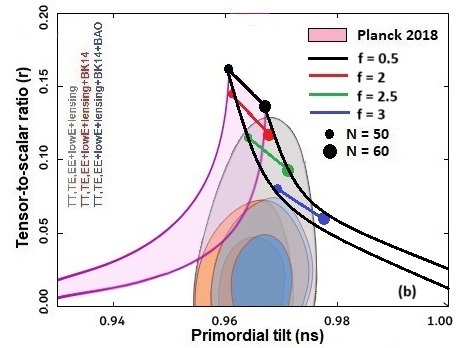}
     \includegraphics[width=.32\textwidth,keepaspectratio]{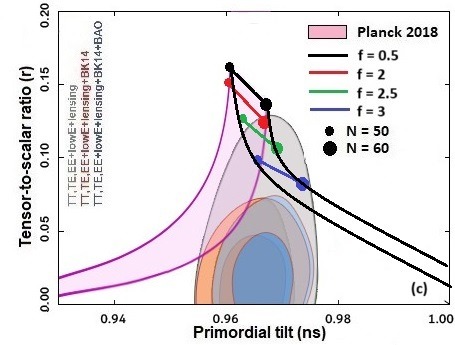}

\caption{The marginalized joint 68\% and 95\% CL regions for $n_{s}$ and $r$ at $k = 0.002$ Mpc$^{-1}$ from Planck alone and in combination with BK14 or BK14+BAO \cite{Planck:2018jri} and the $n_{s}-r$ constraints on the parameters space of NI (\ref{l1}) in $f(R)$ gravity associated to $\beta=1.5$ (a), $\beta=2$ (b) and $\beta=2.5$ (c). The results are obtained for the case that the parameters $C$, $\alpha$ and $V_0$ are considered as the unit.}
\label{fig12}
\end{figure*}
Also, the value of inflaton at the last step of inflation ($\epsilon_H=1$) can be obtained by the equation
\begin{equation}
1=\frac{\beta V_0\tan(\frac{\varphi_f}{2f})\Bigg(24\beta-f^2\Big(2\sin^{-1}\big(\cos(\frac{\varphi_f}{2f})\big)+\cos(\frac{\varphi_f}{2f})\sqrt{2\big(1-\cos(\frac{\varphi_f}{f})\big)}\Big)\Bigg)}{f\Big(f^2\sin^{-1}\big(\cos(\frac{\varphi_f}{2f})\big)-12\beta\Big)^2}.
\label{l7}
\end{equation}
From Eq.(\ref{g5}), the number of \textit{e}-folds of the model is
\begin{equation}
N=\frac{2f^2\Big(12\beta-f^2\sin^{-1}\big(\cos(\frac{\varphi}{2f})\big)\Big)\cot(\frac{\varphi}{2f})-f^4\sqrt{2\big(1-\cos(\frac{\varphi}{f})\big)}\csc(\frac{\varphi}{2f})\ln\Big[\frac{1}{2}\sin(\frac{\varphi}{2f})\Big]}{8\beta V_0}\Bigg|^{\varphi_i}_{\varphi_f}.
\label{l8}    
\end{equation}
From Eq.(\ref{g6}) and by using the slow-roll parameters (\ref{l5}) and (\ref{l6}) and the number of \textit{e}-folds (\ref{l8}), 
the spectral parameters of the model are given by
\begin{eqnarray}
n_s\simeq1-\frac{4\beta V_0\tan(\frac{\mathcal{A}}{2})\Bigg(24\beta-f^2\Big(2\sin^{-1}\big(\cos(\frac{\mathcal{A}}{2})\big)+\cos(\frac{\mathcal{A}}{2})\sqrt{2\big(1-\cos(\mathcal{A})\big)}\Big)\Bigg)}{f\Big(f^2\sin^{-1}\big(\cos(\frac{\mathcal{A}}{2})\big)-12\beta\Big)^2}+\hspace{3cm}\nonumber\\+\frac{4\beta V_0\sin^3(\frac{\mathcal{A}}{2})\Big(f^2\sin^{-1}\big(\cos(\frac{\mathcal{A}}{2})\big)-12\beta\Big)^{-2}}{\sqrt{2\big(1-\cos(\mathcal{A})\big)}\Big(f^2\sin^{-1}\big(\cos(\frac{\mathcal{A}}{2})\big)-12\beta\Big)+f^2\sin(\frac{\mathcal{A}}{2})\sin(\mathcal{A})}\Bigg\{f^2\cos(\frac{\mathcal{A}}{2})\times\hspace{2cm}\nonumber\\
\times\sqrt{2\big(1-\cos(\mathcal{A})\big)}+2\cos(\mathcal{A})\Big(f^2\sin^{-1}\big(\cos(\frac{\mathcal{A}}{2})\big)-12\beta\Big)\Bigg\},\hspace{0.5cm}
\label{l9}    
\end{eqnarray}
\begin{equation}
r\simeq\frac{16\beta V_0\tan(\frac{\mathcal{A}}{2})\Bigg(24\beta-f^2\Big(2\sin^{-1}\big(\cos(\frac{\mathcal{A}}{2})\big)+\cos(\frac{\mathcal{A}}{2})\sqrt{2\big(1-\cos(\mathcal{A})\big)}\Big)\Bigg)}{f\Big(f^2\sin^{-1}\big(\cos(\frac{\mathcal{A}}{2})\big)-12\beta\Big)^2},
\label{l10}    
\end{equation}
where $\mathcal{A}=\frac{f^2(24\beta-\pi f^2)}{4N\beta V_0}$.

In Fig.\ref{fig12}, we present the $n_{s}-r$ constraints coming from the marginalized
joint 68\% and 95\% CL regions of Planck 2018 in combination with the BK14+BAO datasets \cite{Planck:2018jri} on NI (\ref{l1}) studied in the context of $f(R)$ gravity when the parameters $C$, $\alpha$ and $V_0$ are considered as the unit. Panels present the predictions of the model associated to $\beta=1.5$ (a), $\beta=2$ (b) and $\beta=2.5$ (c) in comparison with Planck 2018. From panel (a), one can see that the obtained values of $n_s$ and $r$ in the range $2\leq f\leq3$ are compatible with the CMB observations coming from Planck 2018 alone at $68\%$ CL. Notice that the mentioned consistency for the cases $f=2$ and $f=3$ is valid only for $N=60$ and $N=50$, respectively. Considering the case $\beta=2$ shown in panel (b), the model almost predicts a similar range of $f$ with smaller and larger values of $n_s$ and $r$ (respectively) compared to the case $\beta=1.5$ in panel (a). In panel (c), which is dedicated to the case $\beta=2.5$, the model follows the tendency of the previous panels so that the observational constraint on the parameter $f$ has less consistency with the observations, in particular, in the cases $f=2$ and $f=2.5$. By taking a look at all three panels, we find that the model is not in good agreement with the observations when a combined version of Planck 2018 with BK14 and BAO is considered. Moreover, we find that NI in the context of $f(R)$ gravity reproduces the result of the monomial potential with $n=2$ (solid black line) when we deal with $f\ll m_{pl}$. This result challenges the prediction of NI in GR which says the model reduces to the chaotic inflation for $f\gg m_{pl}$.         
 
\begin{figure*}[!hbtp]
	\centering
	\includegraphics[width=.55\textwidth,keepaspectratio]{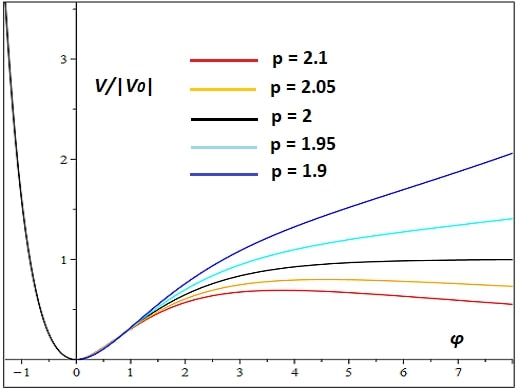}
	\caption{The potential of the $(-T)^{p}$ model (\ref{m1}) for different powers of $p$.}
	\label{fig13}
\end{figure*}
\section{Inflation in $f(T)$ gravity}\label{ft}
Let us now  study inflation  in the framework of $f(T)$ gravity from the two perspectives of PSR and HSR, separately. 
\subsection{The Potential-Slow-Roll Inflation}
We investigate the PSR approach to inflation using two reliable functions of $T$ inspired by the prescriptions presented in \cite{Linder:2010py} that can give a  de Sitter evolution. We shall use the reconstructed potentials derived after  conformal transformation without considering any standard matter $\mathcal{L}_m=0$.
\subsubsection{The $f(T)=T+\alpha(-T)^p$ model}
As a first case, we start with a power-law form of $T$ \cite{Linder:2010py}
\begin{equation}
f(T)=T+\alpha(-T)^p,
\label{m1}    
\end{equation}
where $\alpha$ and $p$ are parameters of the model. This form of $f(T)$ gravity can reproduce some familiar cosmological models depending on the choice of the model parameters \cite{Dvali:2003rk,Chung:1999zs}. For the late-time acceleration, we have $\alpha=f_{0}M_{pl}^{2-2p}$ in which for $p=0$, the model reduces to the $\Lambda$CDM model, while for $p=1$, it recovers the CDM model after re-scaling the gravitational constant as $G\rightarrow G/(1-f_{0})$. Additionally,  it  can reproduce the Dvali-Gabadadze-Porrati (DGP) model \cite{Dvali:2000hr}.   

In the present work, we attempt to consider the model (\ref{m1}) in order to explain the early-time accelerating phase of the universe. Hence, by using the conformal transformation (\ref{a11}), the corresponding potential (\ref{b14}) can be found 
\begin{equation}
V(\varphi)=V_{0}e^{-2\sqrt{\frac{2}{3}}\varphi}(e^{\sqrt{\frac{2}{3}}\varphi}-1)^{\frac{p}{p-1}},
\label{m2}    
\end{equation}
where 
$V_{0}=\frac{p-1}{2}\alpha^{\frac{1}{1-p}}(-\frac{1}{p})^{\frac{p}{p-1}}$. Fig.(\ref{fig13}) shows the behaviour of the potential (\ref{m2}), reconstructed in the Einstein frame, versus the scalar field $\varphi$ for different values of $p$. As in the analogous $f(R)$ case, we have two main classes of potentials depending on the value of $p$. They are:  i) For $p>2$, the potential experiences a maximum value  around $\varphi_m = \sqrt{3/2}\ln{\frac{2p-2}{p-2}}$ and then tends asymptotically to zero for large value of scalar field. In such a case, inflation can occur both for $0 \leq \varphi \leq \varphi_m$ and $\varphi > \varphi_m$. ii) For $p < 2$, the potential increases but its decreasing towards zero is  steeper than the $(-T)^{2}$ model.
iii) For $p=2$ we face with the potential of the $(-T)^{2}$ model asymptotically approaching to a constant value.

From Eqs. (\ref{d3}), the slow-roll parameters of the model, in the Einstein frame, can be calculated as
\begin{equation}
\epsilon_V=\frac{\big((p-2)e^{\sqrt{\frac{2}{3}}\varphi}-2(p-1)\big)^{2}}{3(p-1)^{2}\big(e^{\sqrt{\frac{2}{3}}\varphi}-1\big)^{2}},\hspace{1cm}\eta_V=\frac{(-10p^{2}+26p-16)e^{\sqrt{\frac{2}{3}}\varphi}+2(p-2)^{2}e^{2\sqrt{\frac{2}{3}}\varphi}+8(p-1)^{2}}{3(p-1)^{2}\big(e^{\sqrt{\frac{2}{3}}\varphi}-1\big)^{2}}.
\label{m3}
\end{equation}
\begin{figure*}[!hbtp]
	\centering
	\includegraphics[width=.70\textwidth,keepaspectratio]{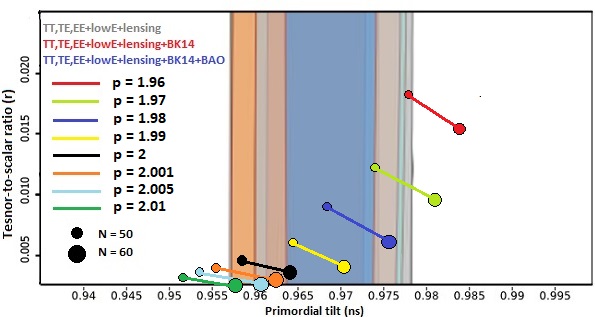}
\caption{The marginalized joint 68\% and 95\% CL regions for $n_{s}$ and $r$ at $k = 0.002$ Mpc$^{-1}$ from Planck alone and in combination with BK14 or BK14+BAO \cite{Planck:2018jri} and the $n_{s}-r$ constraints on the $(-T)^{p}$ model (\ref{m1}) for two cases $p=2$ and $p\neq2$.}
	\label{fig14}
\end{figure*}
By setting $\epsilon_V=1$, the value of the scalar field at the exit of inflation is given by
\begin{equation}
\varphi_{f}=\sqrt{\frac{3}{2}}\ln\bigg[\frac{(p-1)(3-2\sqrt{3})}{\sqrt{3}(2-p)+3(p-1)}\bigg].
\label{m4}
\end{equation}
Using Eq.(\ref{d4}), the number of \textit{e}-folds of the model takes the following form
\begin{equation}
N\simeq-\frac{3p}{4(p-2)}\ln\bigg(\frac{(p-2)e^{\sqrt{\frac{2}{3}}\varphi_{i}}}{2(1-p)}+1\bigg).
\label{m5}
\end{equation}
Finally, using Eq.(\ref{d5}), the spectral parameters of the models can be found as
\begin{equation}
n_{s}\simeq\frac{(4\mathcal{A}^2-4\mathcal{A}-5)p^2+(8\mathcal{A}^2-4\mathcal{A}+16)p-20\mathcal{A}^2}{3(2\mathcal{A}p-2\mathcal{A}-p)^2},\hspace{1cm}r\simeq\frac{64(p-2)^2\mathcal{A}^2}{3(2\mathcal{A}p-2\mathcal{A}- p)^2},
\label{m6}
\end{equation}
where $\mathcal{A}= e^{-\frac{4(p-2)N}{3p}}$. Notice that for the  case of $p=2$, the same analysis of  Eqs.(\ref{d10}) and (\ref{d13}) holds.

In Fig.\ref{fig14}, we show the consistency relation $r=r(n_{s}) $  coming from the marginalized joint 68\% and 95\% CL regions of Planck 2018 alone and in combination with BK14 or the BK14+BAO datasets \cite{Planck:2018jri} on the $(-T)^{p}$ model (\ref{m1}) for two cases $p=2$ and $p\neq2$.

As we see, for the case $p=2$ (solid black line), the model predicts $(n_{s}=0.95815$, $r=0.0049473$) and  $(n_{s}=0.96539$, $r=0.0034183)$ for $N=50$ and $N=60$, respectively. By considering a generalized case $p\neq2$, we find different results for the spectral index $n_{s}$ and the tensor-to-scalar ratio $r$ depending on the value of $p$. From the Planck alone dataset at 68\% CL, we find the obtained values of $n_s$ and $r$ in the range $1.97\leq p\leq2.005$ are in good agreement with the observations. By a combination of Planck 2018 with BK14 and BAO, the observationally favoured range of $p$ reduces to $1.97\leq p\leq2.001$ and $1.98\leq p\leq1.99$ at 68\% and 95\% C.L., respectively. In addition to the above results, the figure reveals that the model $(-T)^{p}$ with a tiny deviation from $p=2$ offers larger values of $n_{s}$ and $r$ which are more compatible with the CMB observations.

\subsubsection{The $f(T)=T+\alpha T(1-e^{\frac{\beta}{T}})$ model}
\begin{figure*}[!hbtp]
	\centering
	\includegraphics[width=.55\textwidth,keepaspectratio]{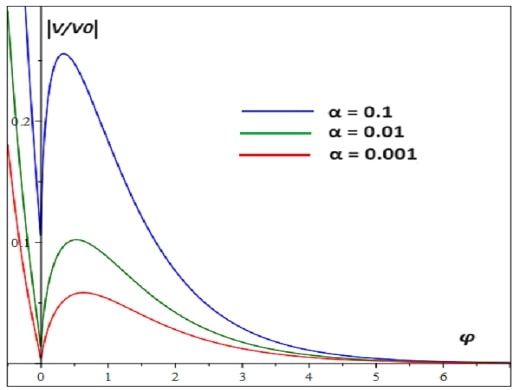}
	\caption{The potential of the $\alpha T(1-e^{\frac{\beta}{T}})$ model (\ref{m7}) for three values $\alpha=0.1, 0.01, 0.001$.}
	\label{fig15}
\end{figure*}

We consider the power-law model discussed in the previous section perturbed by an exponential term, that is \cite{Linder:2010py}
\begin{equation}
f(T)=T+\alpha T(1-e^{\frac{\beta}{T}}),
\label{m7}    
\end{equation}
where $\alpha$ and $\beta$ are the parameters of the model. To describe DE, the parameter $\alpha$ is 
\begin{equation}
\alpha=-\frac{1-\Omega_{m}^{0}}{1-(1-\frac{2\beta}{T_{0}})e^{\frac{\beta}{T_{0}}}},
\label{m8}    
\end{equation}
where $\Omega_{m}^{0}$ and $T_{0}$ are the dimensionless
density parameter of dust matter and the value of the torsion scalar in the present universe, respectively. Thus, this model only involves one single dimensionless parameter. Now let us study such function of $T$ (\ref{m7}) in order to describe the inflationary epoch in $f(T)$ theory.  To perform the inflationary analysis, we use the relation (\ref{b14}) to reconstruct the potential of the model (\ref{m7}) in the Einstein frame as
\begin{equation}
V(\varphi)=\frac{\alpha\beta e^{\frac{\beta}{T}}}{2\Big(1+\alpha\big(1+e^{\frac{\beta}{T}}(\frac{\beta}{T}-1)\big)\Big)^2}.
\label{m9}    
\end{equation}
By inverting $F$ introduced in Eq.(\ref{b14}), the torsion scalar $T$ is found to be 
\begin{equation}
T=\frac{\beta}{W_{k}(X)+1},\hspace{0.5cm}\text{where}\hspace{0.5cm}X\equiv\frac{e^{\sqrt{\frac{2}{3}}\varphi}-(1+\alpha)}{\alpha e},
\label{m10}    
\end{equation}
where $W_{k}$ is the Lambert function with two important branches $k=0, -1$ returning real values for real input. Plugging the above expression into the potential (\ref{m9}), the exact potential in the Einstein frame is 
\begin{equation}
V(\varphi)=\frac{V_{0}e\alpha X}{\big(1+\alpha(1+eX)\big)^2W_k(X)},
\label{m11}    
\end{equation}
where $V_0=\frac{\beta}{2}$. Fig.(\ref{fig15}) shows the behavior of the above potential for three values $\alpha=0.1, 0.01, 0.001$. We have to restrict ourselves to positive values of $\varphi$ in order to have real values of the Lambert function, resulting in real values of the potential. Therefore, one can see that the potential assumes a maximum value and then goes to zero for larger values of $\varphi$. This potential reveals all inflationary predictions of the model. However, for the sake of clarity, we study the potential of the model using the iterative method. By inverting Eq.(\ref{b14}), one can find the torsion scalar as follows
\begin{equation}
T^{(n)}=\frac{\beta}{\frac{X}{e^{\frac{\beta}{T_{n-1}}-1}}+1}.
\label{m12}    
\end{equation}
For some initial values, we find that the zeroth solution $T^{(0)}$ associated to the non-perturbed case (when $\beta\rightarrow-\infty$) is zero, the first solution is $T^{(1)}=\beta$, the second solution $T^{(2)}=\frac{\beta}{X+1}$. By substituting the second solution in Eq.(\ref{m9}), the potential of the model can be obtained as
\begin{equation}
V(\varphi)\simeq\frac{V_0e^{X+1}}{(1+\alpha(1+Xe^{X+1}))^2},
\label{m13}    
\end{equation}
where $V_0=\frac{\alpha\beta}{2}$. By comparing the above potential with Eq.(\ref{m11}), we realize that two expressions show almost similar behaviour if we expand the Lambert function up to the odd powers of $X$. In the following, we present the inflationary analysis for the exact form of the potential (\ref{m11}). By inserting the exact potential (\ref{m11}) into Eq.(\ref{d3}), the slow-roll parameters of the model can be found as
\begin{equation}
\epsilon_V=\frac{\Big(2\alpha eX+\big(-1+(-1+eX)\alpha\big)W_k(X)\Big)^2}{3\alpha^2e^2X^2\Big(1+W_{k}(X)\Big)^2},
\label{m14}    
\end{equation}
\begin{eqnarray}
\eta_{V}=\frac{2}{3\alpha^2e^2X^2(1+W_k(X))^3}\bigg\{4\alpha^2e^2X^2+W_k(X)\bigg(3\alpha eX\big(-1+(-1+3eX)\alpha\big)+\hspace{2cm}\nonumber\\+
W_k(X)\Big(-1+\alpha\big(-2-\alpha+eX(-8-8\alpha+5\alpha eX)\big)+\alpha eX(-3-3\alpha+\alpha eX)W_k(X)\Big)\bigg)\bigg\}.\hspace{0.5cm}
\label{m15}    
\end{eqnarray}
By setting $\epsilon_V=1$, we find the following expression for the end of inflation
\begin{equation}
1=\frac{\Big(2\alpha eX_f+\big(-1+(-1+eX_f)\alpha\big)W_k(X_f)\Big)^2}{3\alpha^2e^2X_f^2\Big(1+W_{k}(X_f)\Big)^2},
\label{m16}    
\end{equation}
where $X_f=\frac{e^{\sqrt{\frac{2}{3}}\varphi_f}-(1+\alpha)}{\alpha e}$. Also, using Eq. (\ref{d4}), the number of \textit{e}-folds of the model can be obtained as
\begin{equation}
N=\sqrt{\frac{3}{2}}\int^{\varphi_{i}}_{\varphi_{f}}\frac{\alpha eX(1+W_{k}(X))}{2\alpha eX+\big(-1+(-1+eX)\alpha\big)W_k(X)}d\varphi.
\label{m17}    
\end{equation}
To present an analytical study, in the following, we restrict our model to the case of $X\ll1$ in which we deal with $\varphi\ll\sqrt{\frac{3}{2}}\ln[1+(1+e)\alpha]$. Now, by using the Taylor series of the Lambert function $W_k$, one can rewrite the potential (\ref{m11}) as
\begin{equation}
V(\varphi)=\frac{V_{0}e\alpha X}{\big(1+\alpha(1+eX)\big)^2\Big[X-X^2+\frac{3}{2}X^3+...\Big]}.
\label{m18}    
\end{equation}
\begin{figure*}[!hbtp]
	\centering
	\includegraphics[width=.70\textwidth,keepaspectratio]{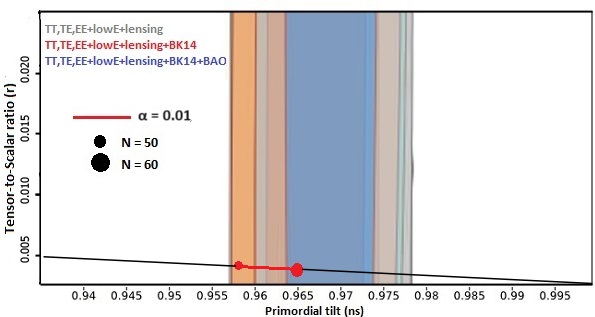}
\caption{The marginalized joint 68\% and 95\% CL regions for $n_{s}$ and $r$ at $k = 0.002$ Mpc$^{-1}$ from Planck alone and in combination with BK14 or BK14+BAO \cite{Planck:2018jri} and the $n_{s}-r$ constraints on the $\alpha T(1-e^{\frac{\beta}{T}})$ model (\ref{m7}) for allowed $\alpha=0.01$.}
	\label{fig16}
\end{figure*}
By neglecting the terms $X^2$ and higher orders of $X$, the slow-roll parameters of the model are derived as
\begin{equation}
\epsilon_V=\frac{(1 +\alpha+eX\alpha)^2(1 +\alpha+e(-2+4 X)\alpha)^2}{3e^2(-1+X)^2\alpha^2(1 +\alpha+2eX\alpha)^2},
\label{m19}    
\end{equation}
\begin{eqnarray}
\eta_V=\frac{1}{3e^2(-1+X)^2\alpha^2(1 +\alpha+2eX\alpha)^2}\Bigg\{2(1+\alpha+eX\alpha)\Big(2+\alpha\big(6+4e^3 X(1+X(-3+4X))\alpha^2+\hspace{2cm}\nonumber\\
+2e^2(3+5X(-2+3 X))\alpha(1+\alpha)+e(-3+13 X)(1+\alpha)^2+2\alpha(3+\alpha)\big)\Big)\Bigg\}.\hspace{1cm}
\label{m20}    
\end{eqnarray}
For the end of inflation ($\epsilon_V=1$), we have the following equation
\begin{equation}
1=\frac{(1 +\alpha+eX_f\alpha)^2(1 +\alpha+e(-2+4X_f)\alpha)^2}{3e^2(-1+X_f)^2\alpha^2(1 +\alpha+2eX_f\alpha)^2},
\label{m21}    
\end{equation}
with the solution
\begin{eqnarray}
X_f=\frac{1}{4e^2(-3\alpha^2+2 \sqrt{3}\alpha^2)}\Bigg\{3e\alpha-5 \sqrt{3}e\alpha+3e\alpha^2-5\sqrt{3} e\alpha^2-6e^2\alpha^2+2\sqrt{3}e^2\alpha^2\pm\hspace{5cm}\nonumber\\
\pm\bigg(36e^2\alpha^2-6\sqrt{3}e^2\alpha^2+72e^2\alpha^3- 
12\sqrt{3}e^2\alpha^3+72e^3\alpha^3 -24\sqrt{3}e^3\alpha^3+36e^2\alpha^4 - 6\sqrt{3}e^2\alpha^4 +72e^3\alpha^4-\hspace{1.5cm}\nonumber\\
-24\sqrt{3}e^3\alpha^4 +48e^4\alpha^4- 24\sqrt{3}e^4\alpha^4\bigg)^{1/2}\bigg\}.\hspace{1cm}
\label{m22}
\end{eqnarray}
Now, using Eq.(\ref{d4}), the number of \textit{e}-folds of the model is given by
\begin{equation}
N=\frac{1}{4}\Bigg\{-3\ln\big[1+\alpha+eX\alpha\big]+\frac{3\Big(-\frac{2(1 +\alpha)(1+\alpha+e\alpha)(3+3\alpha+2e\alpha)}{1+\alpha+eX\alpha}+(1+\alpha+2e\alpha)^2\ln\big[\frac{1 + \alpha+e(-2+4X)\alpha}{1+\alpha+eX \alpha}\big]\Big)}{(3+(3+2e)\alpha)^2}\Bigg\}\Bigg|^{\varphi_i}_{\varphi_f}.
\label{m23}    
\end{equation}
From Eq.(\ref{d5}), the spectral parameters of the model is obtained as
\begin{eqnarray}
n_s\simeq1-\frac{18\Big(6(1 +\alpha)(1+\alpha+e\alpha)+N(3+(3+2e)\alpha)^2\Big)^2}{\Big(3(1+\alpha+\alpha)+N(3+(3+2e)\alpha)\Big)^2\Big(3(1+\alpha)+N(6+(6 +4e)\alpha)\Big)^2}-\hspace{3cm}\nonumber\\
-\frac{1}{N^4(3+(3+2e)\alpha)^4\Big(1+\alpha+e\alpha+\frac{3(1+\alpha)(1+\alpha+e\alpha)}{2N(3+(3+2e)\alpha)}\Big)^2\Big(1 +\alpha+\frac{3(1+\alpha)(1 +\alpha+e\alpha)}{N(3+(3+2e)\alpha)}\Big)^2}\times\hspace{1cm}\nonumber\\
\times\bigg\{(1+\alpha)^2 (1+\alpha+e\alpha)^2\Big(-108(1+\alpha)^2(1+\alpha+e\alpha)^2-27N(1+\alpha)\times\hspace{1cm}\nonumber\\\times(1+\alpha+e\alpha)(3+ (3+2e)\alpha)^2-3N^2(1+\alpha+2e\alpha)^2(3+\hspace{1cm}\nonumber\\+(3+2e)\alpha)^2+2N^3(3+(3+2e)\alpha)^4\Big)\bigg\},
\label{m24}    
\end{eqnarray}
\begin{equation}
r\simeq\frac{48\Big(6(1 +\alpha)(1+\alpha+e\alpha)+N(3+(3+2e)\alpha)^2\Big)^2}{\Big(3(1+\alpha+\alpha)+N(3+(3+2e)\alpha)\Big)^2\Big(3(1+\alpha)+N(6+(6 +4e)\alpha)\Big)^2}.
\label{m25}    
\end{equation}
Fig.\ref{fig16} presents the consistency relation $r=r(n_{s}) $ coming from the marginalized joint 68\% and 95\% CL regions of Planck 2018 alone and in combination with BK14 or the BK14+BAO datasets \cite{Planck:2018jri} on the $T(1-e^{\frac{1}{T}})$ model (\ref{m7}) for $\alpha=0.01$. We find that the model predicts the spectral parameters $n_{s}=0.958505$, $r=0.00464303$ and $n_{s}=0.965618$ and $r=0.0032421$ for $N=50$ and $N=60$, respectively. Note that compared to the case $N=50$, the model for $N=60$ provides more (less) favoured value of $n_s$ ($r$). 


\subsection{The Hubble-Slow-Roll Inflation}
The above analysis can be performed in the Jordan frame without using the conformal transformation for  the $f(T)$ gravity. In such a case, we work with the HSR approach for some conventional inflationary potentials in which the $f(T)$ function takes the well-known form 
\begin{equation}
f(T)=\alpha T+\beta T^{2}\,,
\label{n1}    
\end{equation}
where $\alpha$ and $\beta$ are the constant parameters of the model. This model can be immediately compared with the analogous Starobinsky one. Thus, the dynamical Eqs. (\ref{b3}) can be rewritten as
\begin{equation}
\alpha H^{2}-18\beta H^{4}=\frac{1}{3}\rho,\hspace{1cm}36\beta H^{2}\dot{H}-\alpha\dot{H}=\frac{1}{2}(\rho+p),
\label{n2}
\end{equation}
where dot refers to the time derivative. Also, we used $T=-6H^2$. Using the energy density and pressure of a single scalar field (\ref{g3}) and under the slow-roll approximation $\dot{\varphi}^{2}/2\ll V(\varphi)$, we have
\begin{equation}
H^{2}=\frac{3\alpha\pm\sqrt{9\alpha^{2}-4(54\beta) V}}{2(54\beta)}
\label{n3}.
\end{equation}
Then, by choosing the negative sign, the slow-roll parameters of the model (\ref{g4}) are 
\begin{equation}
\epsilon_H=-\frac{\sqrt{2(54\beta)^{3}}V'\dot{\varphi}}{\sqrt{(9\alpha^{2}-4(54\beta) V)(3\alpha-\sqrt{9\alpha^{2}-4(54\beta) V})^{3}}},
\label{n4}
\end{equation}
\begin{eqnarray}
&\!&\!\eta_H=-\frac{\sqrt{2(54\beta) }}{2V'\dot{\varphi}(9\alpha^{2}-4(54\beta) V)\sqrt{\Big(3\alpha-\sqrt{9\alpha^{2}-4(54\beta) V}\Big)^{3}}}\Bigg\{\bigg((9\alpha^{2}-4(54\beta) V)V''\Big(3\alpha-\sqrt{9\alpha^{2}-4(54\beta) V}\Big)+\nonumber\\&\!&\!
+3(54\beta) V'^{2}\Big(2\alpha-\sqrt{9\alpha^{2}-4(54\beta) V}\Big)\bigg)\dot{\varphi}^{2}+\ddot{\varphi}V'(9\alpha^{2}-4(54\beta) V)\Big(3\alpha-\sqrt{9\alpha^{2}-4(54\beta) V}\Big)\Bigg\},
\label{n5}
\end{eqnarray}
where the prime is  the derivative with respect to the scalar field $\varphi$. Also, from Eq.(\ref{g5}), the number of \textit{e}-folds of the model can be found as
\begin{equation}
N=\int^{t_{e}}_{t_{i}}{Hdt}=\int^{\varphi_{f}}_{\varphi_{i}}{\sqrt{\frac{3\alpha-\sqrt{9\alpha^{2}-4(54\beta) V}}{2(54\beta)\dot{\varphi}^{2}}}d\varphi},
\label{n6}
\end{equation}
where $\varphi_{i}$ and $\varphi_{f}$ are the values of inflaton at the start and the end of inflation, respectively. The spectral index and the tensor-to-scalar ratio are introduced in (\ref{g6}). In the following, we study the model for some usual inflationary potentials.
\begin{figure*}[!hbtp]
	\centering
	\includegraphics[width=.45\textwidth,keepaspectratio]{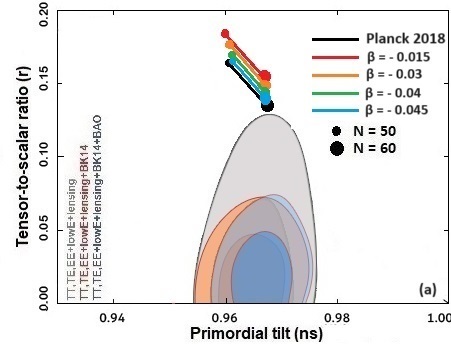}
	\includegraphics[width=.45\textwidth,keepaspectratio]{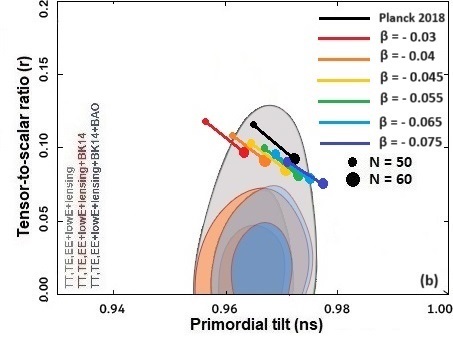}
	\includegraphics[width=.45\textwidth,keepaspectratio]{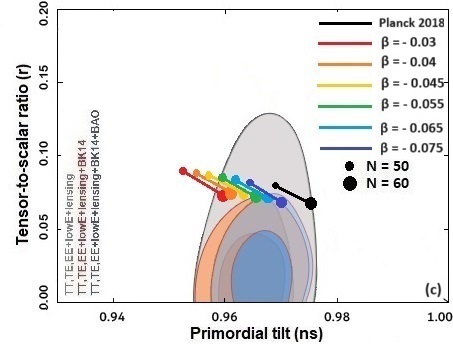}
    \includegraphics[width=.45\textwidth,keepaspectratio]{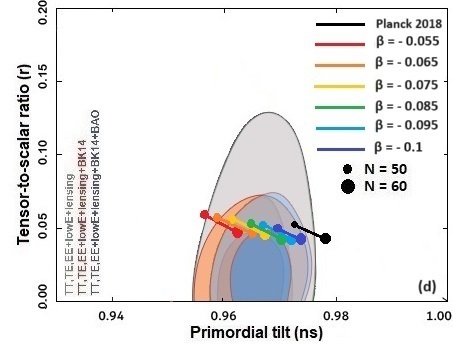}
	\caption{The marginalized joint 68\% and 95\% CL regions for $n_{s}$ and $r$ at $k = 0.002$ Mpc$^{-1}$ from Planck alone and in combination with BK14 or BK14+BAO \cite{Planck:2018jri} and the $n_{s}-r$ constraints on the parameter space of the monomial potential (\ref{g7}) in $f(T)$ gravity associated to the case $n=2$ (a), $n=4/3$ (b), $n=1$ (c) and $n=2/3$ (d). The results are obtained for the case that parameters $\alpha$ and $\lambda$ are considered as the unit.}
	\label{fig17}
\end{figure*}
\subsubsection{Monomial Potential}
For the monomial potential introduced in (\ref{g7}), by using the slow-roll parameters (\ref{n4}) and (\ref{n5}), the reduced Klein-Gordon equation $3H\dot{\varphi}\simeq-V'$ and the number of \textit{e}-folds (\ref{n6}), the spectral parameters (\ref{g6}) for different powers of $n$ can be found as follows.\\
\paragraph{$n=2$.}
In the case  $n=2$, spectral index and tensor-to-scalar ratio are  
\begin{equation}
n_{s}\simeq1+\frac{9\alpha}{4(54\beta)\lambda N^{2}}+\frac{2(16N^{2}(54\beta)^{2}\lambda^{2}-81\alpha^{2})}{N\mathcal{A}^{2}},\hspace{1cm}r\simeq\frac{8(-16N^{2}(54\beta)^{2}\lambda^{2}+81\alpha^{2})}{N\mathcal{A}^{2}},
\label{n7}    
\end{equation}
where $\mathcal{A}=-9\alpha+4(54\beta)\lambda N$.\\
\paragraph{$n=\frac{4}{3}$.}
In the case  $n=\frac{4}{3}$, spectral index and tensor-to-scalar ratio are 
\begin{equation}
n_{s}\simeq1-\frac{1536(54\beta)^2\lambda^2\sqrt{3N}\sqrt[4]{-(54\beta)\lambda}}{\sqrt{\mathcal{A}}(-27\alpha+\sqrt{\mathcal{A}})^{2}}-12(54\beta)\lambda\sqrt{\frac{3}{N}}\frac{\Big(64(54\beta)\lambda N\sqrt{-(54\beta)\lambda}\sqrt{\mathcal{A}}+243\alpha^{2}\sqrt{\mathcal{A}}-6561\alpha^{3}\Big)}{\sqrt[4]{-(54\beta)\lambda}\mathcal{A}(-27\alpha+\sqrt{\mathcal{A}})^{2}},
\label{n8}
\end{equation}
\begin{equation}
r\simeq\frac{6144(54\beta)^2\lambda^2\sqrt{3N}\sqrt[4]{-(54\beta)\lambda}}{\sqrt{\mathcal{A}}(-27\alpha+\sqrt{\mathcal{A}})^{2}},
\label{n9}
\end{equation}
where $\mathcal{A}=192 N\sqrt{(-(54\beta)\lambda)^3}+729\alpha^{2}$.
\\
\paragraph{$n=1$.}
In the case  $n=1$, spectral index and tensor-to-scalar ratio are 
\begin{equation}
n_{s}\simeq1-\frac{8(54\beta)^{2}\lambda^{2}}{3\sqrt{\mathcal{A}}(-3\alpha+\sqrt{\mathcal{A}})^{2}}+\frac{4(54\beta)^{2}\lambda^{2}\Big(-2(54\beta)\lambda\sqrt[3]{16N^{2}((54\beta)\lambda)^4}-27\alpha^{2}+9\alpha\sqrt{\mathcal{A}}\Big)}{3\mathcal{A}(-3\alpha+\sqrt{\mathcal{A}})^{3}},\hspace{0.5cm}
r\simeq\frac{32(54\beta)^{2}\lambda^{2}}{3\sqrt{\mathcal{A}}(-3\alpha+\sqrt{\mathcal{A}})^{2}},
\label{n10}    
\end{equation}
where $\mathcal{A}=\sqrt[3]{16N^{2}((54\beta)\lambda)^4}+9\alpha^{2}$.\\
\paragraph{$n=\frac{2}{3}$.}
In the case $n=\frac{2}{3}$, spectral index and tensor-to-scalar ratio are  
\begin{eqnarray}
&\!&\!n_{s}\simeq1-\frac{48(54\beta)^2\lambda^2\sqrt[5]{3\times10^3}}{5\sqrt{\mathcal{A}}\sqrt[5]{(N\sqrt{-(54\beta)\lambda})^{2}}(-9\alpha+\sqrt{\mathcal{A}})^2}-\frac{4\sqrt[5]{90}(54\beta)\lambda}{5\mathcal{A}\sqrt[5]{(N\sqrt{-(54\beta)\lambda})^4}(-9\alpha+\sqrt{\mathcal{A}})^2}\times\nonumber\\&\!&\!\times\bigg(\sqrt{\mathcal{A}}\big(8(54\beta)\lambda\sqrt[5]{(90N\sqrt{-(54\beta)\lambda})^{2}}-81\alpha^2\big)-54\alpha(54\beta)\lambda\sqrt[5]{(90N\sqrt{-(54\beta)\lambda})^{2}}+729\alpha^3\bigg),
\label{n11}
\end{eqnarray}
\begin{equation}
r\simeq\frac{192(54\beta)^2\lambda^2\sqrt[5]{3\times10^3}}{5\sqrt{\mathcal{A}}\sqrt[5]{(N\sqrt{-(54\beta)\lambda})^{2}}(-9\alpha+\sqrt{\mathcal{A}})^2},
\label{n12}    
\end{equation}
where $\mathcal{A}=-4(54\beta)\lambda\sqrt[5]{(90N\sqrt{-(54\beta)\lambda})^{2}}+81\alpha^2$.

\begin{figure*}[!hbtp]
	\centering
	\includegraphics[width=.45\textwidth,keepaspectratio]{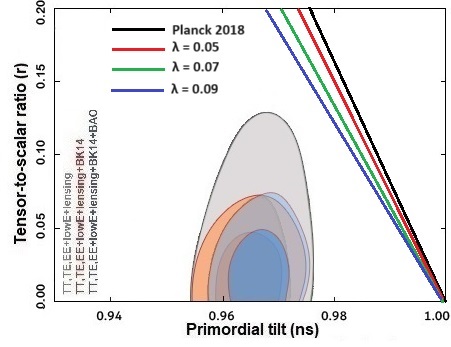}
	\caption{The marginalized joint 68\% and 95\% CL regions for $n_{s}$ and $r$ at $k = 0.002$ Mpc$^{-1}$ from Planck alone and in combination with BK14 or BK14+BAO \cite{Planck:2018jri} and the $n_{s}-r$ constraints on the parameter space of power-law inflation (\ref{k1}) in $f(T)$ gravity for three values $\lambda=0.05, 0.07, 0.09$. The results are obtained for the case that the parameters $\alpha$ and $V_0$ are considered as the unit.}
	\label{fig18}
\end{figure*}
Fig.\ref{fig17} shows the $n_{s}-r$ constraints coming from the marginalized
joint 68\% and 95\% CL regions of Planck 2018 in combination with the BK14+BAO datasets \cite{Planck:2018jri} on the monomial potential studied in the context of $f(T)$ gravity. Panels reveal predictions of the model for the cases $n=2,4/3,1,2/3$ in comparison with the Planck 2018 release (black line) for different values of $\beta$ for $N=50$ (small circle) and $N=60$ (big circle). 

From panel (a), we find that the obtained values of $n_{s}$ and $r$ related to the monomial potential with $n=2$ are situated out of the observational regions. Hence, the model with $n=2$ in $f(T)$ gravity is completely ruled out by CMB observations analogous to its counterpart in GR (black line). By having a look at the Panel (b), one can see that the predicted values of $n_s$ and $r$ for the monomial potential with $n=4/3$ associated to the range $-0.075\leq\beta\leq-0.03$ are in good agreement with Planck alone at 68\% CL, while this observational consistency is put in question at 95\% CL. Moreover, the model does not show any compatibility with CMB observations coming from Planck 2018 combined with BK14 and BAO at both 68\% and 95\% CL. Besides the above information, the panel shows that the monomial potential with $n=4/3$ in $f(T)$ gravity predicts more favoured values of $n_s$ and $r$ in comparison with the model in GR (black line). For the monomial potential with $n=1$ shown in panel (c) and by considering Planck 2018 alone at 68\% CL, we find the obtained values of $n_s$ and $r$ related to $-0.075\leq\beta\leq-0.055$ are fully compatible with the observations for both $N=50$ and $N=60$, while $-0.045\leq\beta\leq-0.03$ predicts favoured values of $n_s$ and $r$ only for $N=60$. Similar to the previous panel, the model is almost incompatible with CMB anisotropies observations when Planck 2018 is combined with BK14 and BAO at 95\% CL. Moreover, the monomial potential with $n=1$ in $f(T)$ gravity is more compatible with the observations compared to the model in GR (black line). Panel (d) shows the observational constraint $-0.1\leq\beta\leq-0.055$ for the monomial potential with $n=2/3$ in the case of Planck 2018 alone at both 68\% and 95\% CL, while this constraint reduces to $-0.1<\beta\leq-0.055$ in the case of Planck 2018 combined with BK14 and BAO at 68\% CL. Moreover, the panel tells us that the predictions of the monomial potential with $n=2/3$ in $f(T)$ gravity are more compatible with CMB observations rather than GR (black line). Remind that all the above results are obtained for the case that parameters $\alpha$ and $\lambda$ are assumed as the unit.

\subsubsection{Power-Law Inflation}
For the exponential potential (\ref{k1}), by combining the slow-roll parameters (\ref{n4}) and (\ref{n5}), the reduced relation $3H\dot{\varphi}\simeq-V'$ and the obtained number of \textit{e}-folds (\ref{n6}), the spectral parameters (\ref{g6}) of the model can be found as 
\begin{eqnarray}
&\!&\!n_{s}\simeq1-\frac{\lambda^2 e^{2\lambda^2\alpha N}\big(e^{\lambda^2\alpha N}+ 6\alpha\big)^2}{6(-3\alpha+ \sqrt{\mathcal{A}})^2\sqrt{\mathcal{A}}}-\nonumber\\&\!&\!
-\frac{e^{\lambda^2\alpha N}(e^{\lambda^2\alpha N}+ 6\alpha)\lambda^2\Big(9e^{\lambda^2\alpha N}(e^{\lambda^2\alpha N}+ 6\alpha)\alpha-2e^{\lambda^2\alpha N}(e^{\lambda^2\alpha N}+ 6\alpha)\sqrt{\mathcal{A}}+108\alpha^3- 36\alpha^2\sqrt{\mathcal{A}}\Big)}{12\mathcal{A}(-3\alpha+\sqrt{\mathcal{A}})^2},
\label{n13}    
\end{eqnarray}
\begin{equation}
r\simeq\frac{2\lambda^2 e^{2\lambda^2\alpha N}\big(e^{\lambda^2\alpha N}+ 6\alpha\big)^2}{3\sqrt{\mathcal{A}}(-3\alpha+ \sqrt{\mathcal{A}})^2},
\label{n14}    
\end{equation}
where $\mathcal{A}=e^{\lambda^2\alpha N}\Big(e^{\lambda^2\alpha N}+6\alpha\Big)+9\alpha^2$.

Fig.\ref{fig18} presents the consistency relation $r=r(n_{s}) $ coming from the marginalized joint 68\% and 95\% CL regions of Planck 2018 alone and in combination with BK14 or the BK14+BAO datasets \cite{Planck:2018jri} on the power-law inflation for three values $\lambda=0.05, 0.07, 0.09$ compared to its counterpart in GR (black line). From the figure, we find that the obtained values of $n_{s}$ and $r$ related to the considered values of $\lambda$ are not compatible with CMB anisotropies observations. Hence, the power-law inflation in the context of $f(T)$ theory is ruled out analogously to Planck 2018 prediction (black line). 
\begin{figure*}[!hbtp]
	\centering
	\includegraphics[width=.55\textwidth,keepaspectratio]{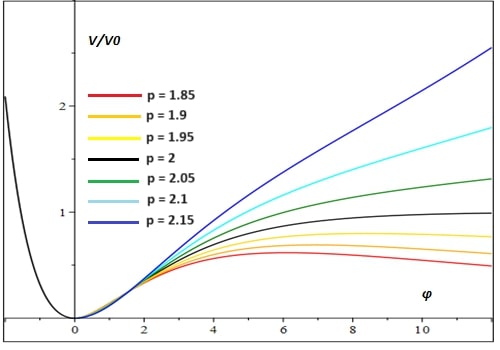}
	\caption{The potential of $Q^{p}$ model (\ref{o1}) for different powers of $p$.}
	\label{fig19}
\end{figure*}
\section{Inflation in $f(Q)$ gravity}\label{fq}
Finally, we investigate  inflation  in the context of $f(Q)$ non-metric gravity from the two perspectives of PSR and HSR.

\subsection{The Potential-Slow-Roll Inflation}
We study the PSR approach to inflation using the potentials associated to different forms of the function $Q$ driven after using the conformal transformation when the role of ordinary matter is neglected $\mathcal{L}_m=0$.
\subsubsection{The $f(Q)=Q+\alpha Q^p$ model}
We first study the power-law function of $Q^{p}$ introduced in Ref.\cite{BeltranJimenez:2019tme}
\begin{equation}
f(Q)=Q+\alpha Q^p,
\label{o1}
\end{equation}
where $\alpha=-\lambda (6M^2)^{1-p}$ while $p$ and $\lambda$ are dimensionless parameters. Here $M$ is a mass scale. Depending on the choice of the parameter $p$, the model mimics the early and late-time accelerating phases of the universe. Thus, $p<1$ is related to the low-curvature regime (suitable for DE) and $p>1$ is relevant to the high-curvature regime (suitable for inflation)  \cite{BeltranJimenez:2019tme}. Adapting  the conformal transformation (\ref{a11}), we obtain the potential corresponding to the above $f(Q)$, that is 
\begin{equation}
V(\varphi)=V_{0}e^{-2\sqrt{\frac{1}{5}}\varphi}(e^{\sqrt{\frac{1}{5}}\varphi}-1)^{\frac{p}{p-1}},
\label{o2}
\end{equation}
where $V_{0}=3M^2(1-p)p^{\frac{p}{1-p}}(-\lambda)^{\frac{1}{1-p}}$. Clearly, we have to consider the constraints discussed in Sec.\ref{emag} for the validity of conformal transformations in non-metric gravity. In Fig.\ref{fig19}, we present the behaviour of the potential (\ref{o2}) for different values of parameter $p$. As above, we have two main classes of models: i) For $p>2$, the potential shows a maximum  around $\varphi_m =\sqrt{5}\ln\big({\frac{2(p-1)}{p-2}}\big)$ so that it is approaching to zero for large value of $\varphi$. ii) For $p<2$, the potential increases gradually but its decreasing  towards zero is  steeper than the $Q^2$ model. iii) For $p=2$, one can obtain the potential of $f(Q)$ gravity proposed in \cite{BeltranJimenez:2019tme} which is relevant for the inflationary era with a given potential $V(\varphi)=\frac{3M^2}{4\lambda}(1-e^{-\sqrt{\frac{1}{5}}\varphi})^2$. 
\begin{figure*}[!hbtp]
	\centering
	\includegraphics[width=.45\textwidth,keepaspectratio]{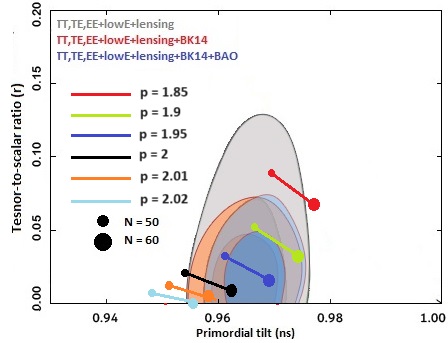}
\caption{The marginalized joint 68\% and 95\% CL regions for $n_{s}$ and $r$ at $k = 0.002$ Mpc$^{-1}$ from Planck alone and in combination with BK14 or BK14+BAO \cite{Planck:2018jri} and the $n_{s}-r$ constraints on the $Q^{p}$ model (\ref{o1}) for two cases $p=2$ and $p\neq2$.}
	\label{fig20}
\end{figure*}

Now, by plugging the potential (\ref{o2}) into Eqs. (\ref{d3}), the slow-roll parameters of the model in the case of $p\neq2$ can be found as 
\begin{equation}
\epsilon_V=\frac{\big((p-2)e^{\sqrt{\frac{1}{5}}\varphi}-2(p-1)\big)^{2}}{10(p-1)^{2}\big(e^{\sqrt{\frac{1}{5}}\varphi}-1\big)^{2}},\hspace{1cm}\eta_V=\frac{(-5p^{2}+13p-8)e^{\sqrt{\frac{1}{5}}\varphi}+(p-2)^{2}e^{2\sqrt{\frac{1}{5}}\varphi}+4(p-1)^{2}}{5(p-1)^{2}\big(e^{\sqrt{\frac{1}{5}}\varphi}-1\big)^{2}}.
\label{o3}
\end{equation}
By setting $\epsilon_V=1$, we find the value of $\varphi$ at the end of inflation as
\begin{equation}
\varphi_{f}=\sqrt{5}\ln\Big(\frac{(p-1)(\sqrt{10}-2)}{\sqrt{10}(p-1)-(p-2)}\Big).
\label{o4}
\end{equation}
Also, the number of \textit{e}-folds (\ref{d4}) of the model takes the form
\begin{equation}
N\simeq-\frac{5p}{2(p-2)}\ln\bigg(\frac{(p-2)e^{\sqrt{\frac{1}{5}}\varphi_{i}}}{2(1-p)}+1\bigg),
\label{o5}
\end{equation}
and the spectral parameters (\ref{d5}) of the model are 
\begin{equation}
n_{s}\simeq\frac{4p(-4p+3)\mathcal{A}+(16p^{2}-24p+4)\mathcal{A}^2+p^{2}+8p}{5\Big((2p-2)\mathcal{A}-p\Big)^2},\hspace{1cm}r\simeq\frac{32(p-2)^{2}\mathcal{A}^2}{5\Big((2p-2)\mathcal{A}-p\Big)^2},
\label{o6}    
\end{equation}
where $\mathcal{A}=e^{-\frac{2(p-2)N}{5p}}$. 

For the case of $p=2$, the slow-roll parameters of the model are reduced to 
\begin{equation}
\epsilon_V=\frac{2}{5\big(e^{\sqrt{\frac{1}{5}}\varphi}-1\big)^{2}},\hspace{1cm}\eta_V=\frac{2\big(2-e^{\sqrt{\frac{1}{5}}\varphi}\big)}{5\big(e^{\sqrt{\frac{1}{5}}\varphi}-1\big)^{2}}.
\label{o7}    
\end{equation}
Then, by setting the condition $\epsilon_V=1$, we have
\begin{equation}
\varphi_{f}=\sqrt{5}\ln\Big(1+\sqrt{\frac{2}{5}}\Big).
\label{o8}
\end{equation}
Also, the number of \textit{e}-folds in this case is given by
\begin{equation}
N\simeq\frac{5}{2}e^{\sqrt{\frac{1}{5}}\varphi_{i}}.
\label{o9}    
\end{equation}
The spectral parameters are found as
\begin{equation}
n_{s}\simeq\frac{4N^{2}-28N+5}{(2N-5)^{2}},\hspace{1cm}r\simeq\frac{160}{(2N-5)^{2}}.
\label{o10}    
\end{equation}
Fig.\ref{fig20} shows the $n_{s}-r$ constraints coming from the marginalized joint 68\% and 95\% CL regions of Planck 2018 data in combination with the BK14+BAO datasets \cite{Planck:2018jri} on the $Q^p$ model (\ref{o1}) for the cases $p=2$ and $p\neq2$. By  a quick look at the plot, we find that the $Q^2$ model shows the match with observational values of $n_{s}$ and $r$ for $N=60$ at both 68\% and 95\% CL of all three CMB observational datasets. Regarding the Planck alone dataset, one can see that the obtained values of $n_{s}$ and $r$ related to the case $1.85\leq p\leq2.02$ are in good agreement with the observations at 68\% CL, while this consistency reduces to the case $1.9\leq p\leq2.01$ at 95\% CL. For Planck 2018 combined with the BK14 dataset, we realize that predictions of the model in the cases $1.9\leq p\leq2.02$ and $1.95\leq p\leq2.01$ are compatible with CMB observations at 68\% and 95\% C.L., respectively. For a full consideration of CMB anisotropy observations Planck+BK14+BAO, the observational constraints on the parameter $p$ are reduced to $1.9\leq p\leq2.01$ and $1.9<p<2.01$ at 68\% and 95\% C.L., respectively. Besides these results, the $Q^p$ model with a tiny deviation from $p=2$ predicts more favoured values of $n_{s}$ and $r$ in comparison with the case $p=2$.
\begin{figure*}[!hbtp]
	\centering
	\includegraphics[width=.55\textwidth,keepaspectratio]{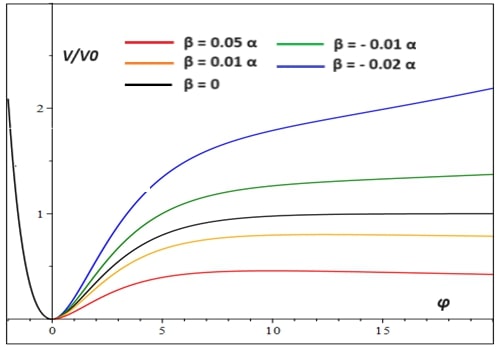}
	\caption{The potential of the logarithmic corrected $f(Q)$ model (\ref{p1}) for different values of $\beta$ and $\alpha\sim10^{-4}$.}
	\label{fig21}
\end{figure*}
\subsubsection{The $f(Q)=Q+\alpha Q^{2}+\beta Q^{2}\ln Q$  model}
As a generalized form of $f(Q)$ quadratic model, we introduce the  logarithmic $f(Q)$ model 
\begin{equation}
f(Q)=Q+\alpha Q^{2}+\beta Q^{2}\ln Q,
\label{p1}
\end{equation}
where $\alpha$, $\beta$ are the parameters of the model. It is clear that we recover the $Q^2$ model when $\alpha=-\frac{\lambda}{6M^2}$ and $\beta=0$. Then, from Eq.(\ref{c13}), the associated potential in the Einstein frame can be found as
\begin{equation}
V(\varphi)=-\frac{(\alpha+\beta)Q^{2}(1+\frac{\beta}{\alpha+\beta}\ln Q)}{2\Big(1+(2\alpha+\beta)Q(1+\frac{2\beta}{2\alpha+\beta}\ln Q)\Big)^{2}}.
\label{p2}    
\end{equation}
By using the definition of $F$ (\ref{c13}), $Q$  is given by
\begin{equation}
Q=\frac{e^{\sqrt{\frac{1}{5}}\varphi}-1}{2\beta W_{k}(X)},\hspace{1cm}\mbox{where}\hspace{1cm}X\equiv\frac{e^{\sqrt{\frac{1}{5}}\varphi}-1}{2\beta}e^{\frac{2\alpha+\beta}{2\beta}},
\label{p3}  
\end{equation}
where $W_{k}$ is the Lambert function with real branches $k=0, -1$. Inserting the obtained expression of the non-metricity scalar (\ref{p3}) into the reconstructed potential (\ref{p2}), the potential of the model takes the following form
\begin{equation}
V(\varphi)=-(1-e^{-\sqrt{\frac{1}{5}}\varphi})^2\frac{1+2W_k(X)}{16\beta W_k(X)^2}.
\label{p4}
\end{equation}
This is the exact potential  compatible with large values of  $\beta$. For small values \ie $|\beta|\ll\alpha$ and by using the iterative method, up to the leading order in $Q$, the potential of the logarithmic $f(Q)$ model is given by
\begin{equation}
V(\varphi)=V_{0}\frac{(1-e^{-\sqrt{\frac{1}{5}}\varphi})^{2}}{1+\frac{\beta}{2\beta}+\frac{\beta}{\alpha}\ln\Big(\frac{e^{\sqrt{\frac{1}{5}}\varphi}-1}{2\alpha}\Big)},
\label{p5}  
\end{equation}
where $V_{0}=-\frac{1}{8\alpha}$. Notice that the potential (\ref{p4}) reduces to the potential (\ref{p5}) when the Lambert function expands in the limit $W_k(X)\gg1$ for $|\beta|\ll\alpha$. In Fig.\ref{fig21}, we present the behaviour of the obtained potential (\ref{p5}) for different values of $\beta$ and $\alpha\sim\mathcal{O}(10^{-4})$ in which $\alpha=0$ recovers the quadratic $f(Q)$ model discussed in the previous section. For $\beta<0$, the potential tilts upward resulting larger values of the scalar-to-tensor ratio $r$, while for $\beta>0$, the potential shows an unstable extremum and then it runs away for larger values of $\varphi$.
\begin{figure*}[!hbtp]
	\centering
	\includegraphics[width=.45\textwidth,keepaspectratio]{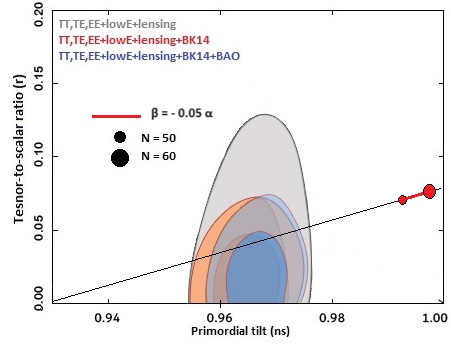}
\caption{The marginalized joint 68\% and 95\% CL regions for $n_{s}$ and $r$ at $k = 0.002$ Mpc$^{-1}$ from Planck alone and in combination with BK14 or BK14+BAO \cite{Planck:2018jri} and the $n_{s}-r$ constraints on the logarithmic corrected $f(Q)$ model (\ref{p1}) for $\beta=-0.05\alpha$ and $\alpha\sim10^{-4}$.}
	\label{fig22}
\end{figure*}
Now using Eq.(\ref{d3}), the slow-roll parameters of the logarithmic $f(Q)$ model can be calculated as
\begin{equation}
\epsilon_V=\frac{2\Big(2\beta\ln(\frac{e^{\sqrt{\frac{1}{5}}\varphi}-1}{2\alpha})+(1-e^{\sqrt{\frac{1}{5}}\varphi})\beta+2\alpha\Big)^{2}}{5\Big(2\beta\ln(\frac{e^{\sqrt{\frac{1}{5}}\varphi}-1}{2\alpha})+\beta+2\alpha\Big)^{2}\big(e^{\sqrt{\frac{1}{5}}\varphi}-1\big)^{2}},
\label{p6}    
\end{equation}
\begin{eqnarray}
&\!&\!\eta_V=\frac{1}{5\Big(2\beta\ln(\frac{e^{\sqrt{\frac{1}{5}}\varphi}-1}{2\alpha})+\beta+2\alpha\Big)^{2}\big(e^{\sqrt{\frac{1}{5}}\varphi}-1\big)^{2}}\Bigg\{-8\beta^{2}(e^{\sqrt{\frac{1}{5}}\varphi}-2)\ln\Big(\frac{e^{\sqrt{\frac{1}{5}}\varphi}-1}{\alpha}\Big)^{2}-4\Big((-4\beta\ln({2})+4\alpha+5\beta)\times
\nonumber\\&\!&\!
\times e^{\sqrt{\frac{1}{5}}\varphi}+8\beta\ln({2})-8\alpha-4\beta\Big)\beta\ln\big(\frac{e^{\sqrt{\frac{1}{5}}\varphi}-1}{\alpha}\big)-4\big(-2\beta\ln(2)+2\alpha+\beta\big)\big(-\beta\ln(2)+\alpha+2\beta\big)e^{\sqrt{\frac{1}{5}}\varphi}+8e^{2\sqrt{\frac{1}{5}}\varphi}\beta^{2}+\nonumber\\&\!&\!
+4\big(2\beta\ln(2)-2\alpha-\beta\big)^{2}\Bigg\}.
\label{p7}    
\end{eqnarray}
By setting $\epsilon_V=1$ and then using the Taylor expansion under the condition $\big|\frac{e^{\sqrt{\frac{1}{5}}\varphi_{f}}-1}{2\alpha}-1\big|\leq1$, we find an estimation for the value of inflaton when the inflationary period ends, in the limit $|\beta|\ll\alpha$, as
\begin{equation}
\varphi_{f}\simeq\sqrt{5}\ln\Bigg\{\frac{1}{10}\Bigg(5-\frac{10\alpha^2}{\beta}+\sqrt{5}\Bigg[\sqrt{2}(1-\alpha)\pm\sqrt{7+2\sqrt{10}+\frac{2\alpha}{\beta^2}\Big(10\alpha^3+\alpha\beta\big(10+2\sqrt{10}(1+\alpha)\big)+\beta^2(\alpha-2-\sqrt{10})\Big)}\Bigg]\Bigg)\Bigg\},
\label{p8}
\end{equation}
and also the number of \textit{e}-folds of the model is found as
\begin{equation}
N\simeq-\frac{3\ln\big(\frac{1-\frac{\sigma^{2}}{3\hat{\epsilon}}}{1-\frac{\sigma^{2}}{3}}\big)-6\tanh^{-1}(\frac{\sigma}{\sqrt{3\hat{\epsilon}}})}{\sigma(2+\sigma)}\hspace{1cm}\text{where}\hspace{1cm}\sigma\equiv\frac{\beta\Big(1+\frac{\beta}{2\alpha}+\frac{\beta}{\alpha}\ln\big(\frac{e^{\sqrt{\frac{1}{5}}\varphi}-1}{2\alpha}\big)\Big)^{-1}}{\alpha},
\label{p9}    
\end{equation}
where $\hat{\epsilon}$ is the first slow-roll parameter of the quadratic $f(Q)$ model shown in Eq.(\ref{o7}). Then, the spectral parameters of the logarithmic $f(Q)$ model can be obtained as
\begin{eqnarray}
&\!&\!n_{s}\simeq1-\frac{12\Big(2\beta\ln(\frac{\mathcal{A}-1}{2\alpha})+(1-\mathcal{A})\beta+2\alpha\Big)^{2}}{5\Big(2\beta\ln(\frac{\mathcal{A}-1}{2\alpha})+\beta+2\alpha\Big)^{2}\big(\mathcal{A}-1\big)^{2}}+\frac{2}{5\Big(2\beta\ln(\frac{\mathcal{A}-1}{2\alpha})+\beta+2\alpha\Big)^{2}\big(\mathcal{A}-1\big)^{2}}\times\nonumber\\&\!&\!
\times\Bigg\{-8\beta^{2}(\mathcal{A}-2)\ln\Big(\frac{\mathcal{A}-1}{\alpha}\Big)^{2}+4\beta\Big((4\beta\ln({2})-4\alpha-5\beta)\mathcal{A}-8\beta\ln({2})+8\alpha+4\beta\Big)\ln\big(\frac{\mathcal{A}-1}{\alpha}\big)-\nonumber\\&\!&\!
+4\big(2\beta\ln(2)-2\alpha-\beta\big)\big(\beta\ln(2)-\alpha-2\beta\big)\mathcal{A}+8\mathcal{A}^2\beta^2+4\big(2\beta\ln(2)-2\alpha-\beta\big)^{2}\Bigg\},
\label{p10}    
\end{eqnarray}
\begin{equation}
r\simeq\frac{32\Big(2\beta\ln(\frac{\mathcal{A}-1}{2\alpha})+(1-\mathcal{A})\beta+2\alpha\Big)^{2}}{5\Big(2\beta\ln(\frac{\mathcal{A}-1}{2\alpha})+\beta+2\alpha\Big)^{2}\big(\mathcal{A}-1\big)^{2}},
\label{p11}    
\end{equation}
where the quantities are defined as
\begin{equation}
\mathcal{A}=\frac{2\alpha}{\beta}\Big(-e^{-\frac{N\beta}{3\alpha}}\mathcal{B}\mp1\Big)+1,\hspace{0.7cm}\mathcal{B}=1+\frac{\beta}{2\alpha}\Big(e^{\sqrt{\frac{1}{5}}\varphi_{f}}-1\Big).
\label{p12}
\end{equation}
Fig.\ref{fig22} shows the $n_{s}-r$ constraints coming from the marginalized joint 68\% and 95\% CL regions of Planck 2018 data in combination with the BK14+BAO datasets \cite{Planck:2018jri} on the logarithmic $f(Q)$ model (\ref{p1}) for $\beta=-0.05\alpha$ and $\alpha\sim10^{-4}$. The figure shows the values $n_{s}=0.99371$, $r=0.062015$ and  $n_{s}=0.99877$, $r=0.074833$ for $N=50$ and $N=60$, respectively. In comparison with the $Q^2$ model studied in the previous section, we see that the presence of the logarithmic correction provides more favoured values of the tensor-to-scalar ratio. However, the price we pay for this is high because we lose the validity of the spectral index.  

\subsection{The Hubble-Slow-Roll Inflation}
Besides the above approach, we can accomplish the inflationary analysis in $f(Q)$ theory in the Jordan frame without using the conformal transformation. To this aim, we work with the Hubble parameter instead of the reconstructed potential in the context of the HSR approach by assuming the function $f(Q)$ 
\begin{equation}
f(Q)=\alpha Q+\beta Q^{2},
\label{q1}    
\end{equation}
where $\alpha$ and $\beta$ are the free parameters of the model. By using $Q=6H^{2}$, \cite{Capozziello:2022wgl}, we can rewrite the dynamical Eqs. (\ref{c3}) as
\begin{equation}
\alpha H^{2}+17\beta H^{4}=\frac{1}{3}\rho,\hspace{1cm}\alpha\dot{H}+36\beta H^{2}\dot{H}=-\frac{1}{2}(\rho+p),
\label{q2}
\end{equation}
where the dot denotes the derivative concerning cosmic time $t$. Recalling energy density and pressure of a single scalar field (\ref{g3})  and using the slow-roll condition $\ddot{\varphi}\ll H\dot{\varphi}$ and $\frac{\dot{\varphi}^{2}}{2}\ll V(\varphi)$, we have
\begin{equation}
H^{2}=\frac{-3\alpha\pm\sqrt{9\alpha^{2}+4(51\beta) V}}{2(51\beta)}.
\label{q3}
\end{equation}
By choosing the positive sign of the above expression, the slow-roll parameters (\ref{g4}) of the model are obtained as
\begin{equation}
\epsilon_H=-\frac{\sqrt{2(51\beta)^{3}}V'\dot{\varphi}}{\sqrt{(9\alpha^{2}+4(51\beta) V)(-3\alpha+\sqrt{9\alpha^{2}+4(51\beta) V})^{3}}},
\label{q4}
\end{equation}
\begin{eqnarray}
&\!&\!\eta_H=-\frac{\sqrt{2(51\beta) }}{2V'\dot{\varphi}(9\alpha^{2}+4(51\beta) V)\sqrt{\Big(-3\alpha+\sqrt{9\alpha^{2}+4(51\beta) V}\Big)^{3}}}\Bigg\{\bigg(-(9\alpha^{2}+4(51\beta) V)V''\Big(3\alpha-\sqrt{9\alpha^{2}+4(51\beta) V}\Big)+\nonumber\\&\!&\!
+3(51\beta) V'^{2}\Big(2\alpha-\sqrt{9\alpha^{2}+4(51\beta) V}\Big)\bigg)\dot{\varphi}^{2}-\ddot{\varphi}V'(9\alpha^{2}+4(51\beta) V)\Big(3\alpha-\sqrt{9\alpha^{2}+4(51\beta) V}\Big)\Bigg\},
\label{q5}
\end{eqnarray}
where the prime represents the derivative with respect to the inflaton field $\varphi$. From Eq.(\ref{g5}), the number of \textit{e}-folds  is 
\begin{equation}
N=\int^{\varphi_{f}}_{\varphi_{i}}{\sqrt{\frac{-3\alpha+\sqrt{9\alpha^{2}+4(51\beta) V}}{2(51\beta)\dot{\varphi}^{2}}}d\varphi},
\label{q6}
\end{equation}
where the subscribes $i$ and $f$ correspond to the values of inflaton at the start and end of inflation, respectively. Also, the spectral parameters $n_s$ and $r$ are introduced in (\ref{g6}). Now, let us study the model for some conventional forms of inflationary potential. 

\begin{figure*}[!hbtp]
	\centering
	\includegraphics[width=.45\textwidth,keepaspectratio]{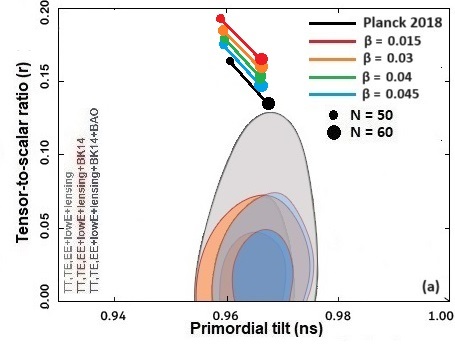}
	\includegraphics[width=.45\textwidth,keepaspectratio]{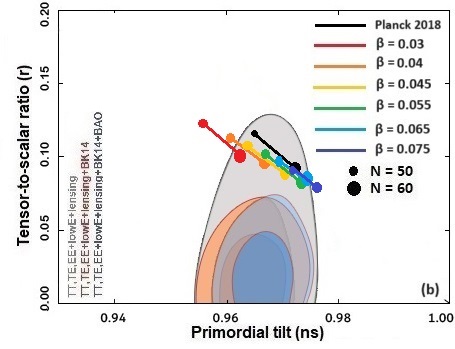}
	\includegraphics[width=.45\textwidth,keepaspectratio]{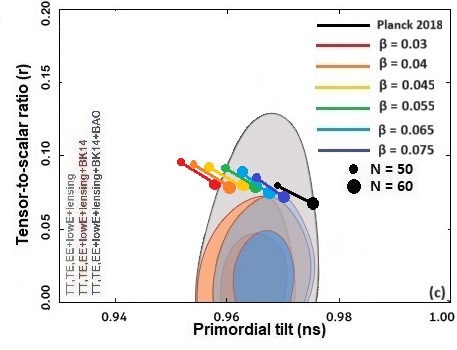}
    \includegraphics[width=.45\textwidth,keepaspectratio]{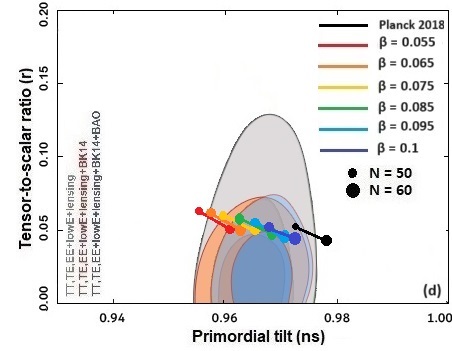}
	\caption{The marginalized joint 68\% and 95\% CL regions for $n_{s}$ and $r$ at $k = 0.002$ Mpc$^{-1}$ from Planck alone and in combination with BK14 or BK14+BAO \cite{Planck:2018jri} and the $n_{s}-r$ constraints on the parameter space of the monomial potential (\ref{g7}) in $f(Q)$ gravity associated to the case $n=2$ (a), $n=4/3$ (b), $n=1$ (c) and $n=2/3$ (d). The results are obtained for the case that parameters $\alpha$ and $\lambda$ are considered as the unit.}
	\label{fig23}
\end{figure*}

\subsubsection{Monomial Potential}
In the case of the monomial potential (\ref{g7}), by
using the slow-roll parameters (\ref{q4}) and (\ref{q5}), the relation $3H\dot{\varphi}\simeq-V'$ and the number of \textit{e}-folds (\ref{q6}),
one can obtain the spectral parameters (\ref{g6}) for different powers of $n$ as follows.\\
\paragraph{$n=2$.}
In the case  $n=2$, spectral index and tensor-to-scalar ratio are  
\begin{equation}
n_{s}\simeq1-\frac{9\alpha}{4(51\beta)\lambda N^{2}}-\frac{2(16N^{2}(51\beta)^{2}\lambda^{2}-81\alpha^{2})}{N\mathcal{A}^{2}},\hspace{1cm}r\simeq\frac{8(16N^{2}(51\beta)^{2}\lambda^{2}-81\alpha^{2})}{N\mathcal{A}^{2}},
\label{q7}    
\end{equation}
where $\mathcal{A}=-9\alpha+4(51\beta)\lambda N$.\\
\paragraph{$n=\frac{4}{3}$.}
In the case  $n=\frac{4}{3}$, spectral index and tensor-to-scalar ratio are 
\begin{equation}
n_{s}\simeq1-\frac{1536\sqrt{3N}\sqrt[4]{((51\beta)\lambda)^{9}}}{\sqrt{\mathcal{A}}(-27\alpha+\sqrt{\mathcal{A}})^{2}}-12\sqrt{\frac{3}{N}}\sqrt[4]{((51\beta)\lambda)^{3}}\frac{\Big(64\sqrt{((51\beta)\lambda)^{3}}N\sqrt{\mathcal{A}}-243\alpha^{2}\sqrt{\mathcal{A}}+6561\alpha^{3}\Big)}{\mathcal{A}(-27\alpha+\sqrt{\mathcal{A}})^{2}},
\label{q8}
\end{equation}
\begin{equation}
r\simeq\frac{6144\sqrt{3N}\sqrt[4]{((51\beta)\lambda)^{9}}}{\sqrt{\mathcal{A}}(-27\alpha+\sqrt{\mathcal{A}})^{2}},
\label{q9}
\end{equation}
where $\mathcal{A}=192N\sqrt{((51\beta)\lambda)^{3}}+729\alpha^{2}$.
\\
\paragraph{$n=1$}
In the case  $n=1$, spectral index and tensor-to-scalar ratio are 
\begin{equation}
n_{s}\simeq1-\frac{8(51\beta)^{2}\lambda^{2}}{3\sqrt{\mathcal{A}}(-3\alpha+\sqrt{\mathcal{A}})^{2}}+\frac{4(51\beta)^{2}\lambda^{2}\Big(-2(51\beta)\lambda\sqrt[3]{16N^{2}(51\beta)\lambda}-27\alpha^{2}+9\alpha\sqrt{\mathcal{A}}\Big)}{3\mathcal{A}(-3\alpha+\sqrt{\mathcal{A}})^{3}},\hspace{1cm}
r\simeq\frac{32(51\beta)^{2}\lambda^{2}}{3\sqrt{\mathcal{A}}(-3\alpha+\sqrt{\mathcal{A}})^{2}},
\label{q10}    
\end{equation}
where $\mathcal{A}=\sqrt[3]{16N^{2}((51\beta)\lambda})^4+9\alpha^{2}$.\\
\paragraph{$n=\frac{2}{3}$}
In the case $n=\frac{2}{3}$, spectral index and tensor-to-scalar ratio are  
\begin{eqnarray}
&\!&\!n_{s}\simeq1-\frac{48(51\beta)^2\lambda^2\sqrt[5]{3\times10^3}}{5\sqrt{\mathcal{A}}\sqrt[5]{(N\sqrt{(51\beta)\lambda})^{2}}(-9\alpha+\sqrt{\mathcal{A}})^2}-\frac{4\sqrt[5]{90}(51\beta)\lambda}{5\mathcal{A}\sqrt[5]{(N\sqrt{(51\beta)\lambda})^4}(-9\alpha+\sqrt{\mathcal{A}})^2}\times\nonumber\\&\!&\!\times\bigg(\sqrt{\mathcal{A}}\big(8(51\beta)\lambda\sqrt[5]{(90N\sqrt{(51\beta)\lambda})^{2}}+81\alpha^2\big)-54\alpha(51\beta)\lambda\sqrt[5]{(90N\sqrt{(51\beta)\lambda})^{2}}-729\alpha^3\bigg),
\label{q11}
\end{eqnarray}
\begin{equation}
r\simeq\frac{192(51\beta)^2\lambda^2\sqrt[5]{3\times10^3}}{5\sqrt{\mathcal{A}}\sqrt[5]{(N\sqrt{(51\beta)\lambda})^{2}}(-9\alpha+\sqrt{\mathcal{A}})^2},
\label{q12}    
\end{equation}
where $\mathcal{A}=4(51\beta)\lambda\sqrt[5]{(90N\sqrt{(51\beta)\lambda})^{2}}+81\alpha^2$.\\
In Fig.\ref{fig23}, we present the $n_{s}-r$ constraints coming from the marginalized
joint 68\% and 95\% CL regions of Planck 2018 in combination with the BK14+BAO datasets \cite{Planck:2018jri} on the monomial potential studied in the context of $f(Q)$ gravity. Panels show predictions of the model for the cases $n=2,4/3,1,2/3$ in comparison with the Planck 2018 release (black line) for different values of $\beta$ for $N=50$ (small circle) and $N=60$ (big circle). From panel (a), we realize that predictions of the model for $n_s$ and $r$ are not in good agreement with CMB observations. Hence, the monomial model with $n=2$ in $f(Q)$ gravity is excluded by the observations analogous to the model in the context of GR (black line). From panel (b), we find that the monomial potential with $n=4/3$ predicts favoured values of $n_s$ and $r$ in the range $0.03<\beta\leq0.075$ in comparison with Planck alone at 68\% CL. Moreover, the model is not consistent with a full package of the CMB observational datasets Planck+BK14+BAO at both 68\% and 95\% CL. Additionally, the panel reveals that the monomial potential with $n=4/3$ in $f(Q)$ gravity provides more favoured values of $n_s$ and $r$ rather than the model in GR (black line). For the monomial potential with $n=1$, panel (c) shows the predicted values of $n_s$ and $r$ associated to the range $\beta\geq0.055$ are in good agreement with Planck 2018 alone at 68\% CL for both $N=50$ and $N=60$, while $0.03<\beta\leq0.045$ predicts favoured values of $n_s$ and $r$ only for $N=60$. Moreover, the model is ruled out by CMB observations coming from the Planck 2018+BK14+BAO datasets at 95\% CL. Note that the monomial potential with $n=1$ in $f(Q)$ gravity is more consistent with CMB anisotropies observations compared to the model in GR (black line). Panel (d) presents the CMB observational constraint $0.055\leq\beta\leq0.1$, coming from Planck 2018 alone at both 68\% and 95\% CL, for the monomial potential with $n=2/3$. Also, the constraint reduces to $0.055<\beta\leq0.1$ in the case of the combined datasets Planck 2018+BAO+BK14 at 68\% CL. Moreover, the panel tells us that the predictions of the monomial potential with $n=2/3$ in $f(Q)$ gravity are more consistent with CMB observations in comparison with the model in GR (black line). Remind that all the above results are obtained for the case that parameters $\alpha$ and $\lambda$ are assumed as the unit.

\begin{table}
\begin{center}
\rotatebox{90}{
\begin{tabular}{c|c|c|c|c|c|c|c}
   \toprule\toprule

   \textbf{Gravity} & \textbf{Model}& \textbf{Condition}&\textbf{CMB Dataset} & $N$ & \textbf{Parameter} &$n_s$& $r_{0.002}$ \\
  \hline
      &  & &Planck 2018 at 68\% CL & 50-60&$0.985\leq p\leq1.002$&$0.95913\leq n_s\leq0.97754$&$0.0036089\leq r\leq0.01159$\\
    \cline{4-8}

      &$R+R^{2p}/M^{4p-2}$  &------& Planck 2018 at 95\% CL & 50-60&$0.99\leq p\leq1.001$&$0.96046\leq n_s\leq0.97308$&$0.0038865\leq r\leq0.0084$\\
   \cline{4-8}
   &     & &Planck+BK14+BAO at 68\% CL&50-60&$0.985\leq p\leq1.001$ &$0.96046\leq n_s\leq0.97754$&$0.0038865\leq r\leq0.01159$\\
       \cline{4-8}
    &   & &Planck+BK14+BAO at 95\% CL&50-60& $0.99\leq p\leq0.995$&$0.96784\leq n_s\leq0.97309$&$0.0059789\leq r\leq0.0084$\\
  \cline{2-8}
   $f(R)$ &   & &&50&$\beta=-0.08\alpha$&0.96324&0.0027253
   \\
   \cline{5-8}
   & $R+R^2(\alpha+\beta\ln R)$ &$\alpha\sim10^{-9}$ &Planck+BK14+BAO at 95\% CL&60&$\beta=-0.08\alpha$&0.97482&0.0049489\\  \cline{2-8}
   &  & &&50&$M=10^{-3}m$&0.96126&0.0028081\\
   \cline{5-8}
   &   $R+(m^4/3M^2)\Big[R/m^2-\ln(1+R/m^2)\Big]$ &$m\sim10^{-3}$& Planck+BK14+BAO at 95\% CL
   &60&$M=10^{-3}m$&0.96612&0.0023928\\
    \hline\hline 
         &  & &Planck 2018 at 68\% CL & 50-60&$1.97\leq p\leq2.005$&$0.95844\leq n_s\leq0.97753$&$0.0034768\leq r\leq0.011595$\\
    \cline{4-8}

      & $T+\alpha(-T)^{p}$  &------& Planck 2018 at 95\% CL & 50-60&$1.975\leq p\leq2.004$&$0.95831\leq n_s\leq0.9731$&$0.0038213\leq r\leq0.010621$\\
   \cline{4-8}
   $f(T)$ &    && Planck+BK14+BAO at 68\% CL&50-60&$1.97\leq p\leq2.001$  &$0.96112\leq n_s\leq0.97753$&$0.0040323\leq r\leq0.011595$\\
       \cline{4-8}
    &   & &Planck+BK14+BAO at 95\% CL&50-60& $1.98\leq p\leq1.99$&$0.96784\leq n_s\leq0.9731$&$0.0059789\leq r\leq0.008401$\\
  \cline{2-8}
   &   & & &50&$\alpha=0.01$&0.958505&0.00464303\\
   \cline{5-8}
   & $T+\alpha T(1-e^{\frac{\beta}{T}})$ & ------&Planck+BK14+BAO at 95\% CL&60&$\alpha=0.01$&0.965618&0.0032421\\ 

    \hline\hline

        &  & &Planck 2018 at 68\% CL & 50-60&$1.85\leq p\leq2.02$&$0.95407\leq n_s\leq0.97411$&$0.011876\leq r\leq0.07536$\\
    \cline{4-8}

      & $Q-\lambda Q^{p}/M^{2p-2}$  & ------&Planck 2018 at 95\% CL & 50-60&$1.9\leq p\leq2.01$&$0.95857\leq n_s\leq0.97137$&$0.013312\leq r\leq0.04486$\\
   \cline{4-8}
   $f(Q)$ &  &  & Planck+BK14+BAO at 68\% CL&50-60&$1.9\leq p\leq2.01$&$0.95857\leq n_s\leq0.97137$&$0.013312\leq r\leq0.04486$\\
       \cline{4-8}
    &   & &Planck+BK14+BAO at 95\% CL&50-60&$1.9<p<2.01$&$0.95857< n_s<0.97137$&$0.013312< r<0.04486$\\
  \cline{2-8}
   &   & &&50&$\beta=-0.05\alpha$&0.99371&0.062015\\
   \cline{5-8}
   & $Q+Q^2(\alpha+\beta\ln Q)$ &$\alpha\sim10^{-4}$ &Planck+BK14+BAO at 95\% CL&60&$\beta=-0.05\alpha$&0.99877&0.074833\\  \cline{2-8}
    \cline{1-7}
       
\end{tabular}
}
 \caption{Comparing the inflationary models in extended metric-affine gravities $f(R)$, $f(T)$ and $f(Q)$ from the PSR perspective.}
  \label{tab1}
\end{center}

\end{table}
\subsubsection{Power-Law Inflation}
For the exponential potential (\ref{k1}), by combining the slow-roll parameters (\ref{q4}) and (\ref{q5}), the reduced Klein-Gordon equation $3H\dot{\varphi}\simeq-V'$ and the obtained number of \textit{e}-folds (\ref{q6}), the spectral parameters (\ref{g6}) of the model are shown in eqs.(\ref{n13}) and (\ref{n14}). We notice that the final expressions of $n_s$ and $r$ are $\beta$-independent, so the prediction of the power-law inflation in $f(Q)$ gravity is similar to its counterpart in $f(T)$ gravity. Consequently, the exponential potential in $f(Q)$ gravity is excluded by CMB observations.    

\section{A Comparison Between Inflationary Models}\label{comparison}
In this section, we present a qualitative comparison between the three considered gravitational theories $f(R)$, $f(T)$ and $f(Q)$ regarding the inflationary results obtained in the previous sections. Let us first review all results summarized in Tables \ref{tab1} and \ref{tab2}. 
\begin{itemize}
    \item \textbf{Inflation in $f(R)$ gravity}
    \begin{itemize}
        \item{\textbf{The PSR Approach.} Based on the reconstructed potential for some accredited functions of the Ricci scalar $R$ in the Einstein frame.}
        \begin{itemize}
         \item{\textit{$f(R)=R+\alpha R^{2p}$.}}\\
         As a generalization of the $R^2$ Starobinsky model, we study the $R^{2p}$ model, with real values of $p$, by comparing predictions of the model with CMB observations. In a non-trivial case $p\neq1$, predictions of the model for $n_{s}$ and $r$, in the range $0.985\leq p\leq1.002$, are compatible with Planck 2018 alone datasets at 68\% CL, while the range reduces to $0.99\leq p\leq1.001$ at 95\% CL. For a combination of Planck 2018 with BK14 and BAO, the observational constraints on the power $p$ turn to $0.985\leq p\leq1.001$ and $0.99\leq p\leq0.995$ at 68\% and 95\% C.L., respectively. Therefore, in order to raise the tensor-to-scalar ratio of the Starobinsky model, a large deviation from $p=1$ is ruled out by the observations.\\
         \item{\textit{$f(R)=R+ R^{2}(\alpha+\beta\ln R)$.}}\\ 
         As an interesting generalized version of the $R^2$ Starobinsky model, we consider a logarithmic correction $R^2\ln R$ to the $R^2$ model. By comparing the results with CMB observations, the model, in the case of $\beta=-0.08\alpha$, predicts a more (less) favoured value of $n_s$ ($r$) compared to the $R^2$ Starobinsky model when the number of \textit{e}-folds is assumed to be $\sim50$, while the model shows an inverse prediction for $n_s$ and $r$ in the case of $N=60$. Moreover, the price we pay for increasing the tensor-to-scalar ratio $r$ of the $R^2$ Starobinsky model by considering a logarithmic term, is high because of losing the validity of spectral index $n_s$. This result is also pursued in the case of $\beta=-0.1\alpha$ through the obtained values $r\sim10^{-2}$ and $n_s\sim0.98$.\\
         \item{\textit{$f(R)=R+\alpha\big(\beta R-\ln(1+\beta R)\big)$.}}\\
         As a viable $f(R)$ inflationary model, we investigate a power-law model corrected with the logarithmic term $R-\ln(1+R)$ and then we compare the results with CMB anisotropies datasets. For allowed values of the two parameters $\alpha$ and $\beta$, the model predicts $n_{s}=0.96126$, $r=0.002808$ and $n_{s}=0.96612$, $r=0.0023928$ for $N=50$ and $N=60$, respectively. Therefore, the smallness of tensor-to-scalar ratio $r$ in $f(R)$ gravity still remains as an ambiguity even in the presence of such logarithmic corrections.
        \end{itemize}
        \item{\textbf{The HSR Approach.} Based on the Hubble parameter corresponding to some usual inflationary potentials for the function $f(R)=\alpha R+\beta R^2$ in the Jordan frame.}
        \begin{itemize}
        \item{\textit{Monomial Potential.}} \\
         We study the monomial potential $\lambda \varphi^n$, with integer or fractional power $n$, in the context of $f(R)$ gravity by comparing the results with CMB observations.\\
         The inflationary analysis of the model with $n=2$ is divided into two parts related to keeping the terms up to the first and second orders of the binomial series representation of the number of \textit{e}-folds. i) By keeping only the first term of the used series, the predicted values of $n_s$ and $r$, in the range $0.026\leq\beta\leq0.03$, are well-consistent with Planck 2018 alone datasets at 68\% CL, while the model is fully excluded by a full consideration Planck+BK14+BAO. ii) By keeping the terms up to the second orders of the series, the predictions of the model are almost similar to the case (i) for $0.023\leq\beta\leq0.03$. Remarkably, the monomial potential with $n=2$, which is ruled out by the observations in GR, is returned to the playground by the effects of $f(R)$ gravity.\\
         The inflationary analysis of the model with $n=4/3$ is divided into two parts related to a non-zero and zero integral constant $C$. i) By considering the non-zero constant $C=1$, the model provides favoured values of $n_s$ and $r$, in the range $0.08\leq\beta\leq0.15$, compatible with Planck 2018 alone at 68\% CL. For a combination of Planck 2018 with BK14 and BAO, the model is not in good agreement with CMB anisotropies datasets. Moreover, we find that the monomial potential with $n=4/3$ in $f(R)$ gravity predicts the smaller (larger) values of $n_s$ ($r$) in the range $\beta\leq0.11$ compared to its counterpart in GR. ii) By considering the integration constant $C$ as zero, the obtained values of $n_s$ and $r$ in the range $0.008\leq\beta\leq0.09$ are more compatible with Planck alone at 68\% CL. Also, the monomial potential with $n=4/3$ in $f(R)$ gravity provides more favoured values of $n_s$ and $r$, in particular for $N=50$, rather than the predictions of the model in GR.\\
         The inflationary analysis of the model with $n=1$ is divided into two parts related to a negative and positive values of $\beta$. i) By assuming the negative values of $\beta$, the produced values of $n_s$ and $r$, in the range $-0.5\leq\beta\leq-0.2$, are well consistent with Planck alone and the full consideration Planck+BK14+BAO at 68\% CL. ii) By assuming the negative values of $\beta$, we find the observational constraint $0.2\leq\beta\leq0.3$ with a less consistency with CMB observations related to the case $\beta<0$. In comparison with GR prediction, the monomial potential with $n=1$ in $f(R)$ gravity shows more (less) favoured values of $n_s$ and $r$ when $\beta<0$ ($\beta>0$) is considered.\\
         The inflationary analysis of the model with $n=2/3$ is divided into two parts related to a non-zero and zero integral constant $C$. i) By considering the non-zero constant $C=1$, the model shows the values of $n_s$ and $r$ associated to the range $0.3\leq\beta\leq0.7$ for $N=50$ are consistent with Planck alone at 68\% CL but inconsistent with the combined datasets Planck+BK14+BAO. ii) By considering the integration constant $C$ as zero, we find the observational constraint $0.03\leq\beta\leq0.2$ for $N=50$ coming from Planck 2018 alone and its combination with BK14 and BAO at 68\% CL. In comparison with GR, the monomial potential with $n=2/3$ in $f(R)$ gravity has less consistency with the observations for both cases (i) and (ii).\\
         \item{\textit{Power-Law Inflation.}}\\
         We study the exponential potential $e^{-\lambda\varphi}$ in the context of $f(R)$ gravity by compering the results with CMB observations. For $\lambda=0.02$, the model is fully ruled out by the observations, while increasing the value of $\lambda$ helps the model to predict observationally-favoured values of $n_s$ and $r$. For $\lambda=0.03$, the predictions of the power-law inflation, in the range $0.05\leq\beta\leq0.1$, are in good agreement with Planck alone and Planck+BK14+BAO at 68\% CL and 95\% CL. Interestingly, although the power-law inflation in GR is excluded by CMB observations, it returns to the inflationary literature in the presence the higher order terms of the curvature.
        
        \end{itemize}
    \end{itemize}
     \item \textbf{Inflation in $f(T)$ gravity}
    \begin{itemize}
        \item{\textbf{The PSR Approach.}} Based on the reconstructed potential for some accredited functions of the torsion scalar $T$ in the Einstein frame.
        \begin{itemize}
        \item{\textit{$f(T)=T+\alpha(-T)^p$.}}\\
        As one of the most conventional models in $f(T)$ gravity, we study the inflation  for a power-law function of the torsion scalar $T$ as $(-T)^{p}$ describing a de Sitter universe. By comparison of the results with Planck 2018 alone release at 68\% CL, the model predicts observationally favoured values of $n_s$ and $r$ in the range $1.97\leq p\leq2.005$, while for a full combination of Planck 2018 with BK14 and BAO, the observational constraint on the power $p$ reduces to $1.97\leq p\leq2.001$ and $1.98\leq p\leq1.99$ at 68\% and 95\% CL, respectively.\\
        
        \begin{table}
\begin{center}
\rotatebox{90}{
\begin{tabular}{c|c|c|c|c|c|c|c}
   \toprule\toprule

   \textbf{Gravity} & \textbf{Model}&\textbf{Condition}&\textbf{CMB Dataset} & $N$ & \textbf{Parameter} &$n_s$& $r_{0.002}$ \\
  \hline
     &  & First Order of Series&Planck 2018 at 68\% CL&50-60&$0.026\leq\beta\leq 0.03$&$0.96744\leq n_s\leq0.98028$&$0.079222\leq r\leq0.12989$\\
    \cline{3-8}
      &$\lambda\varphi^2$ &Second Order of Series &Planck 2018 at 68\% CL&50-60&$0.023\leq\beta\leq0.03$&$0.96367\leq n_s\leq0.98116$&$0.075944\leq r\leq0.14358$\\

  \cline{2-8}
        & &$\lambda=1$, $\alpha=1$, $C=1$ &Planck 2018 at 68\% CL &50-60&$0.08\leq \beta\leq0.15$&$0.96752\leq n_s\leq0.97714$&$0.092503\leq r\leq0.13094$\\

    \cline{3-8}
      &$\lambda\varphi^{4/3}$ &  $\lambda=1$, $\alpha=1$, $C=0$&Planck 2018 at 68\% CL&50-60&$0.008\leq \beta\leq0.09$&$0.97042\leq n_s\leq0.97891$&$0.067637\leq r\leq0.094647$\\

    \cline{2-8}
         $f(R)=\alpha R+\beta R^2$& &  $\lambda=1$, $\alpha=1$, $C=1$, $\beta<0$& Planck+BK14+BAO at 68\% CL&50&$-0.5\leq \beta\leq-0.2$&$0.97323\leq n_s\leq0.97712$&$0.06675\leq r\leq0.071547$\\
    \cline{3-8}
      & $\lambda\varphi$ & $\lambda=1$, $\alpha=1$, $C=1$, $\beta>0$&Planck 2018 at 68\% CL&50&$0.2\leq\beta\leq0.3$&$0.97238\leq n_s\leq0.97765$&$0.093069\leq r\leq0.11383$\\

     \cline{2-8}
       & & $\lambda=1$, $\alpha=1$, $C=1$ &Planck 2018 at 68\% CL&50&$0.3\leq\beta\leq0.7$&$0.97072\leq n_s\leq0.97883$&$0.089541\leq r\leq0.12109$\\
   
    \cline{3-8}
      & $\lambda\varphi^{2/3}$&$\lambda=1$, $\alpha=1$, $C=0$  &Planck+BK14+BAO at 68\% CL&50&$0.03\leq\beta\leq0.2$&$0.97604\leq n_s\leq0.97822$&$0.047922\leq r\leq0.087016$\\
   
   \cline{2-8}
     & &$\lambda=0.01$, $\alpha=1$, $C=1$, $V_0=1$ &Any combination  &50-60&Ruled out&----------&---------\\
     
    \cline{3-8}
      & $e^{-\lambda\varphi}$&$\lambda=0.02$, $\alpha=1$, $C=1$, $V_0=1$ &Any combination&50-60&Ruled out&-----------&---------\\
    
      \cline{3-8}
      & & $\lambda=0.03$, $\alpha=1$, $C=1$, $V_0=1$&Planck+BK14+BAO  at 98\% CL &50-60&$0.05\leq\beta\leq0.1$&$0.96427\leq n_s\leq0.97482$&$0.025871\leq r\leq0.10112$\\
    
   \hline\hline

      & $\lambda\varphi^2$  &$\lambda=1$, $\alpha=1$&Any combination  & 50-60&Ruled out&-----------&---------\\
  \cline{2-8}

      & $\lambda\varphi^{4/3}$ &$\lambda=1$, $\alpha=1$ & Planck 2018 at 68\% CL &50-60&$-0.075\leq\beta\leq-0.03$&$0.96062\leq n_s\leq0.96708$&$0.088015\leq r\leq0.10614$\\

    \cline{2-8}

      & &&&50-60&$-0.075\leq\beta\leq-0.055$&$0.96427\leq n_s\leq0.96635$&$0.070689\leq r\leq0.075305$\\  
    \cline{5-8}
      $f(T)=\alpha T+\beta T^2$& $\lambda\varphi$  & $\lambda=1$, $\alpha=1$&Planck 2018 at 68\% CL & 60&$-0.1<\beta\leq-0.055$&$0.95618\leq n_s\leq0.96233$&$0.080847\leq r\leq0.096208$\\
     \cline{2-8}
      & $\lambda\varphi^{2/3}$ &$\lambda=1$, $\alpha=1$&Planck+BK14+BAO at 68\% CL&50-60&$-0.1<\beta\leq-0.055$&$0.96106\leq n_s\leq0.96676$&$0.050419\leq r\leq0.061746$\\
   \cline{2-8}
     &&$\lambda=0.05$,  $\alpha=1$, $V_0=1$  & Any combination  &50-60&Ruled out&-----------&---------\\
     
    \cline{3-8}
     &$e^{-\lambda\varphi}$  &$\lambda=0.07$, $\alpha=1$, $V_0=1$ & Any combination  &50-60&Ruled out&-----------&---------\\
    \cline{3-8}
      &  & $\lambda=0.09$, $\alpha=1$, $V_0=1$ & Any combination  &50-60&Ruled out&-----------&---------\\
    \hline\hline
    & $\lambda\varphi^2$  &$\lambda=1$, $\alpha=1$&Any combination  & 50-60&Ruled out&-----------&---------\\
  \cline{2-8}
      & $\lambda\varphi^{4/3}$ &$\lambda=1$, $\alpha=1$ & Planck 2018 at 68\% CL &50-60&$0.03<\beta\leq0.075$&$0.95872\leq n_s\leq0.96615$&$0.088428\leq r\leq0.10723$\\
    \cline{2-8}
      & &&&50-60&$\beta\geq0.055$&$0.96356\leq n_s\leq0.96504$&$0.070752\leq r\leq0.075409$\\  
    \cline{5-8}
      $f(Q)=\alpha Q+\beta Q^2$& $\lambda\varphi$  & $\lambda=1$, $\alpha=1$&Planck 2018 at 68\% CL & 60&$0.03<\beta\leq0.045$&$0.95529\leq n_s\leq0.96128$&$0.080987\leq r\leq0.096371$\\
     \cline{2-8}
      & $\lambda\varphi^{2/3}$ &$\lambda=1$, $\alpha=1$&Planck+BK14+BAO at 68\% CL& 50-60&$0.055<\beta\leq0.1$&$0.96002\leq n_s\leq0.96546$&$0.050549\leq r\leq0.061891$\\
      \cline{2-8}
     &&$\lambda=0.05$, $\alpha=1$, $V_0=1$ & Any combination  &50-60&Ruled out&-----------&---------\\
    \cline{3-8}
     &$e^{-\lambda\varphi}$  &$\lambda=0.07$, $\alpha=1$, $V_0=1$ & Any combination  &50-60&Ruled out&-----------&---------\\
    \cline{3-8}
      &  & $\lambda=0.09$, $\alpha=1$, $V_0=1$ & Any combination  &50-60&Ruled out&-----------&---------\\
    \hline

\end{tabular}
}
\caption{Comparing the inflationary models in extended metric-affine gravities $f(R)$, $f(T)$ and $f(Q)$ from the HSR perspective.}
  \label{tab2}
\end{center}
\end{table}

\item{\textit{$f(T)=T+\alpha T(1-e^{\frac{\beta}{T}})$}.}\\
         As the second case, we perform the inflationary analysis for a power-law function of $T$ that is contaminated by an exponential term. By comparing the results with CMB observations, the model predicts the spectral parameters $n_{s}=0.958505$, $r=0.00464303$ and $n_{s}=0.965618$ and $r=0.0032421$ for $N=50$ and $N=60$, respectively. Compared to the case $N=50$, the model for  $N=60$ provides more (less) favoured values of $n_s$ and $r$.
        \end{itemize}
        
        \item{\textbf{The HSR Approach.}} Based on the Hubble parameter corresponding to some usual inflationary potentials for the function $f(T)=\alpha T+\beta T^2$ in the Jordan frame.
          \begin{itemize}
         \item{\textit{Monomial Potential.}}\\
            We study the monomial potential $\lambda \varphi^n$, with integer or fractional power $n$, in the context of $f(T)$ gravity by comparing the results with CMB observations.\\
            For the monomial potential with $n=2$, we find that the model is completely excluded by any combination of CMB observations analogous to its counterpart in GR. For the monomial potential with $n=4/3$, the obtained values of $n_s$ and $r$ related to the range $-0.075\leq\beta\leq-0.03$ are consistent with Planck alone at 68\% CL and inconsistent with CMB observations coming from Planck+BK14+BAO at both 68\% and 95\% CL. As the final result, the monomial potential with $n=4/3$ in $f(T)$ gravity predicts more favoured values of $n_s$ and $r$ compared to the model in GR. For the monomial potential with $n=1$, we find the observational constraint $-0.075\leq\beta\leq-0.055$  ($-0.045\leq\beta\leq-0.03$) by considering Planck 2018 alone at 68\% CL for both $N=50$ and $N=60$ (only for $N=60$). Note that the monomial potential with $n=1$ in $f(T)$ gravity is more compatible with the observations in comparison with the model in GR. For the monomial potential with $n=2/3$, the model predicts favoured values of $n_s$ and $r$ in the range $-0.1\leq\beta\leq-0.055$ compatible with  Planck 2018 alone at both 68\% and 95\% CL, while the constraint reduces to $-0.1<\beta\leq-0.055$ in the case of Planck+BK14+BAO at 68\% CL. Moreover, predictions of the monomial potential with $n=2/3$ in $f(T)$ gravity are more consistent with CMB observations rather than predictions of the model in GR.\\
      
         \item{\textit{Power-Law Inflation.}}\\
          We studied the exponential potential $e^{-\lambda\varphi}$ in the context of $f(T)$ gravity by comparing the results with CMB observations in which the model is fully ruled out by any combination of observational datasets analogously to its counterpart in GR.

        \end{itemize}
    \end{itemize}
     \item \textbf{Inflation in $f(Q)$ gravity}
    \begin{itemize}
        \item{\textbf{The PSR Approach}}
        Based on the reconstructed potential for some accredited functions of the non-metricity scalar $Q$ in the Einstein frame.
        \begin{itemize}
        \item{\textit{$f(Q)=Q+\alpha Q^{p}$.}}\\
         As a suitable prescription for both low and high-curvature regimes depending the power $p$, we engage the power-law function of the non-metricity scalar in order to study the inflationary epoch in the context of $f(Q)$ gravity. Considering Planck alone 2018, the obtained values of $n_{s}$ and $r$, associated to the range $1.85\leq p\leq2.02$, are well-consistent with CMB  observations at 68\% CL, while this consistency reduces to the case $1.9\leq p\leq2.01$ at 95\% CL. For a combined datasets Planck+BK14, the observational constraints on the parameter $p$ are obtained $1.9\leq p\leq2.02$ and $1.95\leq p\leq2.01$ at 68\% and 95\% C.L., respectively. For a full consideration Planck+BK14+BAO datasets, the observational constraints on the parameter $p$ turn to $1.9\leq p\leq2.01$ and $1.9<p<2.01$ at 68\% and 95\% C.L., respectively. \\
         \item{\textit{$f(Q)=Q+Q^{2}(\alpha+\beta\ln Q)$.}}\\
         As a generalized form of the quadratic model $Q^2$, we study the model in the presence of the logarithmic term $Q^2\ln Q$ in order to present the inflationary analysis in the Einstein frame. In comparison with CMB anisotropies observations, for $\beta=-0.05\alpha$ and $\alpha\sim10^{-4}$, the model predicts the values  $n_{s}=0.99371$, $r=0.062015$ and  $n_{s}=0.99877$, $r=0.074833$ for $N=50$ and $N=60$, respectively. Compared to the $Q^2$ model, although the logarithmic correction plays a constructive role in producing more favoured values of the tensor-to-scalar ratio, it has a destructive role in losing the validity of the spectral index. 
        \end{itemize}
        \item{\textbf{The HSR Approach}} Based on the Hubble parameter corresponding to some usual inflationary potentials for the function $f(Q)=\alpha Q+\beta Q^2$ in the Jordan frame.
          \begin{itemize}
         \item{\textit{Monomial Potential.}}\\
         We study the monomial potential $\lambda \varphi^n$, with integer or fractional power $n$, in the context of $f(Q)$ gravity by comparing the results with CMB observations.\\
         
         For the monomial potential with $n=2$, we find that the model is excluded by any combination of CMB observations analogous to the model in the context of GR and $f(T)$ gravity. For the monomial potential with $n=4/3$, predictions of the model, related to the range $0.03<\beta\leq0.075$, are just well-consistent with Planck alone at 68\% CL. Moreover, the monomial potential with $n=4/3$ in $f(Q)$ gravity presents more favoured values of $n_s$ and $r$ rather than the model in GR. For the monomial potential with $n=1$, the predicted values of $n_s$ and $r$ associated to the range $\beta\geq0.055$ are in good agreement with Planck 2018 alone at 68\% CL for both $N=50$ and $N=60$, while $0.03<\beta\leq0.045$ predicts favoured values of $n_s$ and $r$ only for $N=60$. Notice that the monomial potential with $n=1$ in $f(Q)$ gravity is more compatible with CMB observations compared to the model in GR. For the monomial potential with $n=2/3$, we find the observational constraint $0.055\leq\beta\leq0.1$, coming from Planck 2018 alone at both 68\% and 95\% CL, while it reduces to $0.055<\beta\leq0.1$ in the case Planck+BAO+BK14 at 68\% CL. Consequently, the monomial potential with $n=2/3$ in $f(Q)$ gravity is more consistent with CMB observations in comparison with the model in GR.\\

         \item{\textit{Power-Law Inflation.}}\\
         We study the exponential potential $e^{-\lambda\varphi}$ in the context of $f(Q)$ gravity by compering the results with CMB observations in which the model is fully ruled out by any combination of observational datasets analogously to its counterpart in GR and $f(T)$ gravity.
        \end{itemize}
    \end{itemize}
\end{itemize}

The above summary points out the fact that some features of inflation are improved or disproved depending on the the gravity representation.  The  PSR and the HSR approaches mean that the analysis can be performed both in the Einstein frame (PSR) or in the Jordan frame (HSR). Clearly, in the first case,  conformal transformations play a fundamental role for disentangling the degrees of freedom that contribute to the inflationary behavior. In this perspective, it is worth saying that a direct information comes only in the curvature case because the conformal transformation can be clearly performed. In the teleparallel and non-metric representations, conformal transformations are affected by couplings between geometric and scalar fields pointing out violation of the Lorentz invariance. In this situation, conformal transformations work only in given regimes allowing the comparison with the standard curvature case. However, as reported in Ref. \cite{Capozziello:2022tvv},  this fact could give rise to interesting behaviors related to the exit from inflation.

On the other hand, the HSR case does not imply the above ambiguities  related to the validity or not of the conformal transformations. Here cosmological systems remain in the Jordan frame and then conditions for inflation are directly given on how the scalar field potentials are related to the Hubble parameter. For example, power-law inflation seems to hold only in a curvature representation of cosmological dynamics, while monomial potentials seem to be more versatile according to the ranges of viable parameters. 

The above tables sum up the strengths and the weaknesses of the three representations of gravity, if directly compared with data.

\section{Discussion and Conclusions}\label{conclusion}
After more than twenty years, the late-time cosmic acceleration or DE still remains one of the biggest puzzles of modern cosmology at large scale. As a straight attempt, we are faced with the $\Lambda$CDM model dealing with the cosmological constant as a new type of matter added to the present components of the universe. Contrarily to the achievements of the model, the existence of a big difference between the value predicted by theory and its observational value navigates us to pursue some alternatives in order to explain DE. As a highly-reputed candidate, modified gravity can help us to have a more complete vision of the DE issue by considering some appropriate modifications to the Einstein gravity  based on the Lovelock theorem. Besides, solving the DE puzzle, we expect such theories to be able to describe other cosmological eras such as inflation, DM, etc. Even  possible unifications between DE and inflation can be addressed in the context of modified theories of gravity. Depending on gravity representation, we encounter a wide range of modified theories of gravity. Adding higher order derivatives of the metric to the Hilbert-Einstein action leads to a widely-used class of modified gravity with more degrees of freedom. As a well-known model of this type of modified gravity, $f(R)$ gravity deals with an arbitrary function of the Ricci scalar $R$, including  fourth-order derivatives without  Ostrogradski ghosts. Working with other geometrical objects like torsion and non-metricity, instead of curvature, guides us to other types of modified gravity. Although teleparallel and non-metric approaches  are dynamically equivalent to GR, their modifications are not necessarily equivalent. Inspired by $f(R)$ gravity, one can work with $f(T)$ and $f(Q)$ gravities as modified versions of TEGR and STEGR with  new possibilities in order to explain cosmological and astrophysical phenomena. Another approach of modified gravity occurs by embedding our 4-dimensional universe in extra spatial dimensions in the context of braneworld gravity. Even assuming new degrees of freedom by adding scalar, vector, or tensor fields to the standard gravitational action non-minimally coupled to gravity produces some geometrical modification to GR.\\
Because of the existing large number of modified theories of gravity, there is confusion about which one can explain gravitational and cosmological situations more satisfactorily. In the present manuscript, we tried to present a comparison between three widely-used modified theories of gravity $f(R)$, $f(T)$ and $f(Q)$ by studying cosmic inflation in the context of these theories. We performed the inflationary analysis in both  PSR and HSR approaches for the three mentioned gravities, separately. From the PSR perspective, we worked with the potential reconstructed from some accredited forms of the function $f$ associated with the gravitational theory under consideration in the Einstein frame. From the HSR perspective, we worked with the Hubble parameter obtained from an inflationary function $f$ for some usual potentials in the Jordan frame. To fulfill the purpose of the paper, we compared the obtained results with CMB anisotropies observations coming from Planck 2018 and BICEP2/Keck array satellites in order to find the observational constraints on the model parameters and their predictions from the spectral parameters. Results are summarized in the above tables. By a rapid inspection, it is clear that observations can reasonably select realistic models in the three representation of gravity. From this point of view, a combined analysis using also gravitational waves, in particular, and  multimessenger information, in general,  would be extremely useful  in view of an accurate discrimination  of theories and representations of  gravity.

\section*{Acknowledgments}
S.C. acknowledges the financial support of the Istituto Nazionale di Fisica Nucleare (INFN) - Sezione di Napoli, \textit{inizative specifiche} QGSKY and MOONLIGHT2. This paper is based upon work from COST Action CA21136 - Addressing observational tensions in cosmology with systematics and fundamental physics (CosmoVerse), supported by COST (European Cooperation in Science and Technology).

\bibliographystyle{ieeetr}
\bibliography{biblo}
\appendix
\end{document}